\documentclass[preprint,12pt]{elsarticle}

\usepackage{amssymb}
\usepackage{amsthm}
\usepackage{xcolor}
\usepackage{graphicx}
\usepackage{lineno,hyperref}
\modulolinenumbers[5]

\usepackage{inputenc}
\usepackage{amsmath}
\usepackage{graphics}
\usepackage{epsfig}
\usepackage{graphicx}
\usepackage{latexsym}
\usepackage{graphics}
\usepackage{epsfig}
\usepackage{amsmath,amssymb,amsfonts,theorem,color,bm}
\usepackage{multirow}
\usepackage{stfloats} 
\usepackage{float}
\usepackage{subfig}
\usepackage{soul}
\usepackage{caption}
\usepackage{hyperref}
\hypersetup{colorlinks,allcolors=black}

\newcommand\bQ{\boldsymbol Q}

\newcommand\bv{\boldsymbol v}
\newcommand\bV{\boldsymbol V}

\newcommand\bW{\boldsymbol W}

\newcommand\bS{\boldsymbol S}
\newcommand\bT{\boldsymbol T}

\newcommand\bSigma{\boldsymbol{\Sigma}}

\newcommand\bR{\boldsymbol{R}}

\numberwithin{equation}{section}

\newcommand{\beqn}{\begin{equation}}
\newcommand{\eeqn}{\end{equation}}
\newcommand{\beqnarr}{\begin{eqnarray}}
\newcommand{\eeqnarr}{\end{eqnarray}}
\newcommand{\baling}{\begin{alignat}{1}}
\newcommand{\ealing}{\end{alignat}}

\usepackage{xcolor,colortbl}
\definecolor{Gray}{gray}{0.75}
\newcolumntype{a}{>{\columncolor{Gray}}c}

\journal{XXX}

\begin{document}

\begin{frontmatter}



\title{LC-SVD-DLinear: A low-cost physics-based hybrid machine learning model for data forecasting using sparse measurements}



\author[UPM]{Ashton Hetherington}

\author[Boeing]{Javier López Leonés}

\author[UPM]{Soledad Le Clainche\footnote{Correspondence to: soledad.leclainche@upm.es}}

\affiliation[UPM]{organization={ETSI Aeronáutica y del Espacio, Universidad Politécnica de Madrid},
            addressline={Plaza Cardenal Cisneros, 3}, 
            city={Madrid},
            postcode={28040}, 
            country={Spain}}

\affiliation[Boeing]{organization={Boeing Aerospace Spain},
            addressline={Av. Sur del Aeropuerto de Barajas, 38}, 
            city={Madrid},
            postcode={28042}, 
            country={Spain}}
            
\begin{abstract}
This article introduces a novel methodology that integrates singular value decomposition (SVD) with a shallow linear neural network for forecasting high resolution fluid mechanics data. The method, termed {\it LC-SVD-DLinear}, combines a low-cost variant of singular value decomposition (LC-SVD) with the {\it DLinear} architecture, which decomposes the input features—specifically, the temporal coefficients—into trend and seasonality components, enabling a shallow neural network to capture the non-linear dynamics of the temporal data. This methodology uses under-resolved data, which can either be input directly into the hybrid model or downsampled from high resolution using two distinct techniques provided by the methodology. Working with under-resolved cases helps reduce the overall computational cost. Additionally, we present a variant of the method, {\it LC-HOSVD-DLinear}, which combines a low-cost version of the high-order singular value decomposition (LC-HOSVD) algorithm with the DLinear network, designed for high-order data. These approaches have been validated using two datasets: first, a numerical simulation of three-dimensional flow past a circular cylinder at $Re = 220$; and second, an experimental dataset of turbulent flow passing a circular cylinder at $Re = 2600$. The combination of these datasets demonstrates the robustness of the method. The forecasting and reconstruction results are evaluated through various error metrics, including uncertainty quantification. The work developed in this article will be included in the next release of ModelFLOWs-app\footnote{The website of the software is available at \href{https://modelflows.github.io/modelflowsapp/}{https://modelflows.github.io/modelflowsapp/}}
\end{abstract}




\end{frontmatter}



\section{Introduction\label{sec:introduction}}
Computational fluid dynamics is generally used to resolve complex problem defined by high-dimensional systems, which can range up to large quantities of spatio-temporal flow scales. Solving realistic problems, capturing the highly complex underlying physics of the dynamic system being studied, comes with a proportional computational expense, limiting the problem-solving capabilities based on the amount of available resources. This has caused experts to branch out into new research paths, with the objective being to reduce the computational cost while minimizing the effect on the accuracy of results. There are two main sources which contribute to high computational cost, these being the spatial resolution used to define a singular solution, and the temporal resolution, which is the number of equidistantly spaced solutions generated during a certain period of time in order to study the evolution of the dynamic system. 

Starting off with the spatial resolution, this has been an ongoing problem in the field of fluid mechanics given the size of the databases. Spatial dimensionality is important if the goal is to generate data with high granularity. Experts in industry and researchers are sometimes forced to downscale their problems to fit their available computational resources, search for alternative solutions which can ingest large amounts of data, or to apply high performance computing techniques, such as data parallelization, in order to be able to process their data in a distributed manner, to speed up direct numerical simulation solvers \cite{mortensen2016high}, or to employ non-linear optimization on large scale sparse systems \cite{wu2020pyoptsparse}.

Focusing on data dimensionality reduction, which is the most popular and affordable technique, in recent work, low-cost singular value decomposition \cite{hetherington2023low} was developed as a highly accurate reconstruction technique via data assimilation, capable of increasing the resolution of any dataset, independently of its complexity (two- or three-dimensional, laminar or turbulent, etc.), and its nature (numerical or experimental) to a desired resolution, generally dictated by a paired numerical simulation. This algorithm is capable of reconstructing highly complex under-resolved datasets, consuming 37\% less memory, 630 times faster than standard singular value decomposition, allowing users to initially generate under-resolved numerical and experimental data, and apply LC-SVD to and enhance its resolution at a later stage, skipping the complex task of using a large number of sensors during experiments, or performing high resolution numerical simulations.

Regarding the model proposed in this paper, LC-SVD generates a high resolution denoised reconstruction of the input data. More precisely, LC-SVD will apply SVD to the input data and reconstruct the spatial SVD modes and temporal coefficients matrices to the desired resolution using a small number of robust SVD modes, also referred to as retained modes or coherent structures. When the input data is already high resolution, LC-SVD will use optimal sensor placement, with the use of the \textit{pysensors} Python library, or uniformly distributed data points to create an optimally under-resolved version of the input, and then apply the previously described reconstruction process. After completing this step, LC-SVD then feeds the temporal coefficients matrix to the DLinear model to forecast the next values. These prediction will then be multiplied by the reconstructed spatial SVD modes and singular values to create new snapshots. In order to process high-dimensional data, in addition to LC-SVD, in this work we present, for the first time to the authors' knowledge, LC-HOSVD, a low-cost version of the high-order singular value decomposition methodology which applies the same principles.

It is worth mentioning other commonly used feature extraction techniques, such as principal component analysis (PCA), applied in Ref. \cite{parente2013principal} to identify the most active
directions in multivariate datasets. In \cite{scherl2020robust}, PCA was used to filter and extract dominant coherent structures to identify and complete incoherent of missing data from experimental datasets. Another important algorithm used to identify patterns in data is high-order dynamic mode decomposition (HODMD) \cite{LeClaincheVega17}, which can be used alone \cite{le2019new}, with PCA \cite{corrochano2023higher}, and even with neural networks \cite{MataLeon2023,huang2022predictions}.

Additionally, deep learning neural networks have also proven to be capable of learning compressed meaningful representation of a given dataset, as demonstrated in Ref. \cite{munoz2023extraction,ae_modal}. Architectures such as autoencoders and generative adversarial networks (GAN) are commonly used to enhance the resolution of input data.

Regarding temporal resolution, relying solely on computational fluid dynamics (CFD) solvers to generate new data can result, as described before, quite challenging given the high demand of computational resources these frameworks require to complete a singular step in time. It is well known that deep learning is rapidly advancing in the implementation of data generation and forecasting solutions. By combining state-of-the-art data dimensionality reduction techniques, such as LC-SVD, with shallow neural networks (one layer with trainable parameters per POD mode), we are able to tackle both described problems at the same time, with the development of a fully data-driven reduced order model (ROM), which not only enhances the spatial resolution of an input dataset, but also predicts its temporal evolution. This is exactly what has been accomplished in this work, which was initially inspired by \cite{abadia2022predictive}, where proper orthogonal decomposition (POD) was combined with different neural network architectures, precisely a convolutional neural network (CNN) and recurrent neural network (RNN), to forecast the next snapshots of an input fluid dynamics dataset. 

Other interesting studies have explored the integration of deep learning algorithms with modal decomposition in fluid dynamics and other applications. Similarly, this was accomplished by leveraging the properties of DMD or POD modal expansions to break down the flow into spatial modes (DMD or POD modes) and temporal coefficients (also referred to as temporal modes in POD expansions). This decomposition enables the use of deep learning architectures to reconstruct (interpolate) or predict (extrapolate) system dynamics based solely on the information contained in the temporal coefficients.

In ref. \cite{iuliano2013proper} POD was combined with radial basis functions (RBF), including Gaussian, multi-quadratic, and inverse-quadratic, to estimate the time evolution of POD coefficients (limited to interpolation, without temporal prediction) for constructing surrogate models in aerodynamic design optimization. Ref. \cite{freitag2018recurrent} used gappy POD with RBF to reconstruct an incomplete spatial dataset and applied recurrent neural networks for temporal forecasting in mechanized tunneling processes. In the field of fluid dynamics,  ref. \cite{guo2019data} integrated POD temporal coefficients with nonlinear regression techniques (Gaussian process regression) to develop a temporal predictive model, which was successfully applied to forecast the wake of a two-dimensional circular cylinder over time. Building on this approach, but applied to more complex fluid dynamics problems,  ref. \cite{guemes2019sensing} combined empirical POD \cite{discetti2018estimation} with a CNN to reconstruct temporal flow fields in wall-bounded turbulent flows using temporal information from wall sensors, akin to a linear stochastic estimation. Similarly,  ref. \cite{guastoni2021convolutional} employed this methodology to reconstruct two-dimensional velocity-fluctuation fields at various wall-normal locations in a turbulent open channel flow, partitioning the domain into smaller subdomains to enhance overall flow predictions.

This article presents a novel methodology which combines LC-SVD and LC-HOSVD with the DLinear architecture to forecast high-resolution snapshots with low computational cost, forming a hybrid machine learning model based on physical principles. First, the input data, which can be in low- or high-resolution, is passed to either LC-SVD or LC-HOSVD, which generates a denoised high-resolution reconstruction of both the spatial SVD modes and temporal coefficients matrices, while guaranteeing that the underlying physics of the dynamic system being studied are captured, and that all white noise is minimized so that spatio-temporal patterns can be easily recognizable. This is accomplished by calibrating the number of retained modes, keeping only those that represent the most important features (modes of higher energy), and discarding those that contain noise (modes of lower energy). To achieve this, the reconstruction process is done iteratively, until the reconstruction mean absolute error (MAE) between iterations meets a pre-established threshold. Then, the latest sequence formed by the last $L$ values of the reconstructed temporal coefficients are passed to the DLinear model to predict the next $N_{pred}$ values autoregressively, with the use of a rolling window of stride 1. The model forecasts a complete new sequence containing the next $N_{pred}$ temporal coefficient values that correspond to the input sequence, which can be up-to 100 times the length of the input, one snapshot at a time. Finally, the predicted temporal coefficients are transformed into high resolution snapshots using the reconstructed spatial SVD modes matrix. The DLinear architecture gained a vast amount of traction after the publication of "Are Transformers Effective for Time Series Forecasting?" \cite{zeng2023transformers}, where, in the majority of cases, they outperformed complex state-of-the-art transformer models, which are highly expensive to train given their large amount of parameters, on long time series forecasting tasks. These architectures have been previously combined with mathematical or statistical methods to generate hybrid models, leveraging the benefits of both, as seen in \cite{liu2023multi}.

Various techniques have been applied to validate the effectiveness of this method. These include quantitative error metrics, such as mean absolute error and mean squared error (MSE), to evaluate the raw predictions (the temporal coefficient predictions), and relative root mean squared error (RRMSE) to evaluate the snapshot reconstruction. Additionally, a statistics-based approach has been used to evaluate the results generated by this methodology, by evaluating the Wasserstein distance, and comparing the ground truth and forecast snapshot data distributions.

This article is organized in the following way: the LC-SVD and LC-HOSVD methodologies for optimal data resolution enhancement, the DLinear architecture, and the combination of both (therefore, the novel hybrid models LC-SVD-DLinear and LC-HOSVD-DLinear), are described in sec. \ref{sec:methodology}. Then, the different test cases used to validate the previously described methodologies robustness and precision are presented in sec. \ref{sec:database}, followed up by sec. \ref{sec:results}, where the results of this novel method are displayed. These results include the high-resolution predictions of the test cases generated using both LC-SVD-DLinear and LC-HOSVD-DLinear. Finally, the conclusions are presented in sec. \ref{sec:conclusions}.

\section{Methodology \label{sec:methodology}}
The following section explains the details behind this novel methodology. First, a brief breakdown of the LC-SVD algorithm is provided, since an in-depth explanation can be found in the original work \cite{hetherington2023low}. Following up, the novel LC-HOSVD methodology is described, which follows the same principles as LC-SVD, but seems more suitable for high-dimensional data. Next, the DLinear architecture is explained. After presenting both parts of this novel methodology (optimal data resolution enhancement algorithms and the time series forecasting shallow neural network), the final hybrid models, LC-SVD-DLinear and LC-HOSVD-DLinear, are presented. This section finishes off with a description on how the prediction and reconstruction errors have been analysed, using commonly used error evaluation metrics and statistics. 

\subsection{Data organization \label{sec:dataOrganization}}
A matrix-form dataset consists of a ordered group of $K$ snapshots $\bv_k=\bv(t_k)$, where $t_k$ is the time at instant $k$. For convenience, the snapshots are collected in the following {\it snapshot matrix}
  \begin{equation}
  \bV_1^K = [\bv_{1},\bv_{2},\ldots,\bv_{k},\bv_{k+1},\ldots,\bv_{K-1},\bv_{K}].\label{eq:SnapMatrix}
  \end{equation}
  
where $\bV_1^{K}$ refers to the snapshot matrix, and each one of the $K$ temporal samples $\bv_{k}$ is known as snapshot. Each snapshot is formed by a vector with dimension $J=N_{comp} \times N_x \times N_y $ for two-dimensional datasets, and $J=N_{comp} \times N_x \times N_y \times N_z$ for three-dimensional cases, where $N_{comp}$ is the number of components studied in each dataset (i.e., in a dataset of a velocity field formed by streamwise, normal and spanwise velocity, $N_{comp}=3$), and $N_x$, $N_y$ and $N_z$ correspond to the spatial mesh points distributed along the streamwise, normal and spanwise direction, respectively. When the dataset consists of several components, each one is concatenated in columns.   

An under-resolved version of the previous snapshot matrix, $\bar{\bV}_1^{K}$, has dimensions $\bar{J}\times \bar{K}$, where $\bar{J}<J$ and $\bar{K}<K$. The under-resolved snapshot matrix is: 

\begin{equation}
    \bar{\bV}_1^{K} = [\bar{\bv}_{1}, \bar{\bv}_{2}, \dots, \bar{\bv}_{k}, \bar{\bv}_{k+1}, \dots, \bar{\bv}_{K-1}, \bar{\bv}_{K}],
    \label{eq:SnapMatrixRed}
\end{equation}

where $\bar{\bv}_{k}$ represents a modified reduced snapshot, $\bar{\bv}_{k} \in \mathbb{R}^{\bar{J}}$.

There are various ways to under-resolve a snapshot matrix. First, equidistant data selection could be used to generate a downsampled snapshot matrix by selecting $1$ in every $p$ points, so all data points of the snapshot matrix are considered to have the same level of relevancy. This is ideal when performing interpolation to reconstruct the data back to its original resolution, since there is a uniform coverage of the snapshot matrix. The second method consist in using {\it pysensors} \cite{pysensors} to select the optimal positions of a given number of points by which the snapshot matrix must be reduced which, from now on, will be referred to as the number of sensors $N_s$. This python library locates the optimal sensors positions using a singular value decomposition basis. Both of these methods are supported by the low-cost singular value decomposition.

In some applications, databases are reorganized into a tensor format called a {\it snapshot tensor}, where components like velocity or other variables are separated into distinct tensor elements. In this format, matrix snapshots form a multidimensional array with more than two indices, and the {\it fibers} of the tensor are made up of the matrix's columns and rows.

Consider a two-dimensional database of velocity components, $v_x$ and $v_y$, representing the in-plane velocity in a Cartesian coordinate system with dimensions $J_2 \times J_3$. The data can be represented as:

\begin{equation}
\bv(x_{j_2}, y_{j_3}, t_k) \quad \text{for } j_2 = 1, \ldots, J_2, \quad j_3 = 1, \ldots, J_3, \quad k = 1, \ldots, K.
\end{equation}

This data can be organized into a fourth-order tensor, $\bV$, of size $J_1 \times J_2 \times J_3 \times K$, where:

\begin{equation}
V_{1j_2j_3k} = v_x(x_{j_2}, y_{j_3}, t_k), \quad V_{2j_2j_3k} = v_y(x_{j_2}, y_{j_3}, t_k).
\label{eq:4ordertensor}
\end{equation}

Here, $j_1$ corresponds to the velocity components (with $J_1=2$ for $v_x$ and $v_y$), while $j_2$, $j_3$, and $k$ represent the spatial coordinates and time, respectively. The structure can be adapted for different numbers of components depending on the problem.

For three-dimensional datasets, the data is organized into a fifth-order tensor $\bV$ of size $J_1 \times J_2 \times J_3 \times J_4 \times K$, where:
\begin{equation}
V_{1j_2j_3j_4k} = v_1(x_{j_2}, y_{j_3}, z_{j_4}, t_k), \quad \dots, \quad V_{J_1j_2j_3j_4k} = v_{J_1}(x_{j_2}, y_{j_3}, z_{j_4}, t_k).
\label{eq:5ordertensor}
\end{equation}

Here, $J_1$ refers to the number of components (e.g., three velocity components), and $j_2$, $j_3$, and $j_4$ represent the discrete values of the spatial coordinates $x$, $y$, and $z$. The index $k$ represents the time step.

The snapshot matrix and snapshot tensor are closely related, with the tensor indices $j_1$, $j_2$, $j_3$ (and $j_4$ for 3D data) combined into a single index $j$ in the matrix. Thus, $\bV_1^K \in J \times K$, where $J = J_1 \times J_2 \times J_3 (\times J_4)$. Data can be reshaped between matrix and tensor formats using the NumPy {\it reshape} function.

Similar to the reduced snapshot matrix from eq. \eqref{eq:SnapMatrixRed}, reduced versions of the snapshot tensors can be formed. Let's take a fourth-order snapshot tensor formed by the velocity components tensors from eq. \eqref{eq:4ordertensor}, and reduce its dimensionality:

\begin{equation}
\bar{\bV}_{j_1j_2j_3k} = [\bar{\bV}{_1}_{j_2j_3k}, \bar{\bV}{_2}_{j_2j_3k}].
    \label{eq:4orderreduced}
\end{equation}

where $\bar{\bV}$ is the reduced fourth-order snapshot tensor formed by two velocity components under-resolved both in space and time, so $\bar{J_2} < J_2$, $\bar{J_3} < J_3$ and $\bar{K} < K$. 

\subsection{Low-cost singular value decomposition (LC-SVD)\label{sec:lowcostSVD}}

Low-cost singular value decomposition is an extension of SVD, which was first presented in \cite{hetherington2023low}, and is used to analyse reduced datasets and to reconstruct them to a higher resolution. This method allows users to collect minimum data during experiments, or to perform numerical simulations in a reduced grid, using less computational resources, since by using LC-SVD the data resolution can be enhance with minimum computational cost, using SVD to capture the underlying physics while eliminating any noise. SVD is implemented as follows:

\begin{equation}
\bV_1^{K}\simeq\bW\,\bSigma\,\bT^\top,\label{eq:svd}
\end{equation}

where $\bW$ is a matrix that contains the spatial SVD modes (also referred to as POD modes) in the column dimension, $\bT$ is a matrix containing the temporal coefficients in the row dimension, while $\bSigma$ is a diagonal matrix containing the  singular values $\sigma_1,\cdots,\sigma_{N}$, ranked (by the singular values) in decreasing order in the previous factorization, with $N\leq min(J,K)$ as the number of SVD modes retained. The $()^\top$ operator denotes the matrix transpose. 

LC-SVD is applied to a reduced snapshot matrix $\bar{\bV}_1^{K} \in \mathbb{R}^{\bar J\times \bar K}$ eq. (\ref{eq:SnapMatrixRed}), and uses the decomposed dataset—formed by the spatial SVD modes, singular values, and temporal coefficients matrices from eq. \eqref{eq:svd}—to reconstruct the original snapshot matrix   $\bV_1^{K} \in \mathbb{R}^{J\times K}$  eq. (\ref{eq:SnapMatrix}). The LC-SVD algorithm is applied following these steps:
\begin{itemize}
    \item {\em Step 1: apply standard SVD to the under-resolved data.} SVD is applied to the reduced dimension snapshot matrix eq. (\ref{eq:SnapMatrixRed}) as
    \begin{equation}
    \bar{\bV}_1^{K}\simeq\bar{\bW}\,\bar{\bSigma}\,\bar{\bT}^\top,\label{eq:redSVD}
    \end{equation}
    where $\bar{\bW}^\top\bar{\bW} = \bar{\bT}^\top\bar{\bT}=$ are unit matrices of dimension $\bar{N}\times \bar{N}$.  The number of retained SVD modes, $\bar{N}$, is defined by the user, and mode reduction is implemented by applying matrix slicing on the matrices from eq. \eqref{eq:redSVD} in the dimension that contains the number of modes.
    
    \item {\em Step 2: normalize the spatial SVD modes.} Matrix $\bar{\bSigma}$ can be ill-conditioned when small singular values are retained. This can cause SVD modes calculated in $\bar{\bW}$ to be slightly non orthogonal due to round-off errors. QR factorization is applied to re-orthonormalize these modes as $ \bar{\bW}=\bQ^W \bR^W$, leading to 
    \begin{equation}
        \bar{\bW}=\bar{\bW} (\bR_{\bar N}^W)^{-1},
    \end{equation}
    where $\bR_{\bar N}^W \in \mathbb{R}^{\bar N\times \bar K}$. Notice that, similar to SVD, only $\bar N$ modes are retained. 
    
    \item {\em Step 3: normalization of the temporal coefficients.} 
      Similar to Step 2, the temporal coefficients calculated in $\bar \bT$ could also be slightly non orthogonal so, to ensure they are orthogonal, QR factorization is applied as $ \bar{\bT}=\bQ^T \bR^T$, which leads to 
    \begin{equation}
        \bar{\bT}=\bar{\bT} (\bR_{\bar N}^T)^{-1},
    \end{equation}
    where $\bR_{\bar N}^T\in \mathbb{R}^{\bar N\times \bar K}$.
    Some differences in the sign of temporal coefficients, depending on the type of calculations, may affect the reconstruction of the original dataset. To prevent this, an additional step can be applied, where the sings in  $\bar{\bT}$ are updated as:
     \begin{equation}
        \bar{\bT}=\bar{\bT} \text{ sign}(\text{diag} (\bar{\bSigma})), %
    \end{equation}
    where sign($\cdot$) and diag($\cdot$) correspond to the sign and diagonal of a matrix. 
    
    \item {\em Step 4: reconstruction of the spatial SVD modes.} The spatial SVD modes with enhanced spatial dimension (matrix $\bW$ from eq. (\ref{eq:svd})) are reconstructed as follows:
    \begin{equation}
        \bW\simeq \bW^{rec}= (\bar{\bV}_{1}^{K,{J\bar K}})^\top \bar{\bT} (\bar{\bSigma})^{-1},\label{eq:Wrec}
    \end{equation}
    where $\bW^{rec} \in \mathbb{R}^{J \times \bar N}$.
    
    \item {\em Step 5: recover temporal coefficients of SVD modes.} The temporal coefficients of SVD modes with enlarged spatial dimension (matrix $\bT$ from eq. (\ref{eq:svd})) are recovered as:
    \begin{equation}
        \bT\simeq\bT^{rec} = (\bar{\bV}_{1}^{K,{\bar J K}})^\top \bar{\bW} (\bar{\bSigma})^{-1},\label{eq:Trec}
    \end{equation}
    where $\bT^{rec} \in \mathbb{R}^{K \times \bar N}$.

    \item {\em Step 6: reconstruction of the original dataset.}
    Using the reconstructed spatial SVD modes and temporal coefficients, defined in eqs. (\ref{eq:Wrec})-(\ref{eq:Trec}), in addition to the raw (original under-resolved) singular values, it is possible to reconstruct the original dataset $\bar{\bV}_{1}^{K}$ from eq. (\ref{eq:svd}) as
    \begin{equation}
\bV_1^{K}\simeq \bV_1^{K, rec}= \bW^{rec}\,\bar \bSigma\,(\bT^{rec})^\top.\label{eq:svdRec}
\end{equation}
\end{itemize}

In this work, LC-SVD is applied iteratively, to ensure that all noise is filtered out of the reconstruction. To so do, the MSE between iterations is calculated and, with the loop concluding when $(\bV_{1 (i)}^{K, rec} - \bV_{1 (i - 1)}^{K, rec})^2 < 1e^{-6}$, with $i$ being the current iteration number.

\subsection{Low-cost high-order singular value decomposition (LC-HOSVD)\label{sec:lchosvd}}
The HOSVD algorithm was first introduced by Tucker in 1966 \cite{Tucker66}, and has gained popularity in recent years, particularly due to its implementation by de Lathauwer et al. \cite{DeLathawer,DeLathawer0}. 

HOSVD decomposes high-order datasets (also referred to as tensors interchangeably in this work) by applying SVD to each fiber of the tensor. For instance, HOSVD of the fifth-order tensor defined in eq. (\ref{eq:5ordertensor}) is presented as
 \begin{equation}
 \bV_{j_1j_2j_3j_4k}\simeq\sum_{p_1=1}^{P_1}\sum_{p_2=1}^{P_2}\sum_{p_3=1}^{P_3}\sum_{p_4=1}^{P_4}\sum_{n=1}^{N}  \bW^{(1)}_{j_1p_1} \bW^{(2)}_{j_2p_2} \bW^{(3)}_{j_3p_3} \bW^{(4)}_{j_4p_4} \bS_{p_1p_2p_3p_4n} \bT_{kn},    \label{c10}
 \end{equation}

Here, $\bS_{p_1p_2p_3p_4n}$ is the {\em core tensor}, a fifth-order tensor which contains the singular values, while the columns of $\bW^{(1)}$, $\bW^{(2)}$, $\bW^{(3)}$, $\bW^{(4)}$ and $\bT$ are the {\it modes} of the decomposition.

The columns of $\bW^{(l)}$ for $l=1,2,3,4$ represent the spatial HOSVD modes, corresponding to the database components and spatial variables, while the columns of $\bT$ represent the temporal HOSVD modes, corresponding to the time variable. The singular values of the decomposition is now formed by five sets of values, 
\beqn
\sigma^{(1)}_{p_1}, \quad \sigma^{(2)}_{p_2},\quad \sigma^{(3)}_{p_3},\quad \sigma^{(4)}_{p_4},\quad\text{and } \sigma^t_{n},\label{c11}
\eeqn
which are also sorted in descending order. Truncation in HOSVD, similar to SVD, retains only the most relevant modes. While the full HOSVD is exact, selecting a reduced number of modes helps reduce noise, or even lower dimensionality, based on the application's requirements. This number is selected by the user. 
After truncation, HOSVD (\ref{c10}) is written as:
 \beqn
 \bV_{j_1j_2j_3j_4k}\simeq\sum_{n=1}^{N} \bW_{j_1j_2j_3j_4n}\hat \bV_{kn},\label{c15}
\eeqn
where $\bW_{j_1j_2j_3j_4n}$ and $\bV_{kn}$ are the spatial and temporal modes, and $N$ is the spatial complexity or number of retained HOSVD modes. The modes are defined as:

\beqn
 \bW_{j_1j_2j_3J_4n}= \sum_{p_1=1}^{P_1}\sum_{p_2=1}^{P_2}\sum_{p_3=1}^{P_3}\bS_{p_1p_2p_3p_4n}
 \bW^{(1)}_{j_1p_1} \bW^{(2)}_{j_2p_2} \bW^{(3)}_{j_3p_3} \bW^{(4)}_{j_4p_4}/\sigma^t_r,
 \quad \hat \bV_{kn}=\sigma^t_r\bT_{kn}.\label{c16}
\eeqn

After introducing the HOSVD algorithm, we can highlight the innovation of LC-HOSVD. This low-cost variant applies HOSVD to an under-resolved tensor and reconstructs the spatio-temporal modes to a high resolution with minimal computational expense. 

Contrary to LC-SVD, LC-HOSVD optimally reconstructs the spatial modes of each component, improving noise filtering and effectively up-scaling the problem while accurately capturing the underlying physics described in the database.

The LC-HOSVD method follows with similar steps as LC-SVD. These steps are described below, applying LC-HOSVD to a fourth-order snapshot tensor:

\begin{itemize}
    \item {\em Step 1: apply HOSVD to the under-resolved tensor.} HOSVD is applied to the reduced snapshot tensor, eq. (\ref{eq:4orderreduced}) as: 
    \begin{equation}{\bar{\bV}_{j_1j_2j_3k}}\simeq\sum_{p_1=1}^{P_1}\sum_{p_2=1}^{P_2}\sum_{p_3=1}^{P_3}\sum_{n=1}^{N}  \bar{\bW}^{(1)}_{j_1p_1} \bar{\bW}^{(2)}_{j_2p_2} \bar{\bW}^{(3)}_{j_3p_3} \bar{\bS}_{p_1p_2p_3n} \bar{\bT}_{kn}, 
    \label{eq:redHOSVD}
    \end{equation}
    Noise is filtered out by selecting the first $\bar{N}$ modes, determined by the user. This results in filtered spatial and temporal HOSVD matrices. Applying eq. \eqref{c15} is implemented to reconstruct a denoised version of the input snapshot tensor, $\bar{\bV}_{j_1j_2j_3k}^{rec}$. 

    \item {\em Step 2: transform the snapshot tensor into a snapshot matrix.} The denoised snapshot tensor from Step 1 is transformed into a snapshot matrix. This results in $\bar{\bV}_1^K$, with $\bar{\bV}_1^K \in \bar{J} \times \bar{K}$, where $\bar{J} = J_1 \times \bar{J_2} \times \bar{J_3}$. 

    \item {\em Step 3: apply LC-SVD to the transformed reduced snapshot matrix.} In the previous steps, the input reduced snapshot tensor has been denoised and converted to matricial form. This data transformation allows us to approach the reconstruction problem using LC-SVD. Applying this step results in the decomposition of the snapshot matrix, and the reconstruction of the spatial SVD modes ($\bW^{rec}$) and temporal coefficients ($\bT^{rec}$), while also providing a new set of singular values ($\bar{\bSigma}$). It is important to note that after applying this step none of the matrices or tensors from Step 1 are used again.

    \item {\em Step 4: return the reconstructed snapshot matrix to its original form (tensor).} In this last step, $\bW^{rec}$ and $\bV_1^{K, rec}$ are transformed into tensors, to recover the original shape. This is also accomplished by applying the reshape operation, with the new shape being $J_1, J_2, J_3, \bar{N}$ for the spatial modes, resulting in $\bW^{rec}_{j_1j_2j_3\bar{N}}$, and $J_1, J_2, J_3, K$ for the reconstructed snapshot tensor, giving $\bV^{rec}_{j_1j_2j_3j_4k}$. 
  
\end{itemize}

The previously described steps are applied iteratively, in order to maximize noise filtering and achieve optimal reconstruction, using the same threshold as LC-SVD. 

\subsection{Data preprocessing\label{sec:preprocessing}}
Before moving on to the DLinear model, it is important to cover the data preprocessing required for this model to operate. First, the input temporal coefficients matrix is divided into train, validation and test sets, with proportions 0.7, 0.15 and 0.15, respectively. 

\begin{figure}[H]
    \centering
    \includegraphics[width=1\textwidth, angle=0]{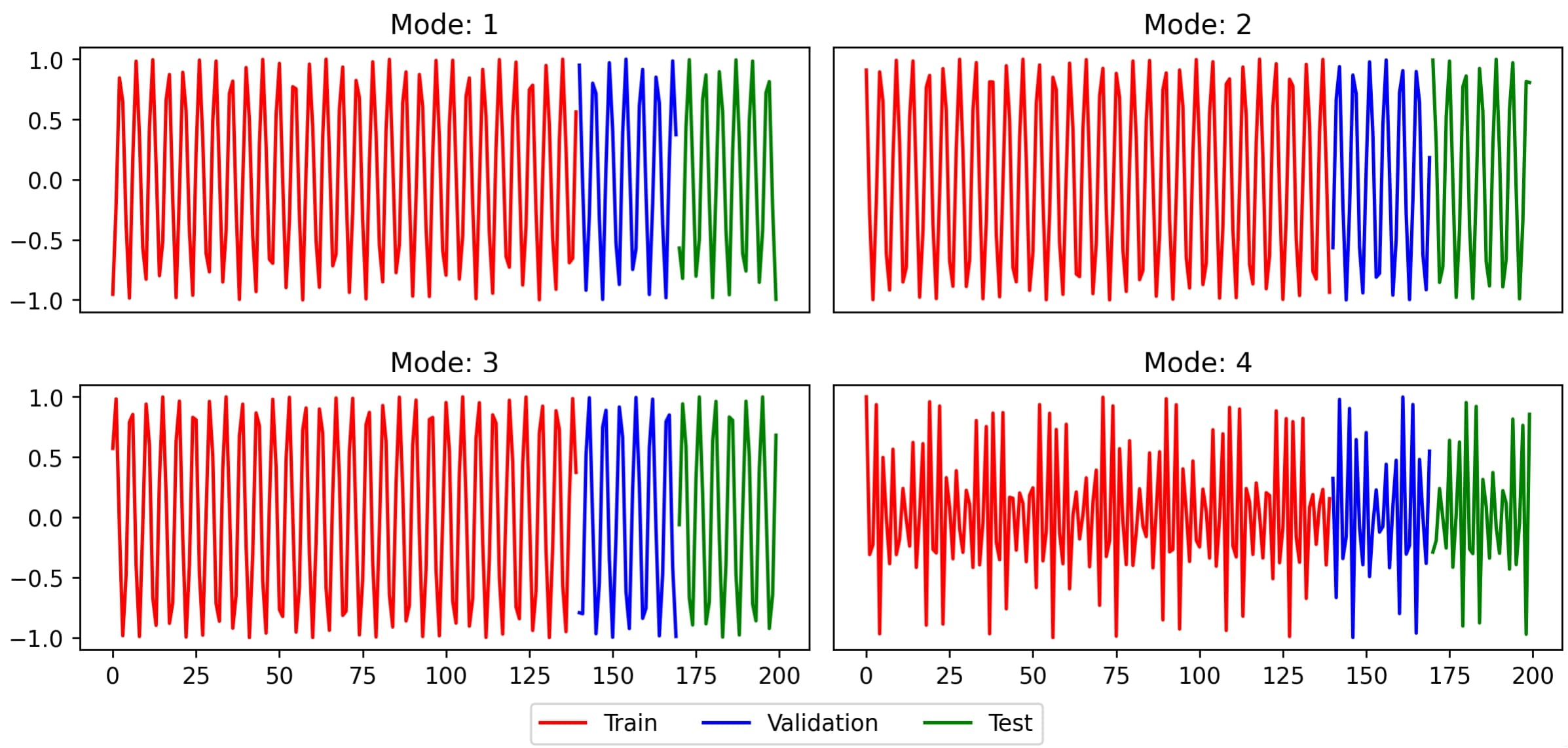}
    \caption{An example of preprocessed data, split into train, validation and test sets, scaled between -1 and 1 with min-max scaling.}
    \label{fig:traintestsplit}
\end{figure}

After splitting the data, the temporal coefficients are normalized using min-max scaling, with the data range being $[-1, 1]$, since DLinear solves a simple linear regression where negative values are possible. Min-max scaling is applied as follows:

\begin{equation}
    \bT_{scaled} = a + \frac{(\bT - \bT_{min})(b - a)}{\bT_{max} - \bT_{min}},
    \label{eq:minmaxscaling}
\end{equation}
where $\bT_{max}$ and $\bT_{min}$ are the maximum and minimum values of $\bT$, while $a$ and $b$ are the bottom and top limits of the data range, so -1 and 1, respectively. An example of the split normalized temporal coefficients can be seen in fig. \ref{fig:traintestsplit}

The final preprocessing step is the generation of sequences. Given a sequence $z$ with sequence length equal to $L$, and a horizon $H$ equal to 1 (single-step forecast), $W$ sequences are formed by applying a sliding window to the data, with step size equal to the horizon. The notation for a single time step $z$ is as follows:

\begin{equation}
    \bT_z = [\bT(z - L + 1,:) \dots \bT(z, :)],
    \label{eq:timestep}
\end{equation}

where $z \geq L$. The horizon ($T_h$) for this sequence is $T(z + $H$, :)$, which is the next time step value. It is important to note that $z$ refers to both a sequence and time step, since both terms represent the same. The sliding window mechanism can be seen in fig. \ref{fig:rollingwindow}:

\begin{figure}[H]
    \centering
    \includegraphics[width=1\textwidth, angle=0]{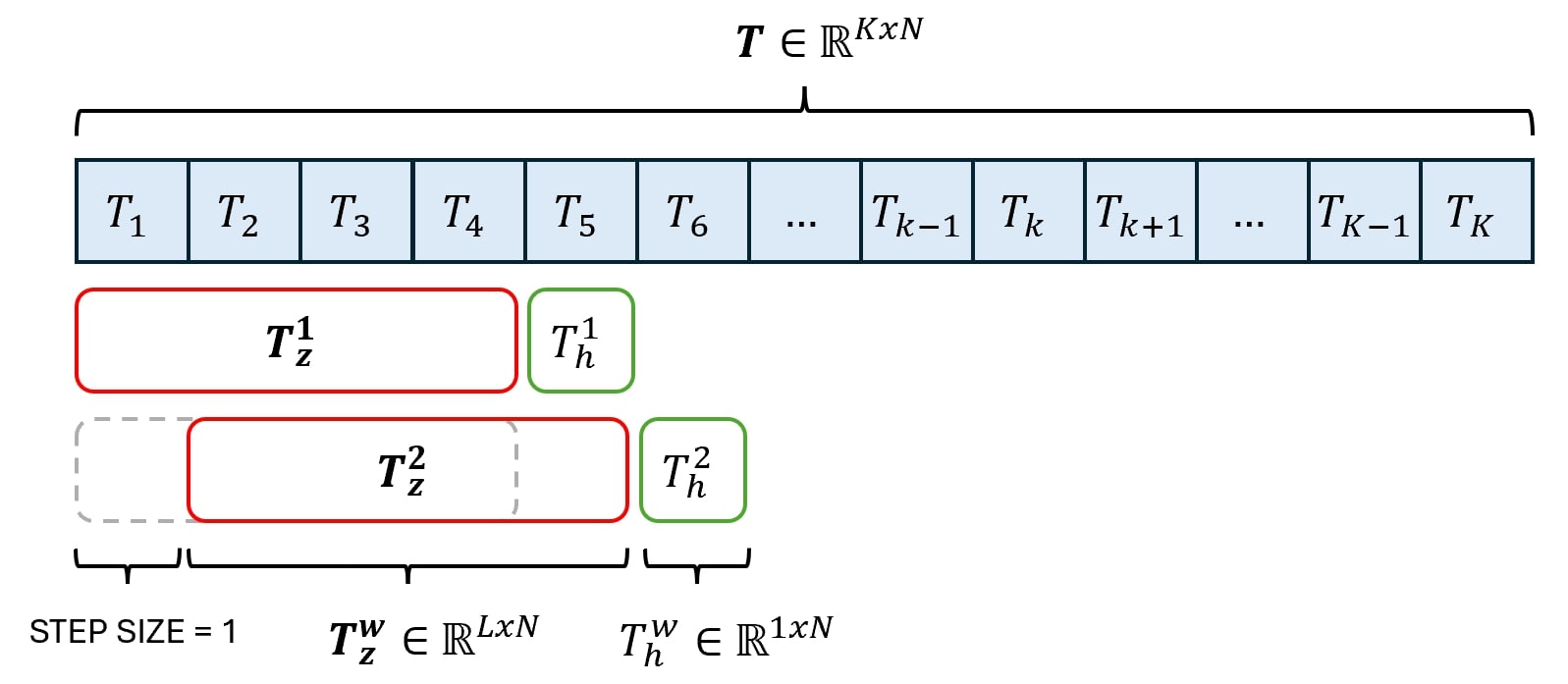}
    \caption{Illustration of a sliding window applied to the temporal coefficients \( \mathbf{T} \), with dimensions \( K \times N \). This process generates \( W \) sequences \( (\mathbf{T}_z) \), each of size \( L \times N \), along with a horizon \( (T_h) \) of size \( H \times N \), with$H = 1$). Here, \( w \) the sequence number, \( L \) is the sequence length, \( H \) denotes the horizon length, and \( N \) is the number of SVD modes.}
    \label{fig:rollingwindow}
\end{figure}

The red box represents the window or sequence $\bT^w_z$, where $\bT_z$ represents a single sequence of length $L$, with $w$ indicating the sequence number, for $w = 1, 2, \dots W$, while the green box is the horizon, $T^w_h$, with length $H = 1$, both for all $N$ modes. As explained above, after the generation of a sequence and its corresponding horizon, a singular step forward in time is taken to generate the next sequence. Usually, the step size corresponds to the horizon length.

\subsection{Decomposition linear neural network (DLinear) \label{sec:dlinear}}
The neural network used in this work was first presented in \cite{zeng2023transformers}, where its performance was compared to the achieved by a variety of transformers used for long time series forecasting (LTSF). The model is a shallow neural network, since it only has a depth of one layer with trainable parameters, but varies in width (based on the number of modes $\bar{N}$). 

\begin{figure}[H]
    \centering
    \includegraphics[width=1\textwidth, angle=0]{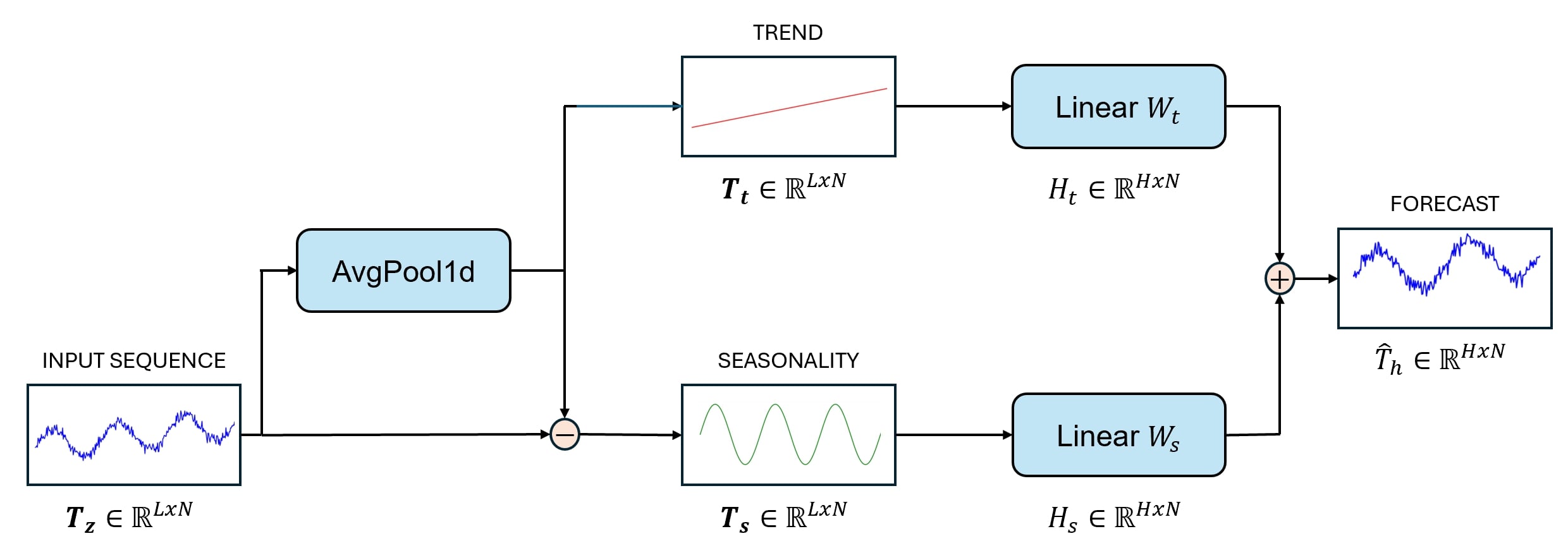}
    \caption{Representation of the DLinear architecture, showing how an input sequence $T_z$ is decomposed into trend $T_t$ and seasonality $T_s$ with the use of an average pooling layer, and both components are passed to individual linear layers which predict the next trend $H_t$ and seasonality $H_s$ values, which are summed, resulting in the prediction $\hat{T}_h$.}
    \label{fig:dlinear}
\end{figure}

The DLinear model adopts the decomposition approach utilized in the Autoformer \cite{autoformer} and FEDformer \cite{fedformer} architectures. A single step $z$ consists of DLinear applying an average pooling layer ({\it AvgPool1d} in fig. \ref{fig:dlinear}), with a kernel size equal to the sequence length $L$ and a stride (step size) of 1, to the input data, which is structured as a matrix containing the temporal coefficients of each retained POD mode in its last dimension, $\bT_z$, with $\bT_z \in L \times \bar{N}$. This process decomposes each mode into trend and seasonality components.  

Given an input sequence $T_z$, the average pooling layer extract the sequence trend as so:

\begin{equation}
T_{avg} = \frac{1}{L} \sum_{i = 1}^{L} \bT_z(i),
    \label{eq:avg}
\end{equation}

where $\bT_z$ and $T_avg$ represent the input data and the average value, respectively, while $i$ denotes the temporal position. After calculating the mean value of the sequence, the trend matrix is formed:

\begin{equation}
\bT_{t} = [T_{avg}, T_{avg}, T_{avg}, \dots, T_{avg}], 
    \label{eq:trend}
\end{equation}

with $\bT_t \in \mathbb{R}^{L \times N}$. The trend 
$T_t$ represents the overall level of the sequence, giving a low-frequency approximation of the data. The seasonality, which can also be referred to as the residual, can be calculated as follows:

\begin{equation}
\bT_{s} = \bT_z - \bT_t. 
    \label{eq:seasonality}
\end{equation}

The seasonality $\bT_s$ represents the high-frequency variations, such as cycles or repeating patterns, which become clearer once the trend is subtracted.

Both components are then fed into the neural network, where parallel linear layers perform a simple linear regression to predict the next values of the trend and seasonality of each mode simultaneously (one layer per component and, therefore, two layers per mode). These are $H_t$ and $H_s$ in fig. \ref{fig:dlinear}, respectively. This means that the model consists of $2 \times N$ parallel linear layers. 

After this, the predicted trend and seasonality values of each mode are summed, and the results for all modes are stacked together to form the output prediction, which contains the forecast temporal coefficient value for each mode, so $\hat{T}_h(z + 1) = f(\bT_z)$, where $f(\dot)$ represents the model. 

The prediction $\hat{T}_h(z + 1)$ is stacked to the end of the temporal coefficients matrix $\bT$, which results in $\hat{\bT}$ with $\hat{\bT} \in \hat{K} \times \bar{N}$, where $\hat{K} = \hat{K} + H$ (for step $z = 1$, $\hat{K} = K$). A new updated sequence $\bT_{(z + 1)}$ is created by applying the sliding window from fig. \ref{fig:rollingwindow} to advance one step in time. The new sequence will include $\hat{T}_h(z + 1)$, making the model autoregressive. 

The model parameters $W_t$ and $W_s$, which are the weights for the trend and seasonality linear layers, are initialized to 0, since each linear layer is independently resolving a linear regression for either the trend or seasonality of each mode. The model shows exceptional results well on datasets where there is a clear trend and seasonality in the temporal coefficients of all retained modes. 

The DLinear model has been developed using the deep learning framework {\it PyTorch} \cite{PyTorch}. The optimizer used to train these models is Adam, with the optimal learning rate and batch size values being tuned using the {\it Optuna} \cite{Optuna} hyperparameter tuner. The range of eligible values for the learning rate is $\alpha \in [1e^{-4}, 1e^{-2}]$, while the batch size options are $B_s \in [4, 8, 16, 32]$. These batch sizes are small, but this is due to the number of sequence samples in the training set being relatively low. Given the simplicity of the model, no regularization is required.

\subsection{Hybrid models: LC-SVD-DLinear and LC-HOSVD-DLinear\label{sec:SVD_NN}}
The combination of the methodologies described in this section allow for the fusion of the advantages of both, resulting in the forecast of new high-resolution snapshots from an input under-resolved dataset, by using a size-variant sequence of data.

As introduced in sec. \ref{sec:introduction} and explained throughout this section, LC-SVD and LC-HOSVD input either a low or high resolution dataset and create a clean high resolution version of the data. When the input data is low resolution LC-SVD will first apply SVD to the dataset and its paired high resolution target and, after filtering out all noise by fixing the number of modes to retain, uses  data assimilation to up-sample the spatial SVD modes and temporal coefficient matrices of the under-resolved dataset up to high resolution, using the spatial SVD modes and temporal coefficient matrices of the high resolution dataset. The resulting high resolution spatial SVD modes and temporal coefficient matrices represent the underlying physics captured in the input data, up-sampled, without any noise. 

When the input data is already in the desired resolution, LC-SVD will downsample the data optimally, and then perform the previously described process. This is done to ensure that the high resolution spatial SVD modes and temporal coefficients contain the most important information describing the physics, while removing all white noise, making any spatio-temporal patterns in the data easily identifiable by the DLinear model.

The DLinear model initially uses the clean high resolution temporal coefficient matrix, $\bT^{rec}$, selects the last $L$ values to form a sequence $\bT^{rec}_z$, and predicts the next temporal coefficient value for each mode, as described in sec. \ref{sec:dlinear}. This is done iteratively until a predefined number of snapshots $N_{snap}$ is generated. Once this desired temporal resolution is reached, then the predicted temporal coefficients are transformed into high resolution snapshots using the up-sampled spatial SVD modes ($\bW^{rec}$), and the original singular values ($\bar{\bSigma}$), by applying eq. \eqref{eq:svd}. 

\begin{figure}[H]
    \centering
    \includegraphics[width=1\textwidth, angle=0]{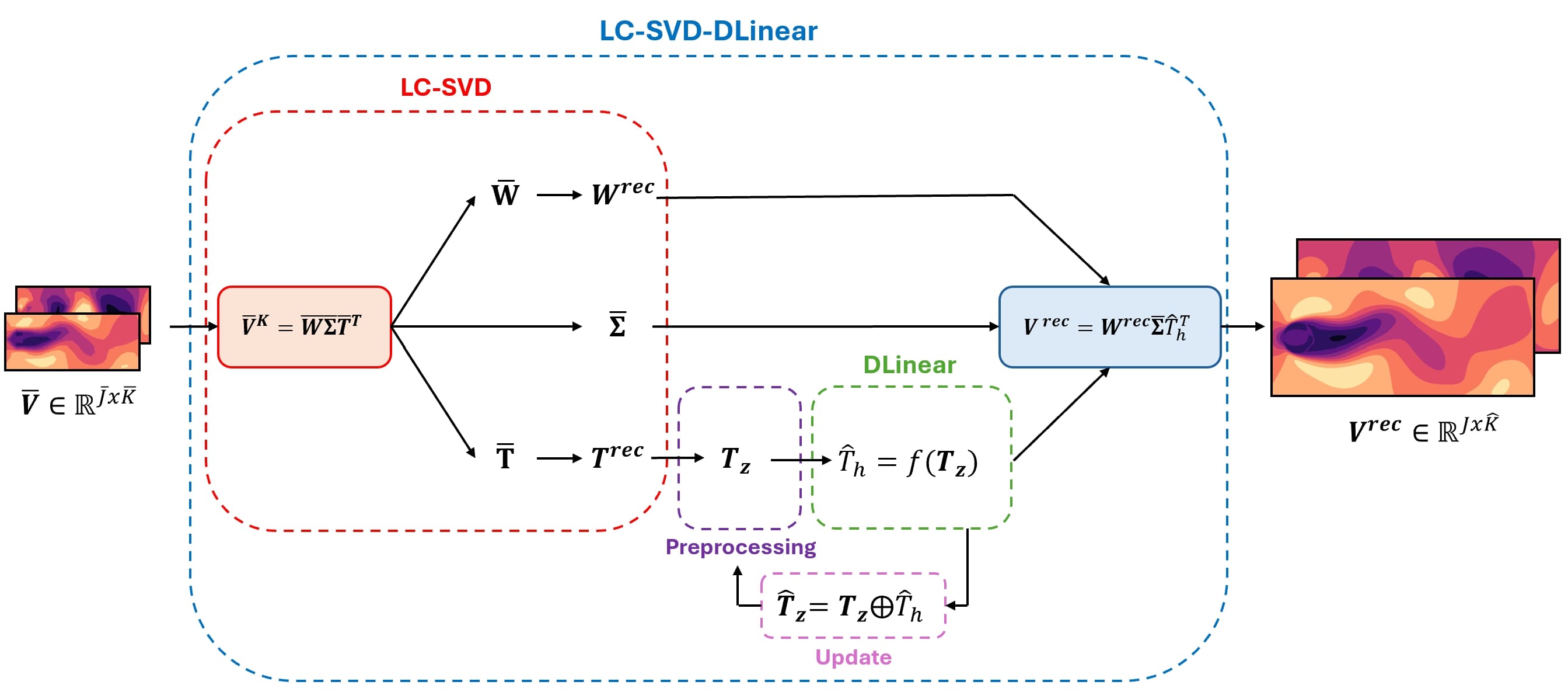}
    \caption{Outline of the LC-SVD-DLinear model, showing how the different parts of the hybrid model are interconnected to forecast high-resolution snapshots, given a low-resolution dataset.}
    \label{fig:lcsvddlineararch}
\end{figure}

Fig. \ref{fig:lcsvddlineararch} represents the architecture of the LC-SVD-DLinear model. In this image, the complete pipeline can be seen of the process explained in this section, from the input of low resolution data, on which LC-SVD is applied to enhance the resolution of the spatial SVD modes and temporal coefficients, to the autoregressive forecasting process, where the temporal modes are preprocessed (see sec. \ref{sec:preprocessing}, a sequence is generated for the model to forecast the next temporal modes values, which are then added to the input sequence, and a new window is created. The outputs are generated by reconstructing the forecast snapshot by multiplying the singular values by the reconstructed SVD modes and the forecast temporal coefficients.  

\subsection{Error analysis\label{sec:error}}
A variety of error metrics have been carefully selected to analyse the robustness and precision of the methodology. The first two metrics presented in this section are applied to the ground truth temporal coefficients ($y$) and the forecast values ($\hat{y}$) generated by the model. The remaining metrics are used to evaluate the error of the reconstructed snapshots. This is, the difference between the ground truth snapshots ($\bV_1^{K}$) and the reconstructed snapshots, ($\hat{\bV_1^{K}}$), generated using the forecast temporal coefficients ($\hat{y}$).

Starting off with the evaluation of the temporal coefficients predictions, the Mean Squared Error (MSE) is used to measure the average of the squared differences between the predicted and observed values, and is expressed as follows:

\begin{equation}
 MSE= {\frac{1}{n}\sum^n_{i = 1}(y_i - \hat{y}_i)^2}, \label{eq:MSE}   
\end{equation}

where $\hat{y}$ are the predicted temporal coefficient values, $y$ are the observed (ground truth) values, and $n$ is the number of observations. The MSE error is robust to outliers, meaning that larger errors are heavily penalized with the square operator. Due to this advantage, this error metrics is used to train the neural network. The MSE units are the observed value units squared. 

The second error metric used to analyse the forecast temporal coefficients is the Mean Absolute Error (MAE). This metric measures the average value of the absolute differences between observed and predicted values, and is expressed as:

\begin{equation}
 MAE= \frac{1}{n}\sum^n_{i = 1} |y_i - \hat{y}_i|, 
 \label{eq:MAE}   
\end{equation}

$\hat{y}$ and $y$ are the predicted and the observed temporal coefficient values, respectively, and $n$ is the number of observations. MAE is expressed in the same units as the observed values. This error metric also comes in useful when measuring the snapshots reconstruction error, and is used to find the average reconstruction error over each data point of a snapshot.

Regarding the reconstruction process, the main metric used to determine the exactitude of the reconstructed snapshots it the relative root mean squared error (RRMSE), which is represented as a percentage and is expressed as follows:

\begin{equation}
RRMSE=\frac{\|\bV_1^{K}-\hat{\bV_1}^{K,rec}\|_2}{\| \bV_1^{K}\|_2},
 \label{eq:RRMSE}   
\end{equation}

where $\bV_1^{K}$ refers to the observed snapshot matrix, $\hat{\bV_1}^{K, rec}$ is the reconstructed snapshots matrix from the predicted temporal coefficients, and $\| \cdot \|_2$ corresponds to the L2-norm.

To support this error metric, we use statistical analysis. First, we compute the Wasserstein distance for each snapshot and locate the snapshot with the highest value. The Wasserstein distance, commonly referred to as the Earth motion distance (EMD), is a metric that measures the distance between two probability distributions, which allows for an easy detection of a drift in data. 

More precisely, the Wasserstein distance represents the "effort" required to transform one distribution into another. This effort is defined as the amount of probability mass that needs to be transferred, scaled by the distance over which it is moved. In other words, the Wasserstein distance between two distributions is the minimum cost to transform one distribution into the other, where the cost is computed as the product of the moved probability mass and the distance it is moved.

\begin{equation}
W(p; V_1, \hat{V_1}^{rec}) = \left( \inf_{\gamma \in \Gamma(V_1, \hat{V_1}^{rec})} \int_{\mathcal{X} \times \mathcal{Y}} d(x, y)^p \, d\gamma(x, y) \right)^{\frac{1}{p}},
\label{eq:wasserstein}
\end{equation}

where $V_1$ and $\hat{V_1}^{rec}$ are the ground truth and predicted snapshot for a given time instance, respectively. $W(p; V_1, \hat{V_1}^{rec})$ is the p-th Wasserstein distance. $\Gamma(V_1, \hat{V_1}^{rec})$ is the set of all joint distributions (couplings) between $V_1$ and $\hat{V_1}^{rec}$. $d(x, y)$ is the distance between points $x$ and $y$. $\gamma(x, y)$ is the probability mass assigned to the pair $(x, y)$. $p \geq 1$ specifies the order of the distance (commonly $p = 1$ or $p = 2$). In this case, we use $p = 1$.

Histograms are used in support of the previous metrics. By locating the snapshot where the highest Wasserstein distance and mean absolute error occur, we can further analyse the drift in the data, comparing the ground truth and forecast snapshot data distributions.

\section{Test cases\label{sec:database}}
The precision, robustness and computational cheapness of LC-SVD-DLinear has been demonstrated by using diverse databases. The selected test cases are a mixture between two- and three-dimensional, laminar and turbulent, experimental and numerical fluid dynamics datasets. The first dataset consists of a three-dimensional cylinder at $Re = 220$, used as an initial benchmark, and an experimental circular cylinder at $Re = 2600$, to demonstrate the methods robustness when applied to real turbulent datasets.

As previously mentioned, these test cases are fluid dynamics datasets and, therefore, are governed by the Navier-Stokes equations. These equations for a viscous, incompressible and Newtonian flow are:

\begin{equation}
\nabla \cdot \vec{\bV} = 0,
\end{equation}

\begin{equation}
\frac{\partial u}{\partial t} + (\vec{\bV} \cdot \nabla) \vec{\bV} = -\nabla p + \frac{1}{Re} \Delta \vec{\bV},
\end{equation}

where $\vec{\bV}$ is the velocity vector, $p$ is the pressure, and $Re$ is the Reynolds number, which varies for each test case. These equations are non-dimensionalised using the characteristic length $L$ and time $L/U$, where $U$ is the characteristic or free stream velocity for each case.

\subsection{Three-dimensional numerical laminar cylinder (laminar cylinder)\label{Cyl2D3D}}
The first dataset consists in a numerical database solving a three-dimensional flow passing a circular cylinder, which is presented in Ref. \cite{VegaLeClaincheBook20}. This dataset is commonly used as a benchmark problem to validate methodologies, given its simplicity. The cylinder dynamics are closely related to the Reynolds number concept, defined with the cylinder diameter $D$. The initial Reynolds number is low and, therefore, the flow is steady. Upon reaching Re $ \approx 46$, a Hofp bifurcation that produces an unsteady flow, which is conducted by a von Karman vortex street \cite{jackson1987finite}, emerges. After reaching Re $ \approx 189$ the oscillations transition from two- to three-dimensional, for specific wavelengths in the spanwise direction \cite{barkley1996three}, with the development of a second bifurcation.

Numerical simulations have been carried out  using the open-source solver Nek5000 \cite{Nek5000} to solve the incompressible Navier-Stokes equations which define the behaviour of the flow. This solver uses spectral elements methods as spatial discretization.

The boundary conditions configured in the simulation for the cylinder surface are Dirichlet for velocity ($u = v = w = 0$) and Neumann for pressure. The conditions in the inlet, upper and lower boundaries of the domain are the same: $u = 1$, $v = w = 0$ for the streamwise, normal and spanwise velocities, respectively, and Neumann condition for pressure. The conditions in the outlet are Dirichlet for pressure and Neumann for velocity. The domain of the computational simulations is composed by 600 rectangular elements, each one of these is discretized using the polynomial order $\Pi = 9$. The dimensions of the computational domain are non-dimensionalized with the diameter of the cylinder. The size of the domain in the normal direction is constant $L_{y} = 15D$, and extends in the streamwise direction from $L_{x} = 15D$ upstream of the cylinder to $L_{x} = 50D$ downstream.

\begin{figure}[H]
    \centering
    \includegraphics[width=1\textwidth, angle=0]{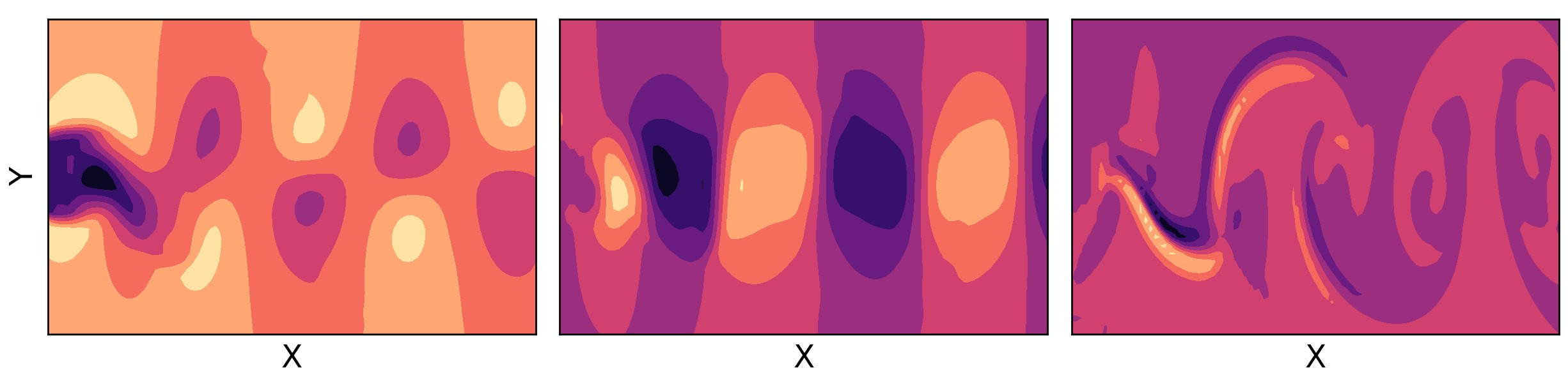}
    \caption{Streamwise (left), normal (middle) and spanwise (right) velocities of a snapshot of the three-dimensional Re $ = 220$ cylinder dataset from Ref. \cite{VegaLeClaincheBook20} in the $XY$ plane, for $z = 32$.}
    \label{fig:cyl3ddataset}
\end{figure}

Only the last $200$ snapshots have been selected, since they correspond to the saturated regime of the numerical simulation, where the flow is fully three-dimensional. The snapshots are equidistant in time with step size $\Delta t = 1$. The flow field velocity components of this dataset are defined by $U_1$ for the streamwise velocity, $U_2$ for the normal velocity, and $U_3$ for the spanwise velocity ($N_{comp} = 3$), which are enclosed in a domain of $N_{x} = 100$ points in the streamwise direction, $N_{y} = 40$ in the normal direction, and $N_{z} = 64$ in the spanwise direction.  Figure  \ref{fig:cyl3ddataset} shows a representative snapshot of this database. 

\subsection {Experimental turbulent cylinder  (turbulent cylinder)  \label{VKIdatasets}}
This second dataset has been extracted from Ref. \cite{mendez2020multiscale}, and consists in a turbulent flow passing a circular cylinder, which is $D = 5 mm$ in diameter and $L = 20 cm$ in length, at two different Reynolds numbers with a transient state in between. For this case, the second steady state, where the flow reaches $Re = 2600$ has been selected.

The experiment was conducted in the low-speed wind tunnel at the Von Karman Institute. The dataset captures a turbulent flow over a circular cylinder at two distinct Reynolds numbers, along with the transition between these states. The Reynolds numbers are Re $ \approx 4000$ and Re $ = 2600$, with vortex shedding frequencies of 450 $Hz$ and 303 $Hz$, respectively, yielding a Strouhal number of approximately $St = fd/U_{\infty} \approx 0.19$ in both regimes. The full range of Reynolds numbers explored falls within the domain of three-dimensional vortex shedding \cite{williamson1996vortex}. During the test, the free stream velocity $U_{\infty}$ shifts between two steady-state conditions, specifically from $U_{\infty} = 12.1 \pm 3$\% to $U_{\infty} = 7.9 \pm 3$\% $m/s$, with the transition occurring smoothly over about 1 second.

\begin{figure}[H]
    \centering
    \includegraphics[width=1\textwidth, angle=0]{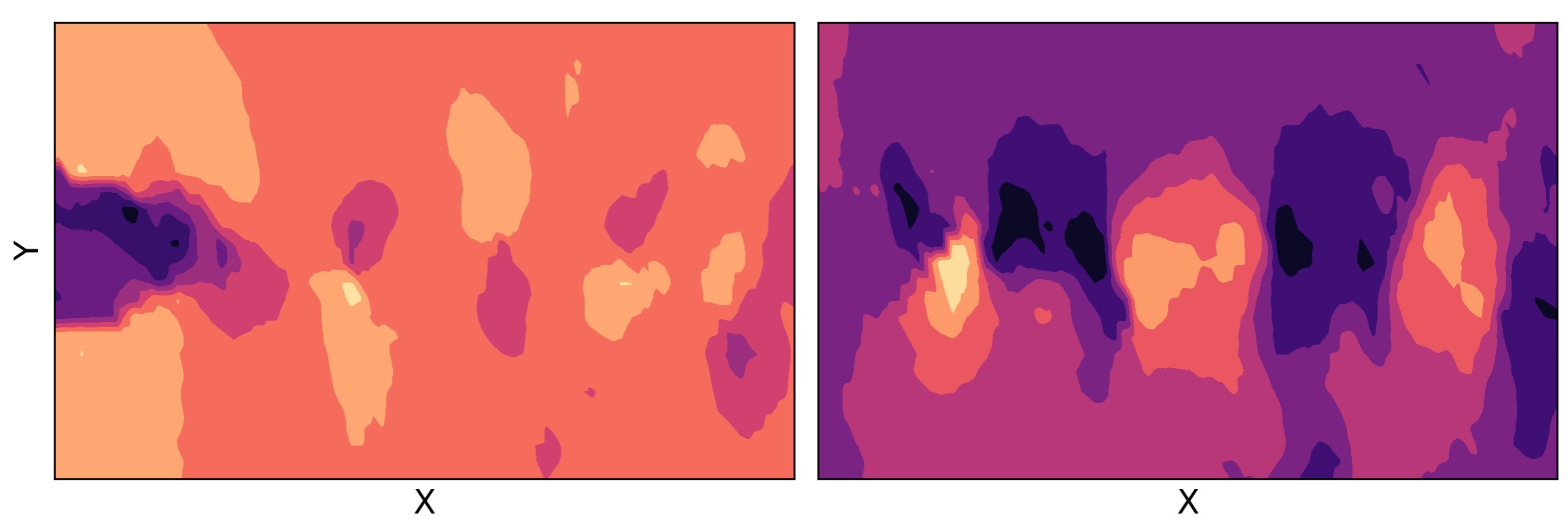}
    \caption{Streamwise (left) and normal (right) velocities of a representative snapshot of the turbulent cylinder at Re $ = 2600$ dataset from Ref. \cite{mendez2020multiscale}.}
    \label{fig:vki2600dataset}
\end{figure}

The experimental domain is defined by $N_{x}$ = 301 points in the streamwise direction and $N_{y}$ = 111 in the normal direction. The flow velocity field consists of two velocity components ($N_{comp} = 2$): streamwise velocity $U_1$ and normal velocity $U_2$, which were measured during 5200 snapshots, with time step $\Delta t = 0.33$. fig. \ref{fig:vki2600dataset} illustrates a representative snapshot of the previously described dataset.

\section{Results\label{sec:results}}
The following section gathers the results obtained after applying both LC-SVD-DLinear and LC-HOSVD-DLinear to the test cases, using an optimal number of POD modes, and down-sampling the datasets according to the optimal number of sensors calculated in Ref. \cite{hetherington2023low}. These are, $N_s = 45$ sensors for the three-dimensional laminar cylinder, and $N_s = 40$ for the turbulent cylinder. The data compression rates are $C_r = 17066$ and $C_r = 835$, respectively, meaning that the original data of each test case dataset is reduced by this proportion.

The first part of the results section illustrates the results obtained when applying both models to the laminar cylinder test case, while the second part is dedicated to results achieved when applied these to the turbulent cylinder. This way, a clear comparison can be made between the performance of both models.

The tables below summarize the parameters of each case. Table \ref{tab:LC-SVD-DLinear} presents the parameters for both test cases when using the LC-SVD-DLinear model, while table \ref{tab:LC-HOSVD-DLinear} illustrates the parameters when applying LC-HOSVD-DLinear. 

\begin{table}[H]
\centering
\begin{tabular}{|l|c|c|c|c|c|}
\hline
\textbf{Test case}          & \textbf{$N_s$} & \textbf{$\bar{N}$} & \textbf{$\alpha$} & \textbf{$B_s$}  & \textbf{$L$}\\ \hline
laminar cylinder        & 45             & 12                & $6.23e^{-4}$                & 16 & 15    \\ \hline
turbulent cylinder      & 40             & 6                 & $1.013e^{-4}$    & 4             & 100   \\ \hline
\end{tabular} 
\caption{Test cases with corresponding sensor counts $N_s$, SVD modes $\bar{N}$, learning rate $\alpha$, batch size $B_s$, and sequence length $L$ for the LC-SVD-DLinear model.}
\label{tab:LC-SVD-DLinear}
\end{table}

\begin{table}[H]
\centering
\begin{tabular}{|l|c|c|c|c|c|}
\hline
\textbf{Test case}          & \textbf{$N_s$} & \textbf{$\bar{N}$} & \textbf{$\alpha$} & \textbf{$B_s$} & \textbf{$L$}\\ \hline
laminar cylinder        & 45            & 12                 & $2.1e^{-4}$                & 4         & 15      \\ \hline
turbulent cylinder      & 40             & 6                 & $1.35e^{-4}$                 & 4        & 100       \\ \hline
\end{tabular}
\caption{Test cases with corresponding sensor counts $N_s$, SVD modes $\bar{N}$, learning rate $\alpha$, batch size $B_s$, and sequence length $L$ for the LC-HOSVD-DLinear model.}
\label{tab:LC-HOSVD-DLinear}
\end{table}

\subsection{Laminar cylinder forecast \label{solLC}}
\begin{figure}[H]
    \centering
    \includegraphics[width=0.8\textwidth, angle=0]{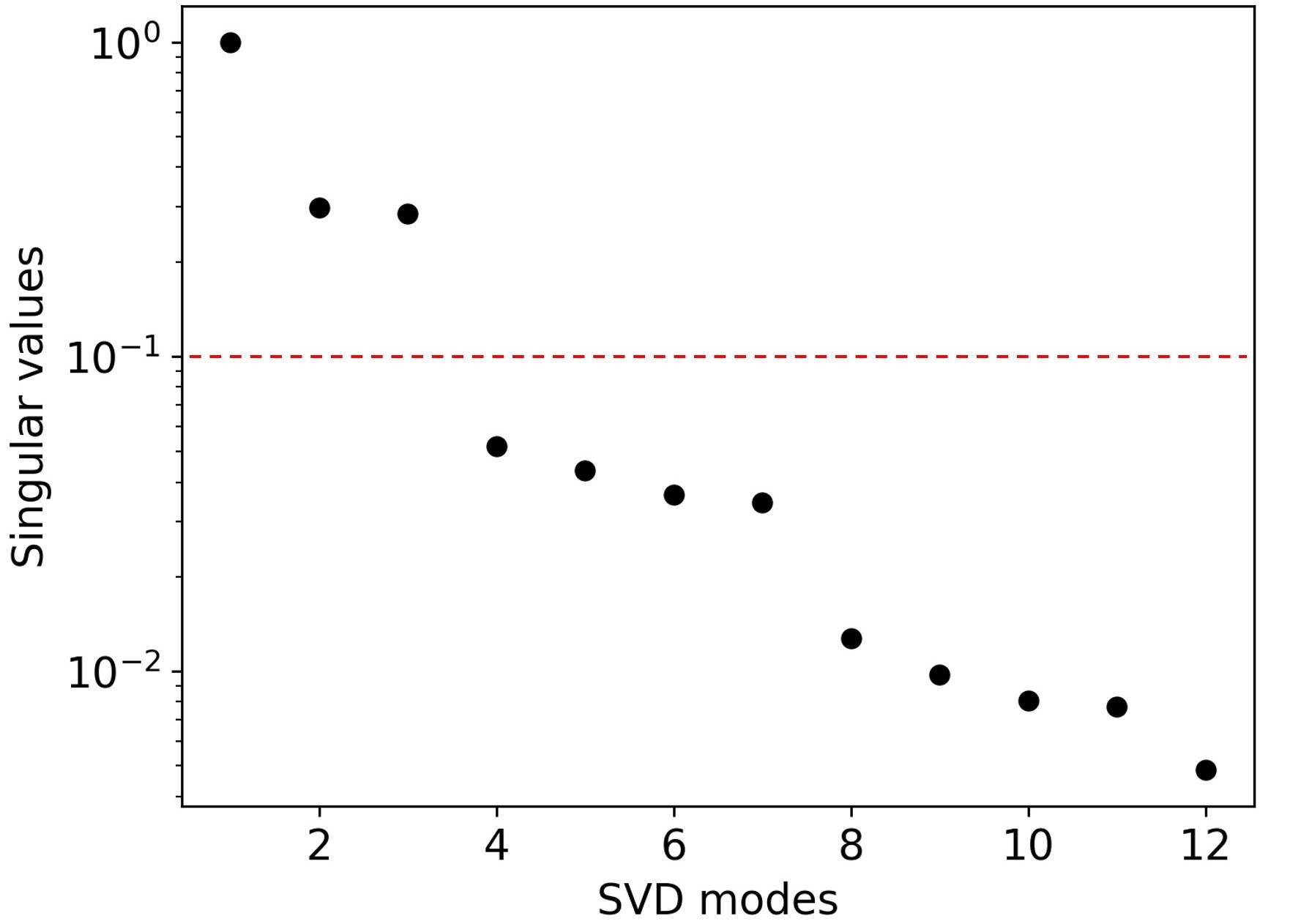}
    \caption{Decay of the retained singular values of the laminar cylinder after applying LC-SVD-DLinear.}
    \label{fig:lcsvd3DCYLmodes}
\end{figure}

\begin{figure}[H]
    \centering
    \includegraphics[width=1\textwidth, angle=0]{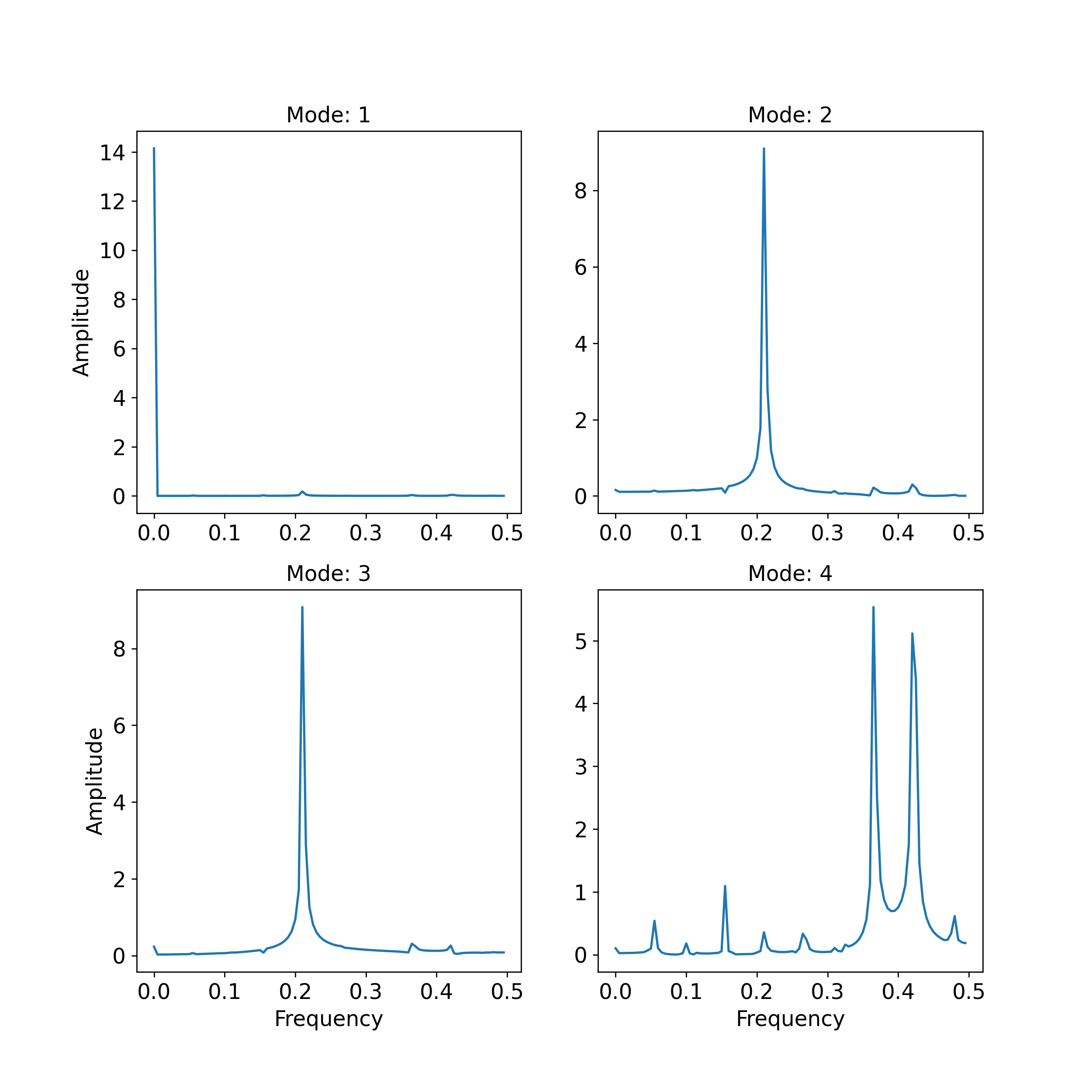}
    \caption{Frequencies of the first 4 temporal modes of the laminar cylinder when applying LC-SVD-DLinear.}
    \label{fig:lcsvd3DCYLfrequencies}
\end{figure}

The application of the LC-SVD-DLinear forecasting model on the numerical three-dimensional laminar cylinder test case starts with the optimal downsampling of the dataset, reducing the spatial dimensionality down to $N_s = 45$ points. After this, noise is filtered out by selecting the first $\bar{N} = 12$ modes, and the reduced tensor is then reconstructed to a clean high resolution version. After this, the reconstructed temporal coefficients, $\bT^{rec}$, are prepared for the forecasting task. Figure \ref{fig:lcsvd3DCYLmodes} displays the singular values decay, which are in descending order based on their energy, while fig. \ref{fig:lcsvd3DCYLfrequencies} shows the frequencies of each of the most important modes after applying a Fourier fast transform (FFT).

Notice in Fig \ref{fig:lcsvd3DCYLmodes} that the first three modes are fundamental based on their amplitudes, since they separate themselves from the rest of the modes by a large decay at $1e^{-1}$. meaning that this remaining group of modes holds less weight in representing the cylinder. More precisely, the first mode represents the mean flow, while modes two and three represent the wake, and the consecutive modes represent smaller scales of the velocity field, progressively becoming less relevant as the mode number increases, until mode 12, where the singular value is of the order of $1e^{-2}$. This is the order of magnitude of the error we assume in the reconstruction. In fig. \ref{fig:lcsvd3DCYLfrequencies} clear dominant frequencies can be identified in the first three modes, these being an immediate frequency for the mean flow and frequencies at harmonic 0.2 for the modes representing the wake, while the fourth mode starts to present some noise in the form of sub-frequencies. This analysis is critical, since identifying the least number of modes necessary to represent the velocity field, and the amount of inherent noise in each one of these will determine the quality of the forecast results. 

\begin{figure}[H]
    \centering
    \includegraphics[width=1\textwidth, angle=0]{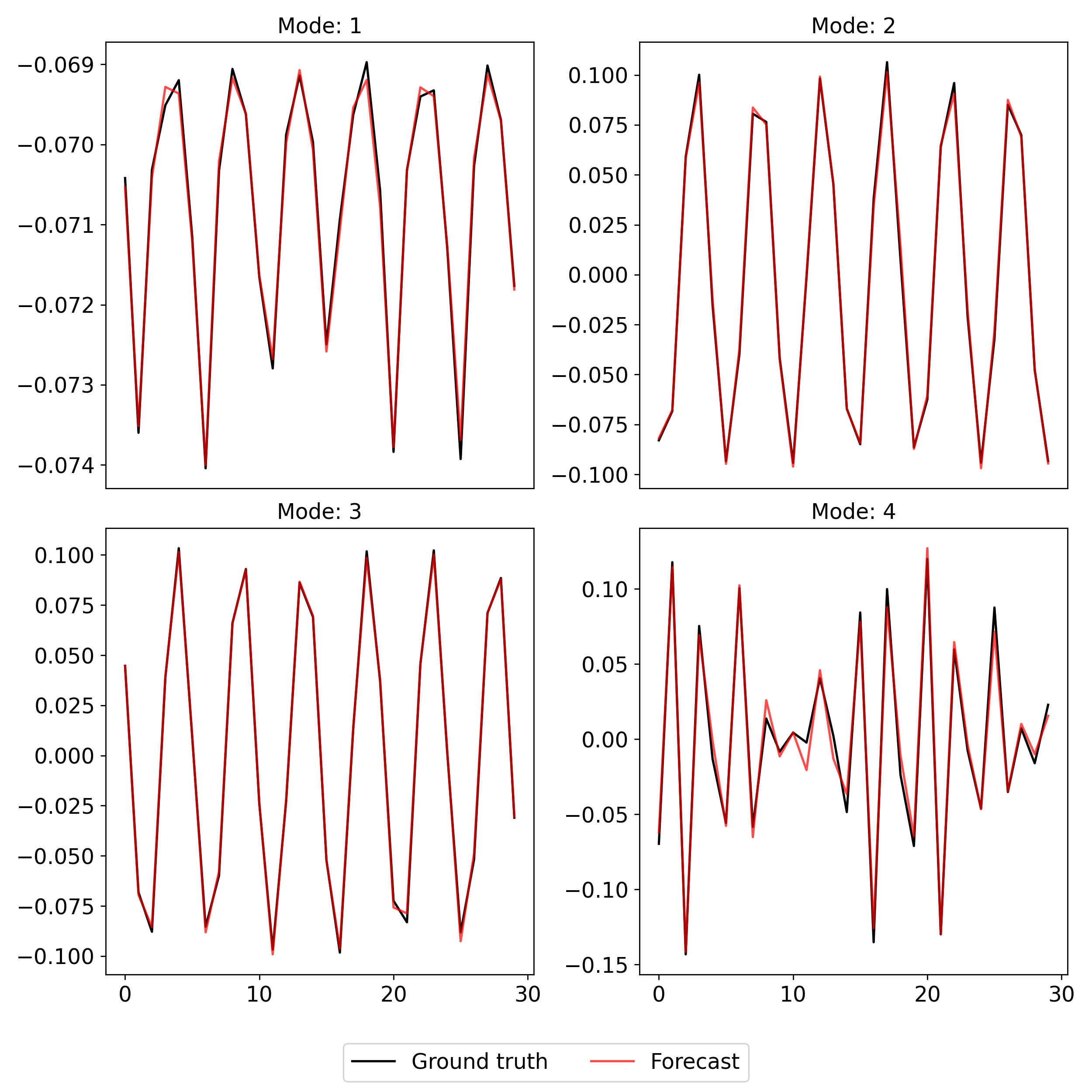}
    \caption{Comparison between the ground truth and forecast temporal coefficient values for the first 4 modes of the three-dimensional laminar cylinder test case when forecasting with LC-SVD-DLinear.}
    \label{fig:testlcsvd3DCYLINDER}
\end{figure}

Figure \ref{fig:testlcsvd3DCYLINDER} illustrates a comparison between the ground truth and the forecast values for the first 4 SVD modes when predicting the test set data. The temporal modes forecast error results are $MAE = 0.454$, and $MSE = 0.282$.

We can see in fig. \ref{fig:testlcsvd3DCYLINDER} that the forecast and ground truth test set temporal modes practically overlap, demonstrating that the model is capable of identifying patterns in each individual mode and predicting the next value in the sequence with high precision. This is mainly due to the model up-sampling and cleaning the data, making temporal patterns easy to learn. The forecast temporal modes are used to reconstruct the test snapshots, with a reconstruction error of $RRMSE = 1.384\%$.  

\begin{figure}[H]
    \centering
    \includegraphics[width=1\textwidth, angle=0]{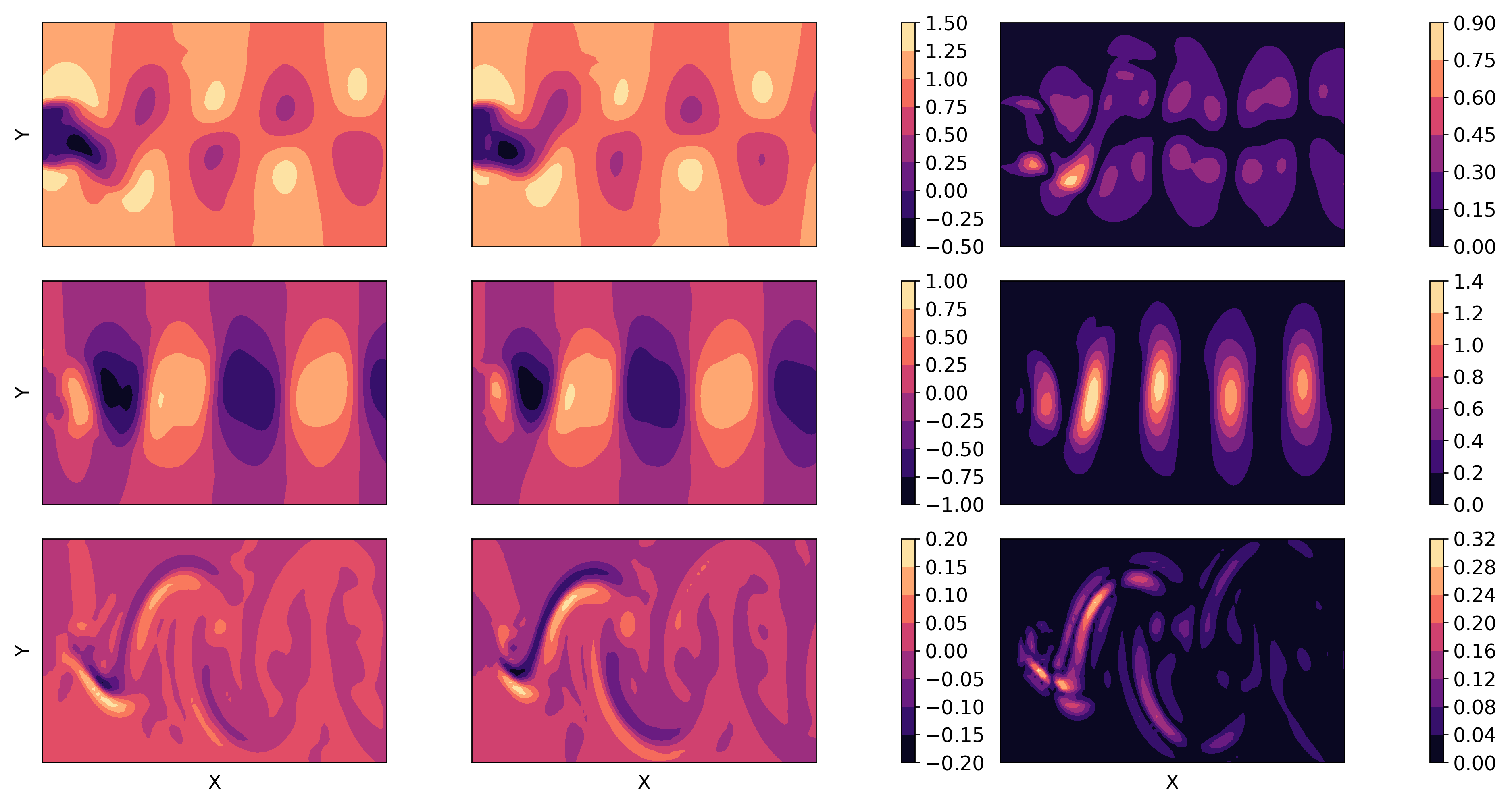}
    \caption{From left to right and top to bottom: the ground truth, forecast and relative error of both, of the streamwise, normal and spanwise velocity components of the three-dimensional cylinder at $Re = 220$, for snapshot $z = 19$, after applying LC-SVD-DLinear.}
    \label{fig:lcsvd3DCYLcont}
\end{figure}

\begin{figure}[H]
    \centering
    \includegraphics[width=1\textwidth, angle=0]{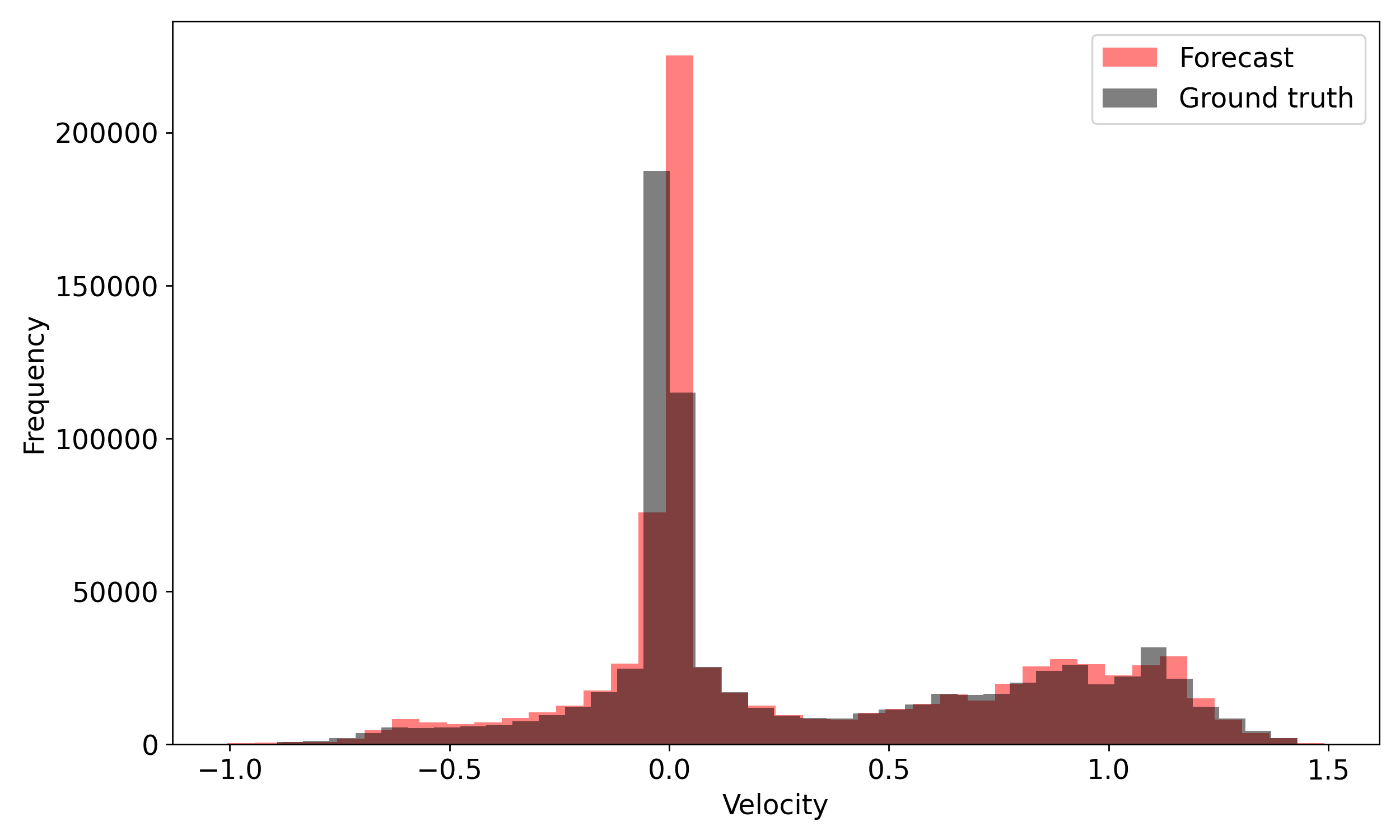}
    \caption{Comparison between the ground truth and forecast data distributions for snapshot $z = 19$ of the laminar cylinder after applying LC-SVD-DLinear.}
    \label{fig:lcsvd3DCYLdist}
\end{figure}

When it comes to snapshot reconstruction, the maximum error is given at snapshot $z = 19$, with a Wasserstein distance of $w = 0.0105$ and a reconstruction MAE of $MAE_{rec} = 0.135m/s$. Figure \ref{fig:lcsvd3DCYLcont} illustrates a comparison between the ground truth and the forecast for each component of the highest error snapshot, while fig. \ref{fig:lcsvd3DCYLdist} compares their data distributions.

The relative error contour maps from fig. \ref{fig:lcsvd3DCYLcont} reveal a slight drift between the ground truth and forecast data for the snapshot where maximum error is encountered. This drift can also be seen in \ref{fig:lcsvd3DCYLdist}, where it is apparent that the mode (most frequent velocity value) has a slight positive shift. Despite the shift, the overall reconstruction of the test data is highly accurate.

\begin{figure}[H]
    \centering
    \includegraphics[width=1\textwidth, angle=0]{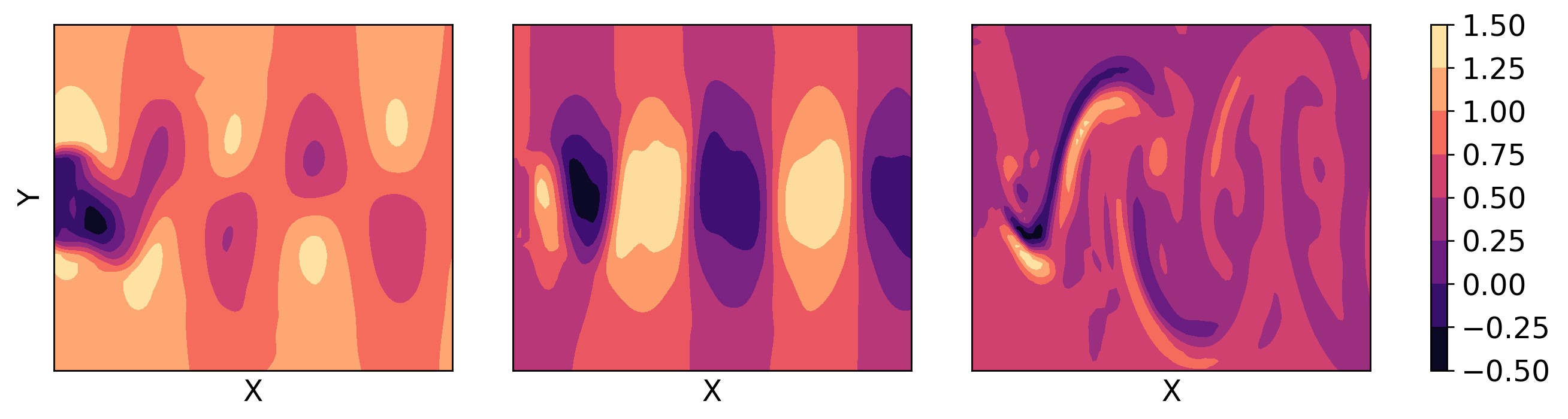}
    \caption{From left to right: forecast of the streamwise, normal and spanwise velocity components of the three-dimensional cylinder at $Re = 220$, for snapshot $z = 500$, forecast by the LC-SVD-DLinear model.}
    \label{fig:lcsvd3DCYLforecast}
\end{figure}

Upon studying the result for the test set, the LC-SVD-DLinear model is used to forecast the next $N_{snap} = 1000$ values for each of the temporal modes, extending the test data. These forecast values are then used to reconstruct the snapshots, with fig. \ref{fig:lcsvd3DCYLforecast} illustrating the reconstruction for snapshot 500.

\begin{figure}[H]
    \centering
    \includegraphics[width=1\textwidth, angle=0]{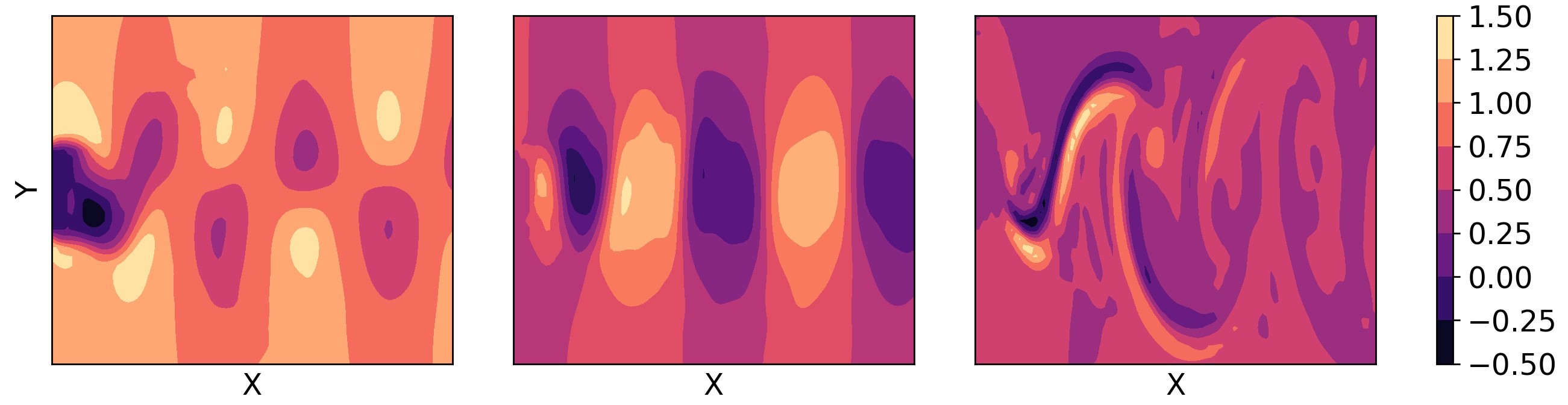}
    \caption{From left to right: forecast of the streamwise, normal and spanwise velocity components of the three-dimensional cylinder at $Re = 220$, for snapshot $z = 1000$, forecast by the LC-SVD-DLinear model.}
    \label{fig:lcsvd3DCYLlatest}
\end{figure}

The last reconstructed snapshot from the new subset of data can be seen in fig. \ref{fig:lcsvd3DCYLlatest}.

The forecast results shown in both Fig. \ref{fig:lcsvd3DCYLforecast} and \ref{fig:lcsvd3DCYLlatest} demonstrate that the LC-SVD-DLinear model has been able to learn the trend and seasonality of the temporal SVD modes, making it capable of forecasting a large number of snapshots from a small input sequence. Despite mode 4 presenting some noise in its spectral space (fig. \ref{fig:lcsvd3DCYLfrequencies}), this has not had any impact on the forecast. This can be seen directly when comparing both plots, since there is no significant drift in the reconstructed velocity field, with the range of velocity values remaining within the original interval.  



Next, the LC-HOSVD-DLinear model is used to forecast the temporal modes. One of the main advantages of this model is that HOSVD allows for a component-based decomposition, meaning that noise is filtered for each component and not generalized, like in SVD. 

\begin{figure}[H]
    \centering
    \includegraphics[width=1\textwidth, angle=0]{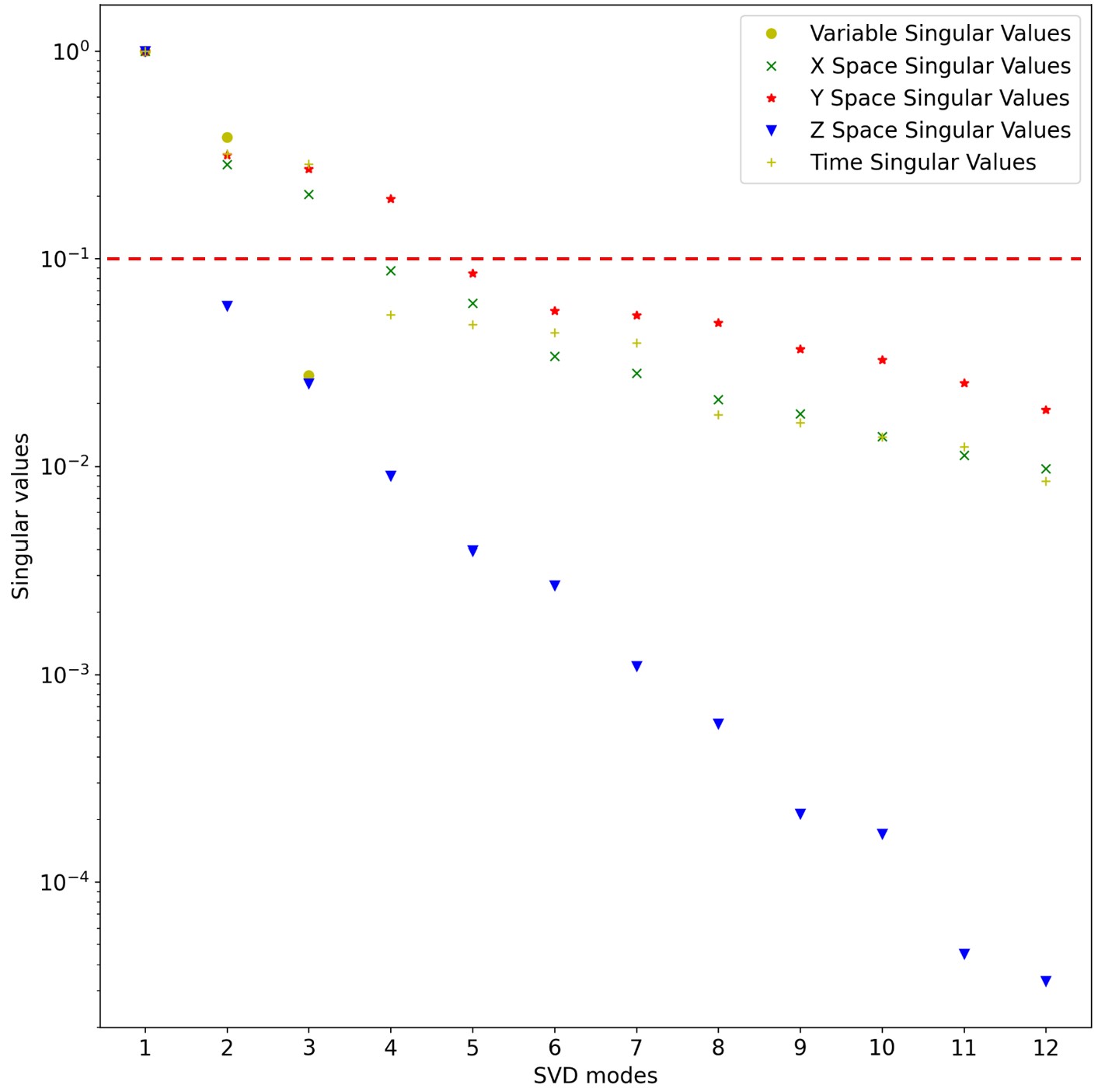}
    \caption{Decay of the retained singular values of the laminar cylinder when applying LC-HOSVD-DLinear.}
    \label{fig:lchosvd3DCYLmodes}
\end{figure}

\begin{figure}[H]
    \centering
    \includegraphics[width=1\textwidth, angle=0]{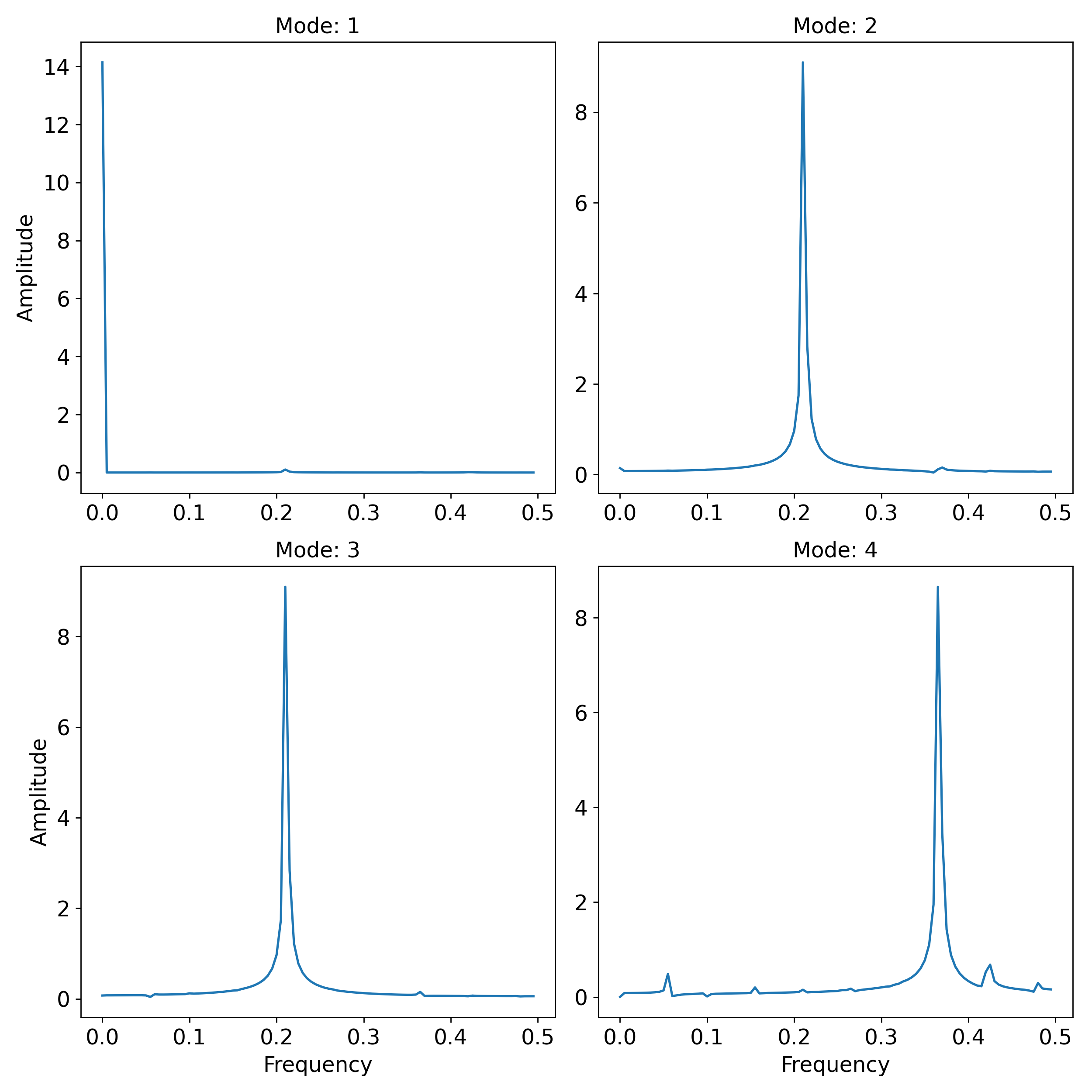}
    \caption{Frequencies of the first 4 temporal modes of the laminar cylinder upon applying LC-HOSVD-DLinear.}
    \label{fig:lchosvd3DCYLfrequencies}
\end{figure}

Figure \ref{fig:lchosvd3DCYLmodes} shows the singular values decay for each component, and fig. \ref{fig:lchosvd3DCYLfrequencies} displays the frequencies of the most relevant modes.

 A steadier decay of the singular values of the X and Y spaces, as well as the time data can be seen in fig. \ref{fig:lchosvd3DCYLmodes}. On the other hand, the singular values Z space decay rapidly, meaning that the data contained in this component is less relevant than the others. Meanwhile, studying the frequency profiles, observe in fig. \ref{fig:lchosvd3DCYLfrequencies} that the frequencies for the first three modes are practically identical to those observed in fig. \ref{fig:lcsvd3DCYLfrequencies}, while the fourth mode appears to have cleaner frequencies, meaning that they contain less noise (mode noise has been filtered out during the HOSVD loop).

\begin{figure}[H]
    \centering
    \includegraphics[width=1\textwidth, angle=0]{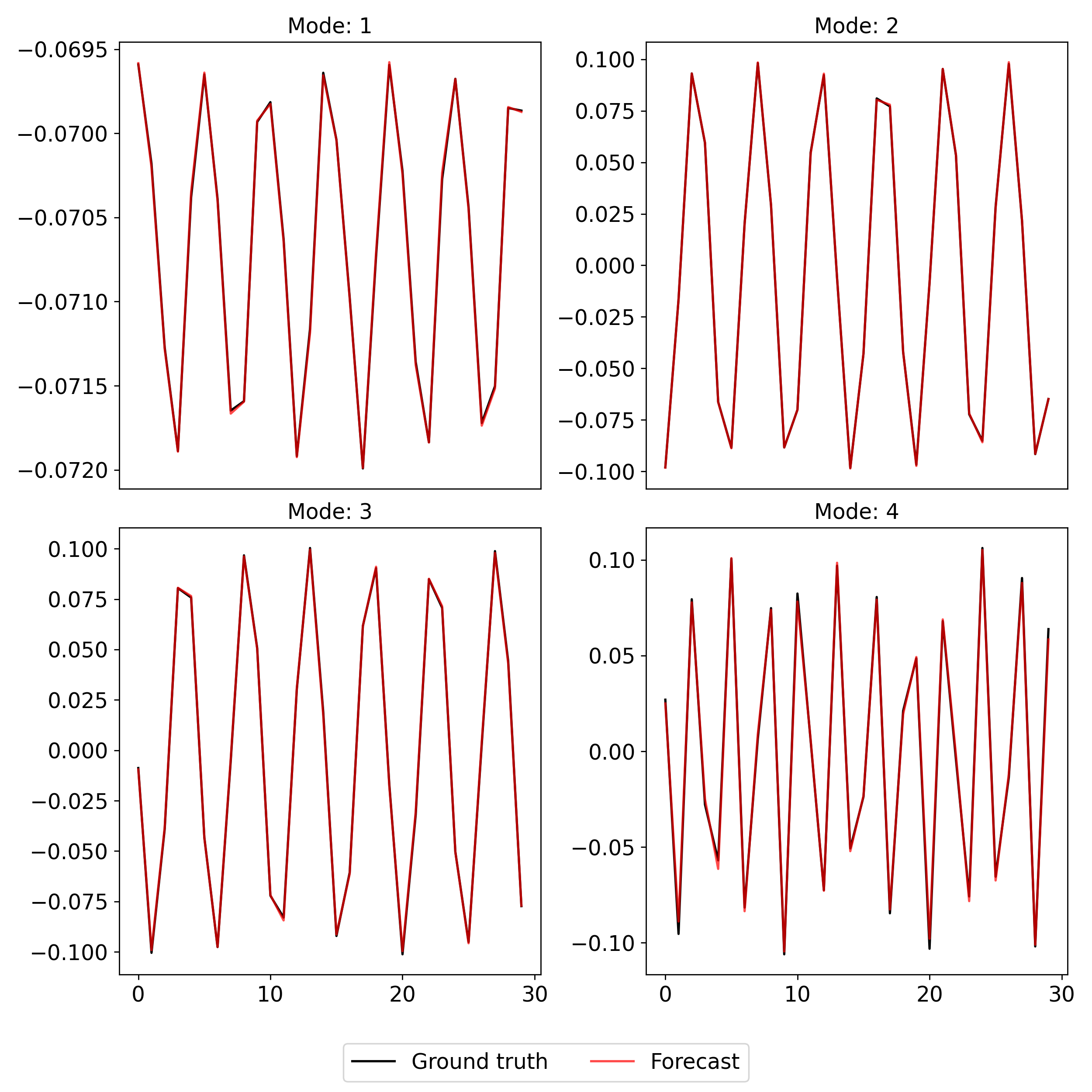}
    \caption{Comparison between the ground truth and forecast temporal coefficient values for the first 4 modes of the three-dimensional laminar cylinder test case, using LC-HOSVD-DLinear.}
    \label{fig:testlchosvd3DCYLINDER}
\end{figure}

In fig. \ref{fig:testlchosvd3DCYLINDER} the temporal coefficients forecast results are compared to the ground truth for the first 4 SVD modes. The forecast error metrics achieved using LC-HOSVD-DLinear are $MAE = 0.491$, and $MSE = 0.315$. 

\begin{figure}[H]
    \centering
    \includegraphics[width=1\textwidth, angle=0]{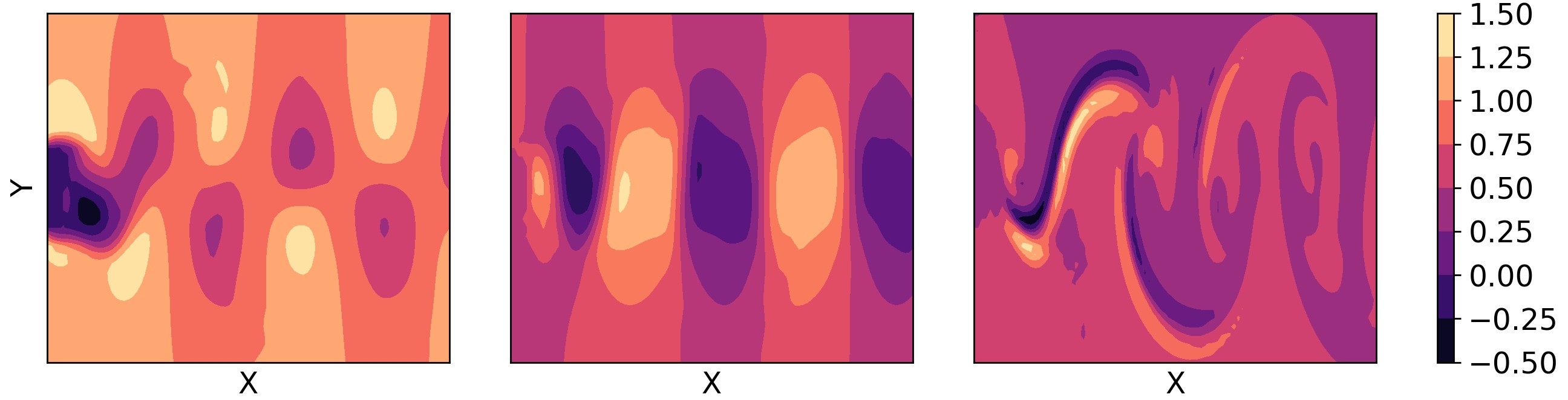}
    \caption{From left to right: forecast of the streamwise, normal and spanwise velocity components of the three-dimensional cylinder at $Re = 220$, for snapshot $z = 500$, applying LC-HOSVD-DLinear.}
    \label{fig:lchosvd3DCYLforecast}
\end{figure}

The forecast metrics obtained using LC-HOSVD-DLinear are similar to those observed to LC-SVD-DLinear, hence why the test data predictions in fig. \ref{fig:lcsvd3DCYLforecast} and fig. \ref{fig:lchosvd3DCYLforecast} are similar. The temporal coefficients forecast by LC-HOSVD-DLinear are then used to reconstruct the test snapshots, resulting in a reconstruction error of $RRMSE = 0.571\%$, which is lower than the accomplished by LC-SVD-DLinear. 

\begin{figure}[H]
    \centering
    \includegraphics[width=1\textwidth, angle=0]{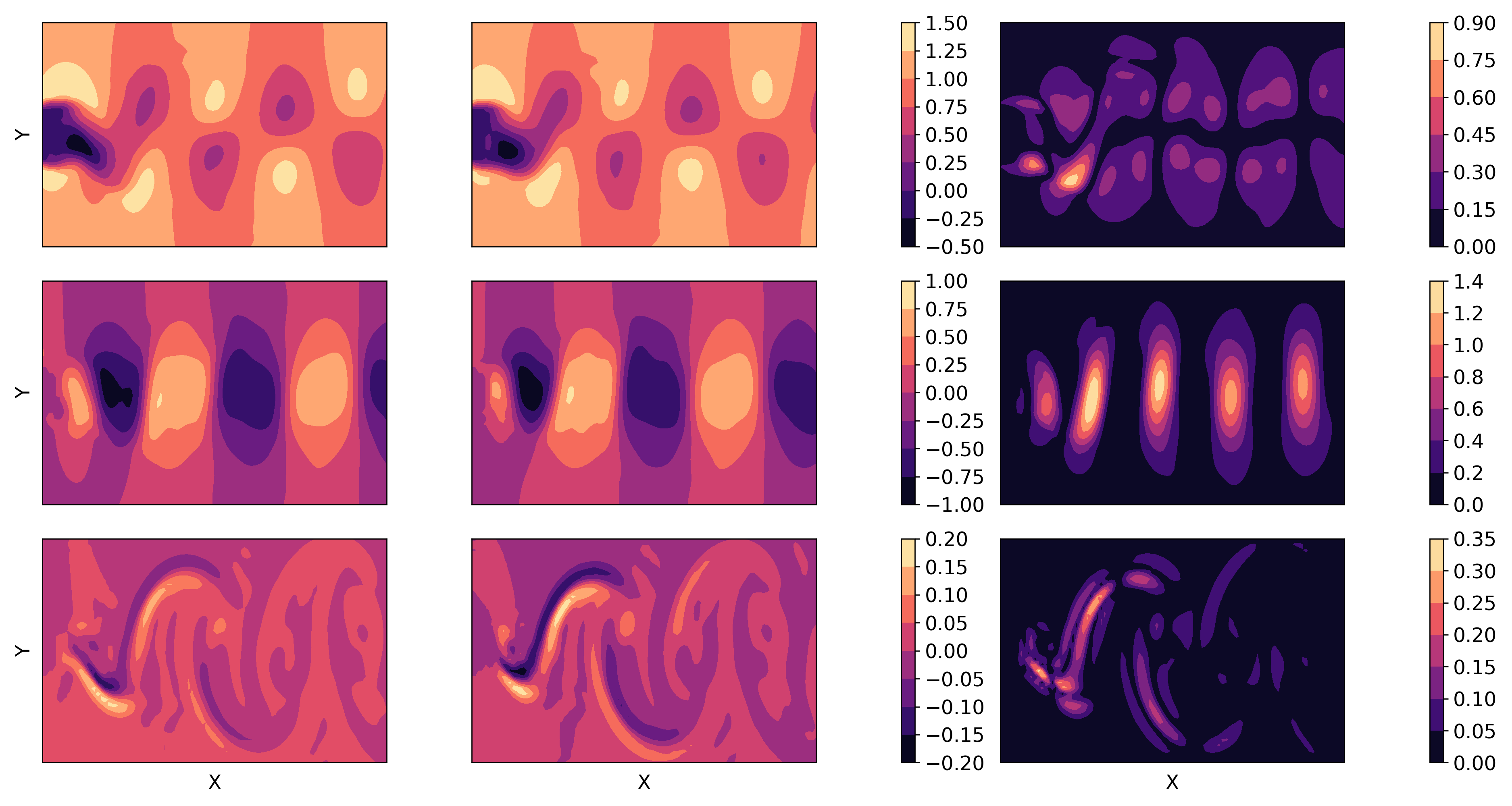}
    \caption{From left to right and top to bottom: the ground truth, forecast and relative error of both, of the streamwise, normal and spanwise velocity components of the three-dimensional cylinder at $Re = 220$, for snapshot $z = 19$.}
    \label{fig:lchosvd3DCYLcont}
\end{figure}

\begin{figure}[H]
    \centering
    \includegraphics[width=1\textwidth, angle=0]{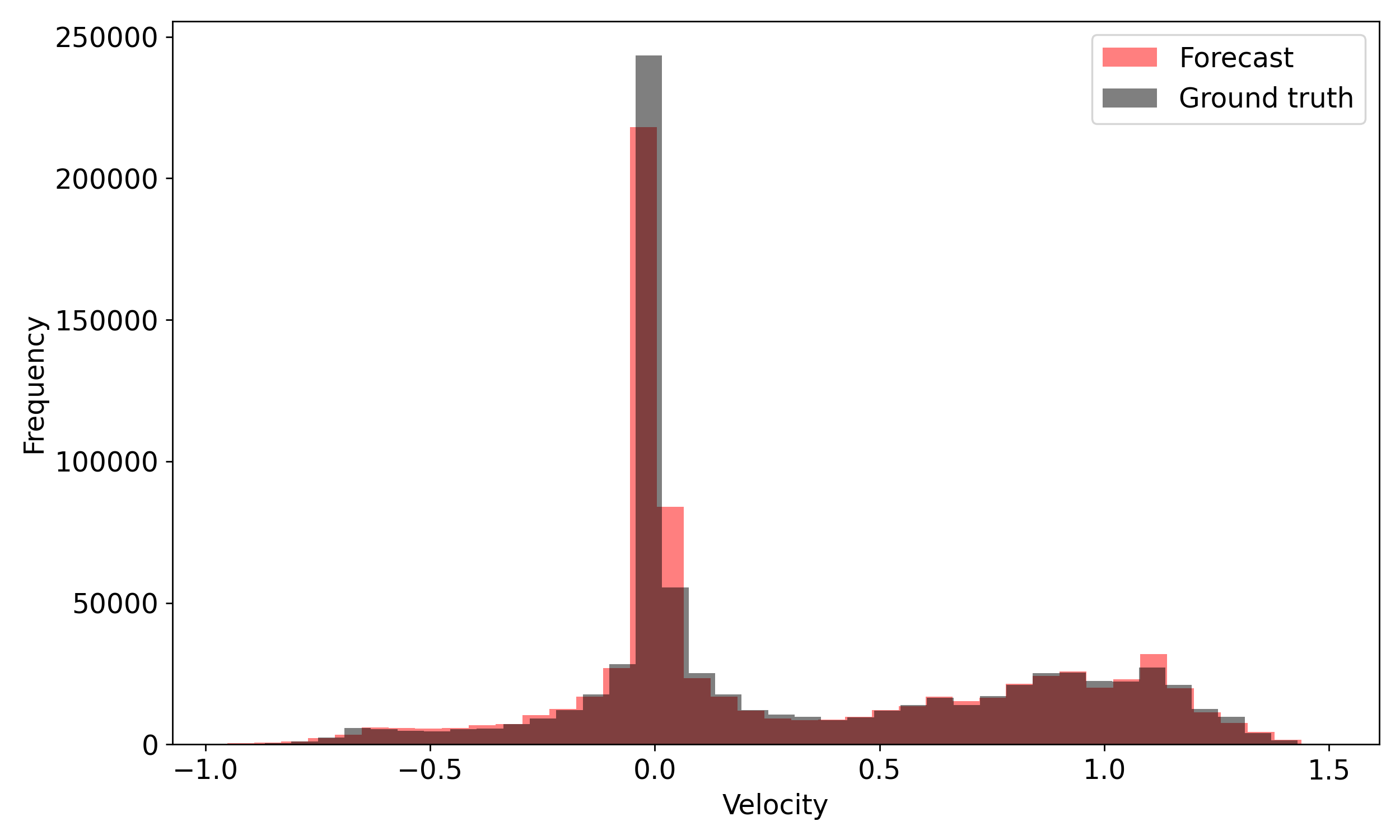}
    \caption{Comparison between the ground truth and forecast data distributions for snapshot $z = 19$.}
    \label{fig:lchosvd3DCYLdist}
\end{figure}

The Wasserstein distance and reconstruction MAE values, which are in the same range as for LC-SVD-DLinear, reveal that the highest error snapshot is, once again, $z = 19$, with a Wasserstein distance equal to $w = 0.0102$, and a reconstruction MAE equal to $MAE_{rec} = 0.1355 m/s$, which can be expected given the simplicity of the data. Figures \ref{fig:lchosvd3DCYLcont} and \ref{fig:lchosvd3DCYLdist} analyze this snapshot in detail. 

The relative error contour maps from fig. \ref{fig:lchosvd3DCYLcont} reveal a very discrete drift between data distributions for the highest error snapshot, which stands out mostly for the normal velocity. When analyzing fig. \ref{fig:lchosvd3DCYLdist}, it is clear that, where the ground truth normal velocity is 0, the LC-HOSVD-DLinear has predicted a slightly larger velocity value, in the order of 0.1 $m/s$ approximately. When comparing these distributions to those observed for the same snapshot ($z = 19$) when using the LC-SVD-DLinear model in fig. \ref{fig:lcsvd3DCYLdist}, it is apparent that this drift is more subtile, which is attributed to a mode robust noise filtering.

After studying the results from the test set, the LC-HOSVD-DLinear model one again predicts the following $N_{snap} = 1000$ temporal coefficient values for each retained mode, and uses these to reconstruct new snapshots. Figure \ref{fig:lchosvd3DCYLforecast} displays snapshot 500 from the new forecast subset. 

\begin{figure}[H]
    \centering
    \includegraphics[width=1\textwidth, angle=0]{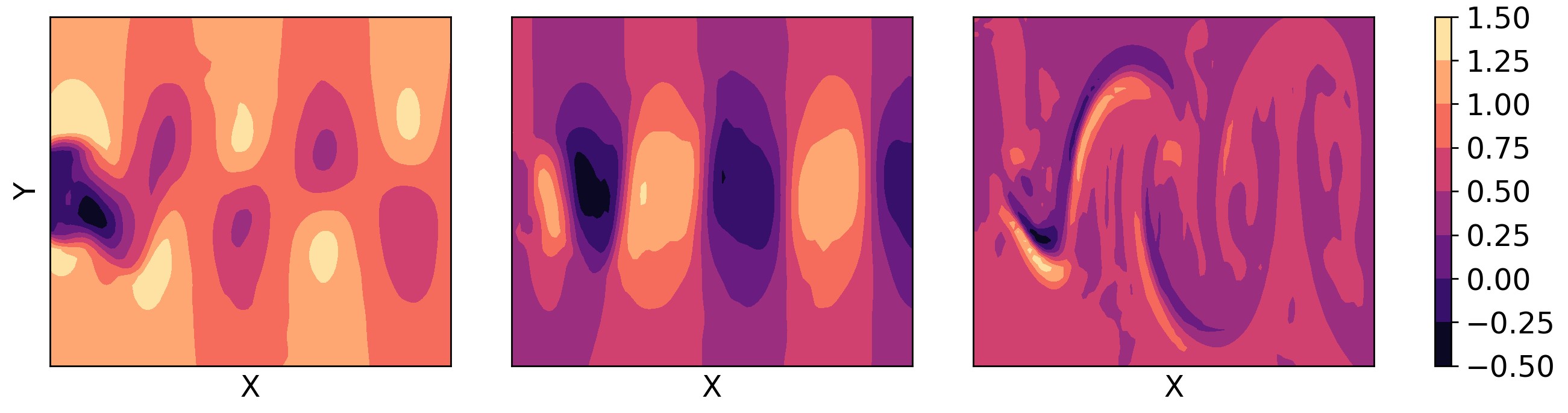}
    \caption{From left to right: forecast of the streamwise, normal and spanwise velocity components of the three-dimensional cylinder at $Re = 220$, for snapshot $z = 1000$, applying LC-HOSVD-DLinear.}
    \label{fig:lchosvd3DCYLlatest}
\end{figure}

The last snapshot corresponding to the forecast set can be seen in fig. \ref{fig:lchosvd3DCYLlatest}.

Once again, the forecast results in fig. \ref{fig:lchosvd3DCYLforecast} and \ref{fig:lchosvd3DCYLlatest} are outstanding, similar to those achieved with the LC-SVD-DLinear model (fig. \ref{fig:lcsvd3DCYLforecast} and \ref{fig:lcsvd3DCYLlatest}). Yet again, there has been no significant drift in the data, since the limit velocity values have not varied from the range of values observed for the ground truth test data, which can be seen in fig. \ref{fig:lchosvd3DCYLcont}. This is expected given the results achieved for LC-SVD-DLinear, since HOSVD performs better at cleaning noise from data.


\subsection{Turbulent cylinder forecast \label{solTC}}
\begin{figure}[H]
    \centering
    \includegraphics[width=0.8\textwidth, angle=0]{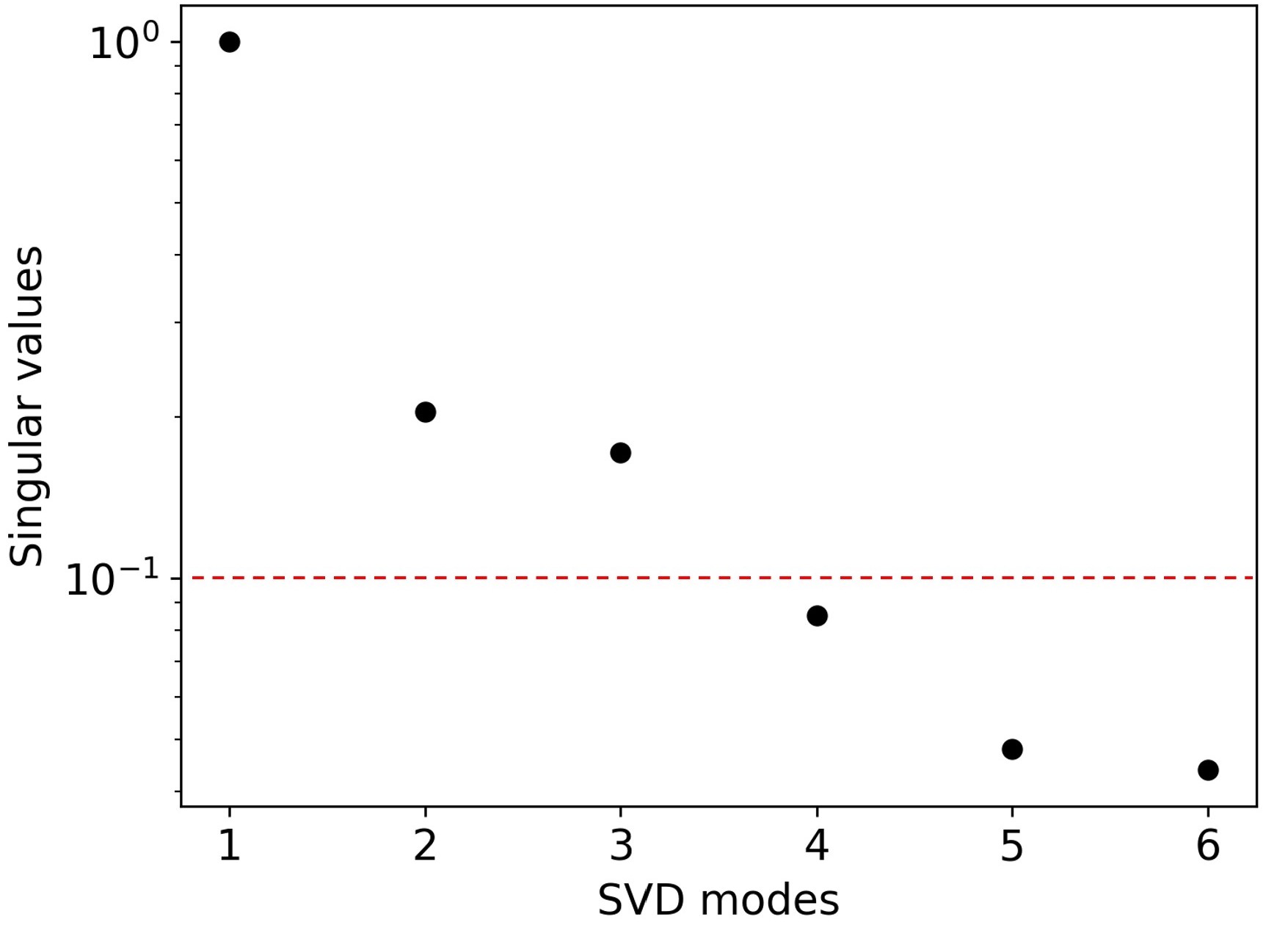}
    \caption{Decay of the retained singular values of the turbulent cylinder after applying LC-SVD-DLinear.}
    \label{fig:lcsvd2DCYLmodes}
\end{figure}

\begin{figure}[H]
    \centering
    \includegraphics[width=1\textwidth, angle=0]{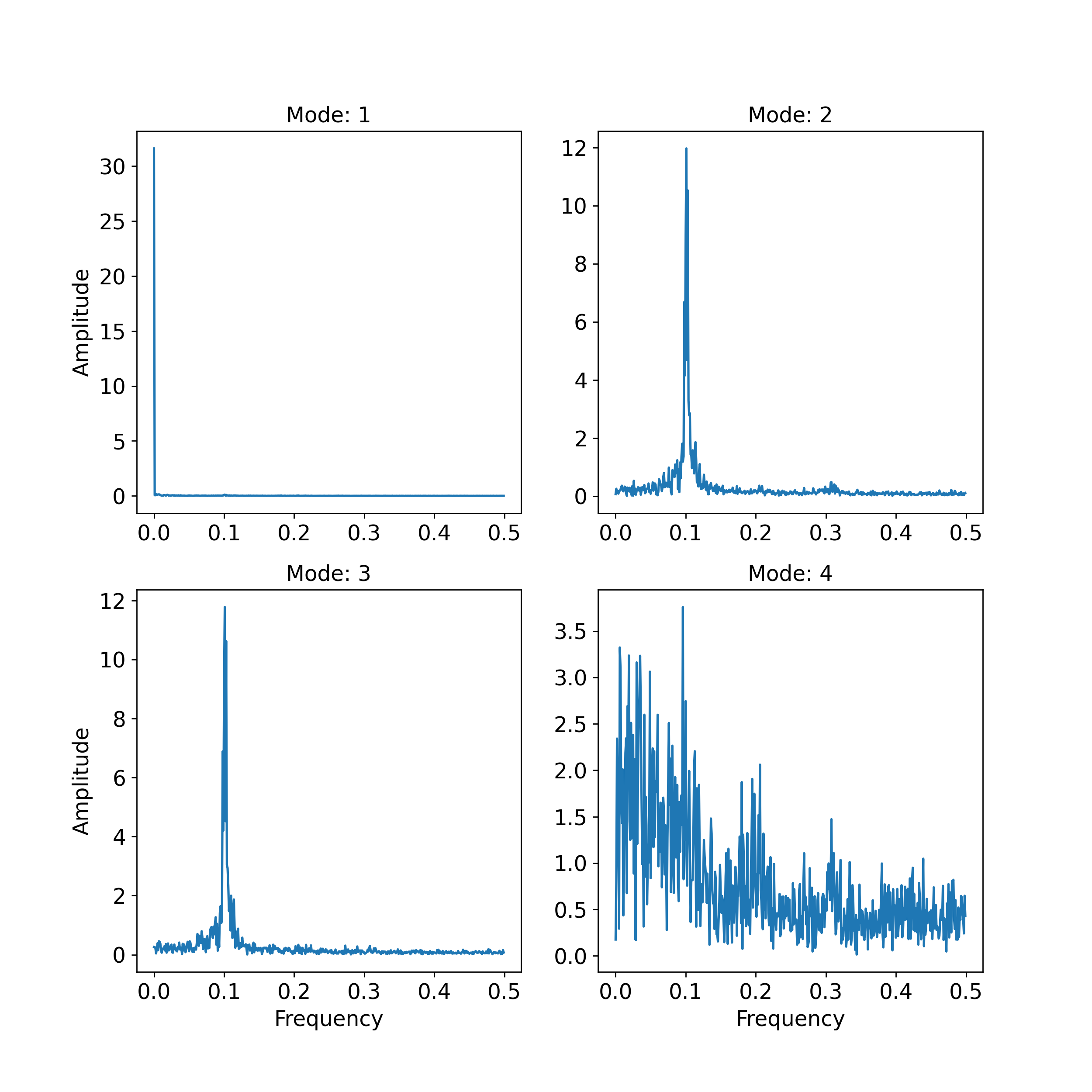}
    \caption{Frequencies of the first 4 temporal modes of the turbulent cylinder after using LC-SVD-DLinear.}
    \label{fig:lcsvd2DCYLfrequencies}
\end{figure}

When it comes to the turbulent cylinder at $Re = 2600$ test case, LC-SVD-DLinear starts by optimally down-sampling, de-noising and reconstructing the dataset using $N_s = 40$ sensors and $\bar{N} = 6$ SVD modes. By selecting a low number of singular values, therefore, keeping only the robust modes, we are able to make spatio-temporal patterns easier to learn. As done for the laminar cylinder test case, the singular values and temporal modes frequencies are analyzed in fig. \ref{fig:lcsvd2DCYLmodes} and fig. \ref{fig:lcsvd2DCYLfrequencies}, respectively.

In fig. \ref{fig:lcsvd2DCYLmodes}, two groups of modes can be identified, with the threshold between groups located at $1e^{-1}$. The first and most energetic mode represents the mean flow of the velocity field, while modes 2 and 3 represent the turbulent wake. The last three modes represent substructures, such as small scales or non-linear dynamics, with values close to the $1e^{-2}$ region. This claim is ratified in fig. \ref{fig:lcsvd2DCYLfrequencies}, where an analysis of the frequencies when applying a FFT to the first 4 temporal modes shows that the first three modes have unique frequencies, immediate for mode 1, and at harmonic 0.1 for modes 2 and 3, while mode 4, which represents non-linear dynamics, has a high spectral complexity, with dominant frequencies at harmonic 0.1, 0.2 and 0.3 (approximately). Given its complexity, this mode can be misinterpreted as noise but, despite this, the model is capable of identifying the periodic pattern in the data, defined by the dominant modes.

\begin{figure}[H]
    \centering
    \includegraphics[width=1\textwidth, angle=0]{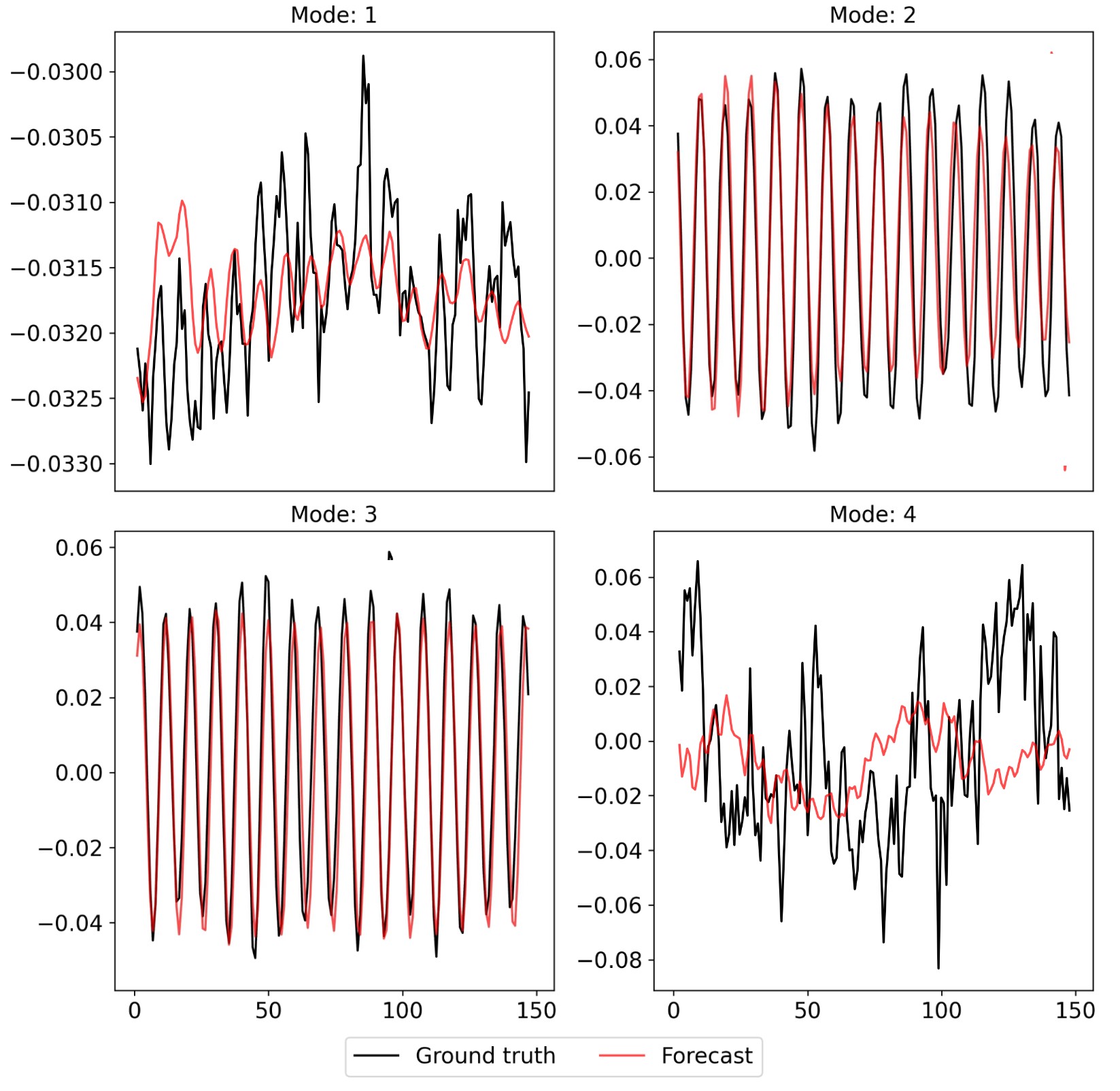}
    \caption{Comparison between the ground truth and forecast temporal coefficient values for the first 4 modes of the turbulent cylinder test case, when applying LC-SVD-DLinear.}
    \label{fig:lcsvd2600test}
\end{figure}

Figure \ref{fig:lcsvd2600test} presents the forecast results when predicting the first 3 temporal modes from the test set. The test data forecast errors metrics are $MAE = 0.339$ and $MSE = 0.167$.

\begin{figure}[H]
    \centering
    \includegraphics[width=1\textwidth, angle=0]{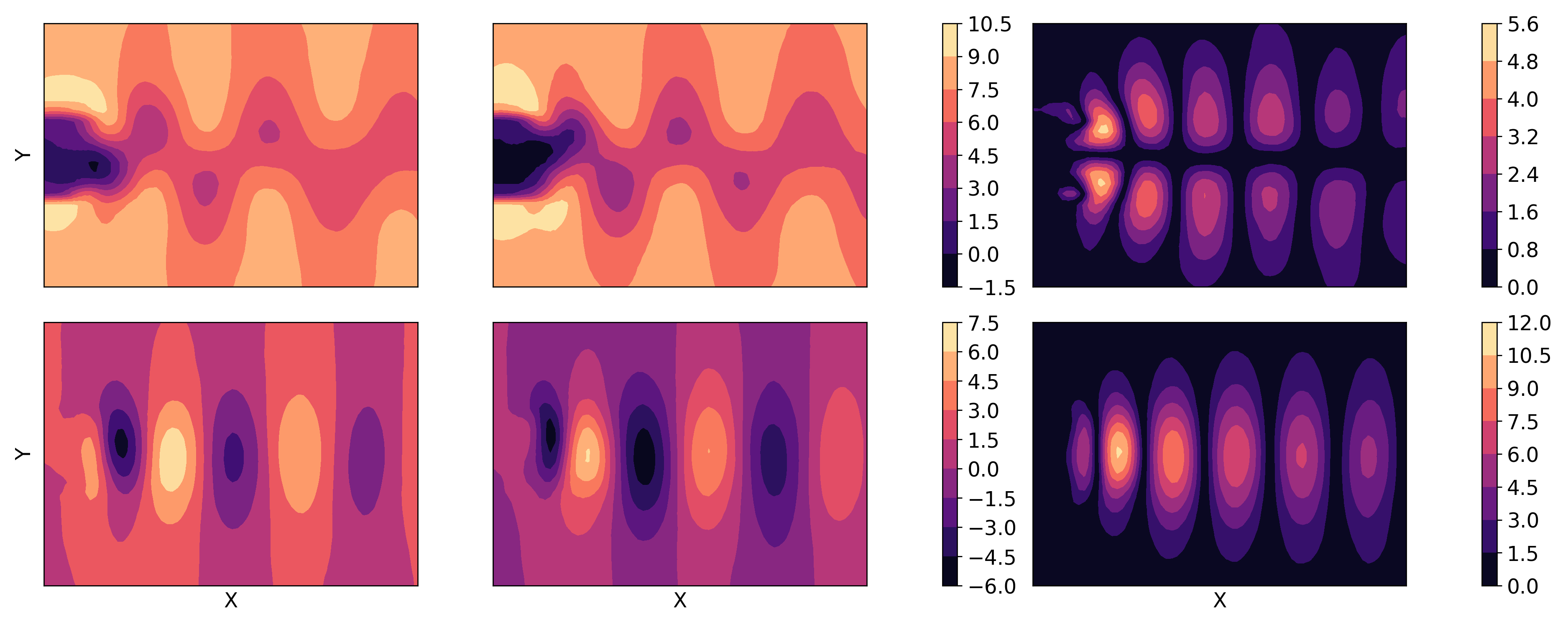}
    \caption{From left to right and top to bottom: the ground truth, forecast and relative error of both, of the streamwise and normal velocity components of the turbulent cylinder at $Re = 2600$, for snapshot $z = 146$, after applying LC-SVD-DLinear.}
    \label{fig:lcsvd2DCYLcont}
\end{figure}

It is apparent that, since the flow is turbulent and, therefore, highly complex, patterns in the some of the temporal coefficients are not as visible, or may even be non-existent. Take the first mode from fig. \ref{fig:lcsvd2600test}, which represents the mean flow, where the model is capable of identifying variations of order of  $\sim O(10^{-4})$, representing noise. However, in the fourth mode, the model has a harder time to identify a clear pattern so well, since the temporal evolution is mostly white noise, caused by low frequencies, hiding the real dynamics of the flow. Nevertheless, the model finds a periodic pattern. Modes 2 and 3 represent the temporal evolution of the wake, which is driven by the main flow instabilities, the von Karman vortex stream, that the model has easily learned and predicted. The reconstruction of the test set snapshots results in a reconstruction error of $RRMSE = 11.554\%$. Figure \ref{fig:lcsvd2DCYLcont} shows the highest error snapshot, $z = 146$ with a Wasserstein distance equal to $w = 0.4$ and $MAE_{rec} = 1.3m/s$.

\begin{figure}[H]
    \centering
    \includegraphics[width=1\textwidth, angle=0]{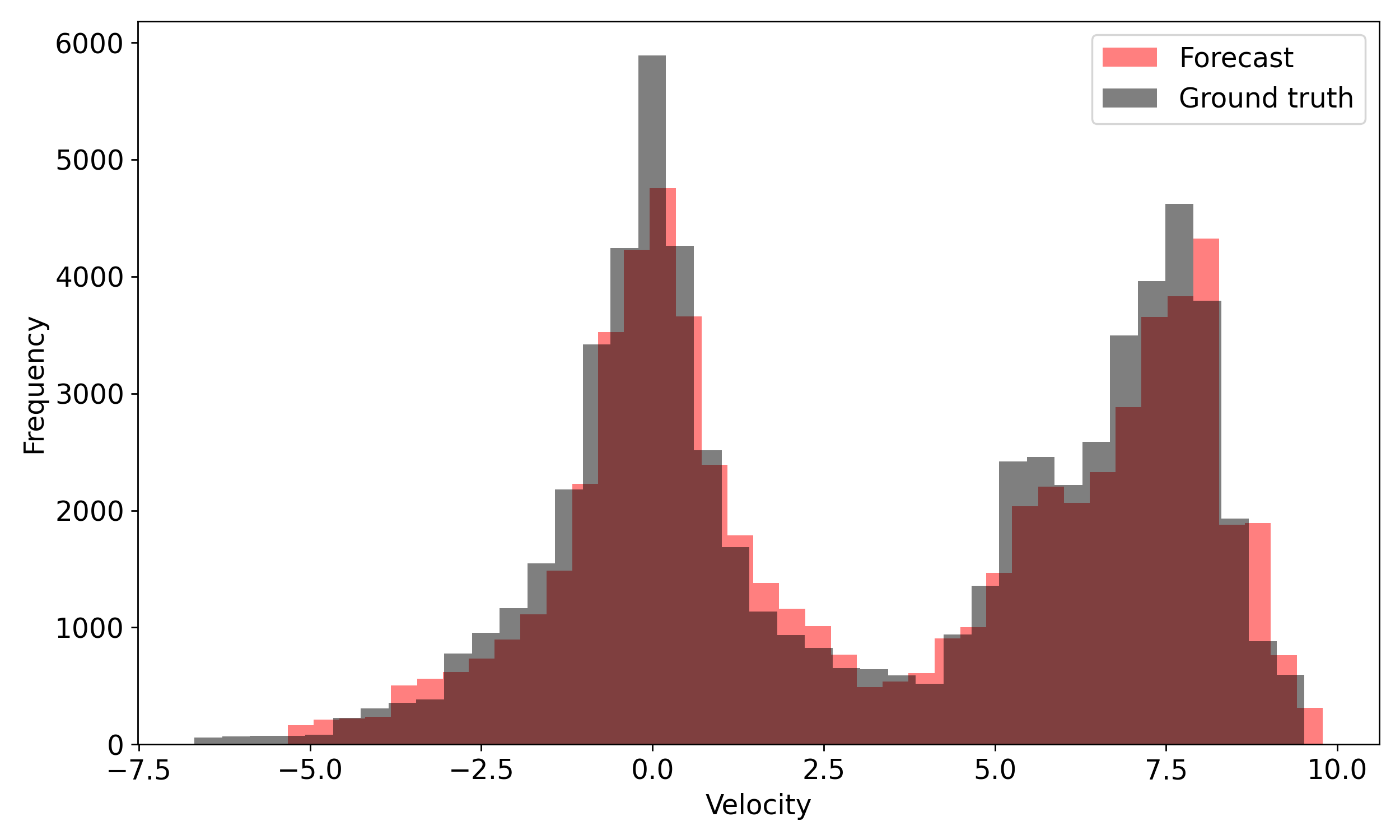}
    \caption{Comparison between the ground truth and forecast data distributions for snapshot $z = 146$ of the turbulent cylinder at $Re = 2600$, upon applying LC-SVD-DLinear.}
    \label{fig:lcsvd2DCYLdist}
\end{figure}

The contour maps of fig. \ref{fig:lcsvd2DCYLcont} show that the relative error follows a similar pattern as the velocity fields, which is caused by the drift in data. This drift between the ground truth and forecast data, for snapshot $z = 146$, can be better explained in support of fig. \ref{fig:lcsvd2DCYLdist}.

\begin{figure}[H]
    \centering
    \includegraphics[width=1\textwidth, angle=0]{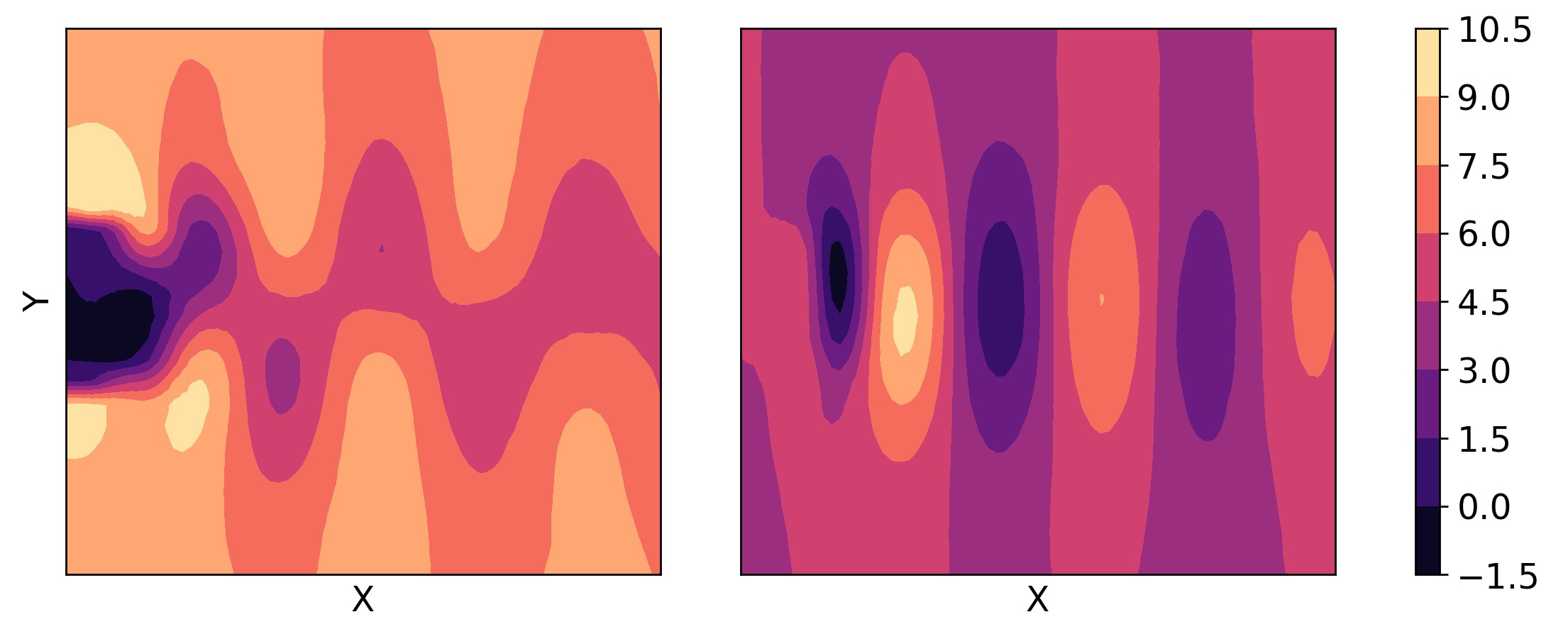}
    \caption{From left to right: forecast of the streamwise and normal velocity components of the turbulent cylinder at $Re = 2600$, for snapshot $z = 100$, using LC-SVD-DLinear.}
    \label{fig:lcsvd2DCYLforecast}
\end{figure}

Given the complexity of the dataset, the drift in data distributions is more apparent. Notice that the forecast data does accomplish to match the data distribution in most of the cases, with a light shift in the positive direction. The small differences found are consistent with the results visible in fig. \ref{fig:lcsvd2DCYLcont}, where the forecast snapshot velocity fields have magnitudes slightly larger than the actual data. After this analysis, the next 200 temporal coefficients for each mode are predicted, and the snapshots are reconstructed. Figure \ref{fig:lcsvd2DCYLforecast}, displays snapshot 100 of the new sequence.

\begin{figure}[H]
    \centering
    \includegraphics[width=1\textwidth, angle=0]{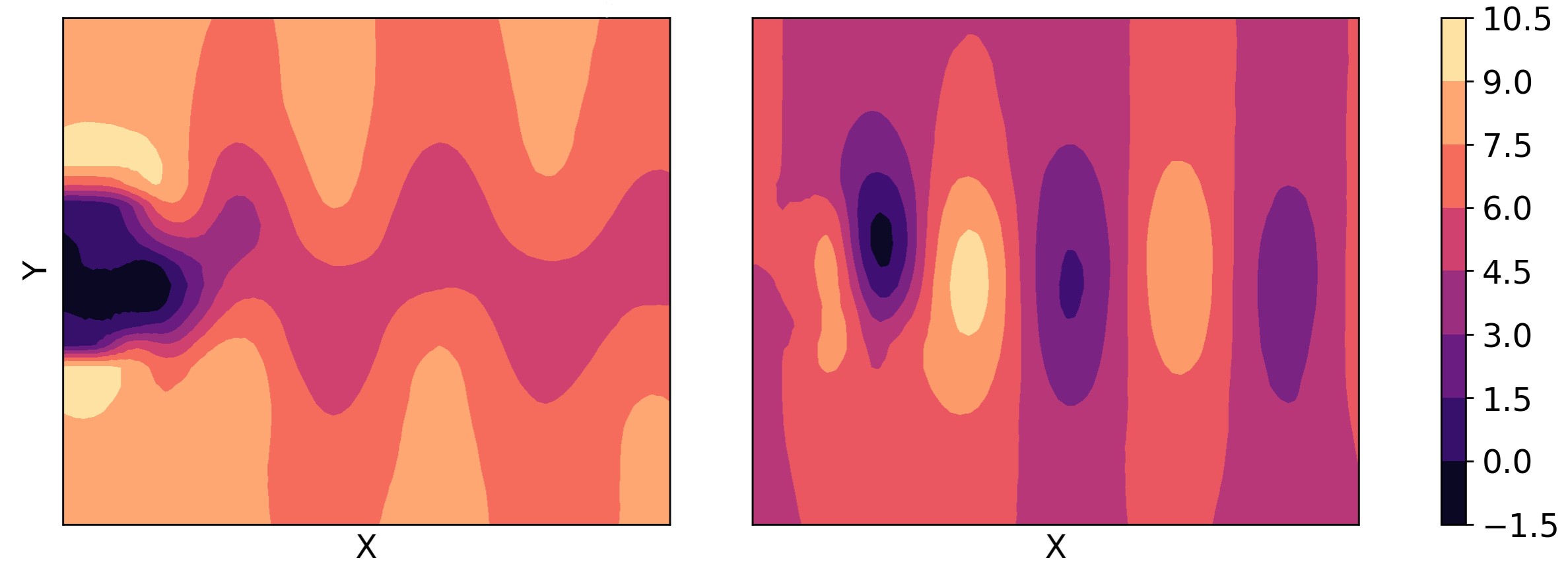}
    \caption{From left to right: forecast of the streamwise and normal velocity components of the turbulent cylinder at $Re = 2600$, for snapshot $z = 200$, using LC-SVD-DLinear.}
    \label{fig:lcsvd2DCYLlatest}
\end{figure}

Figure \ref{fig:lcsvd2DCYLlatest} shows the last reconstructed snapshot ($z = 200$) of this newly generated set of snapshots.  

Despite the complexity of the data, in both fig. \ref{fig:lcsvd2DCYLforecast} and \ref{fig:lcsvd2DCYLlatest} we can see that the small drift has not disrupted the velocity scale, since the range of values fall in line with those observed in the test set (fig. \ref{fig:lcsvd2600test}), 200 snapshots later. It is also important to note that the quality of the forecast results will vary depending on the sequence length, since larger lengths may help capture more frequencies. However, for this case, the optimal length was found to be 100.

\begin{figure}[H]
    \centering
    \includegraphics[width=1\textwidth, angle=0]{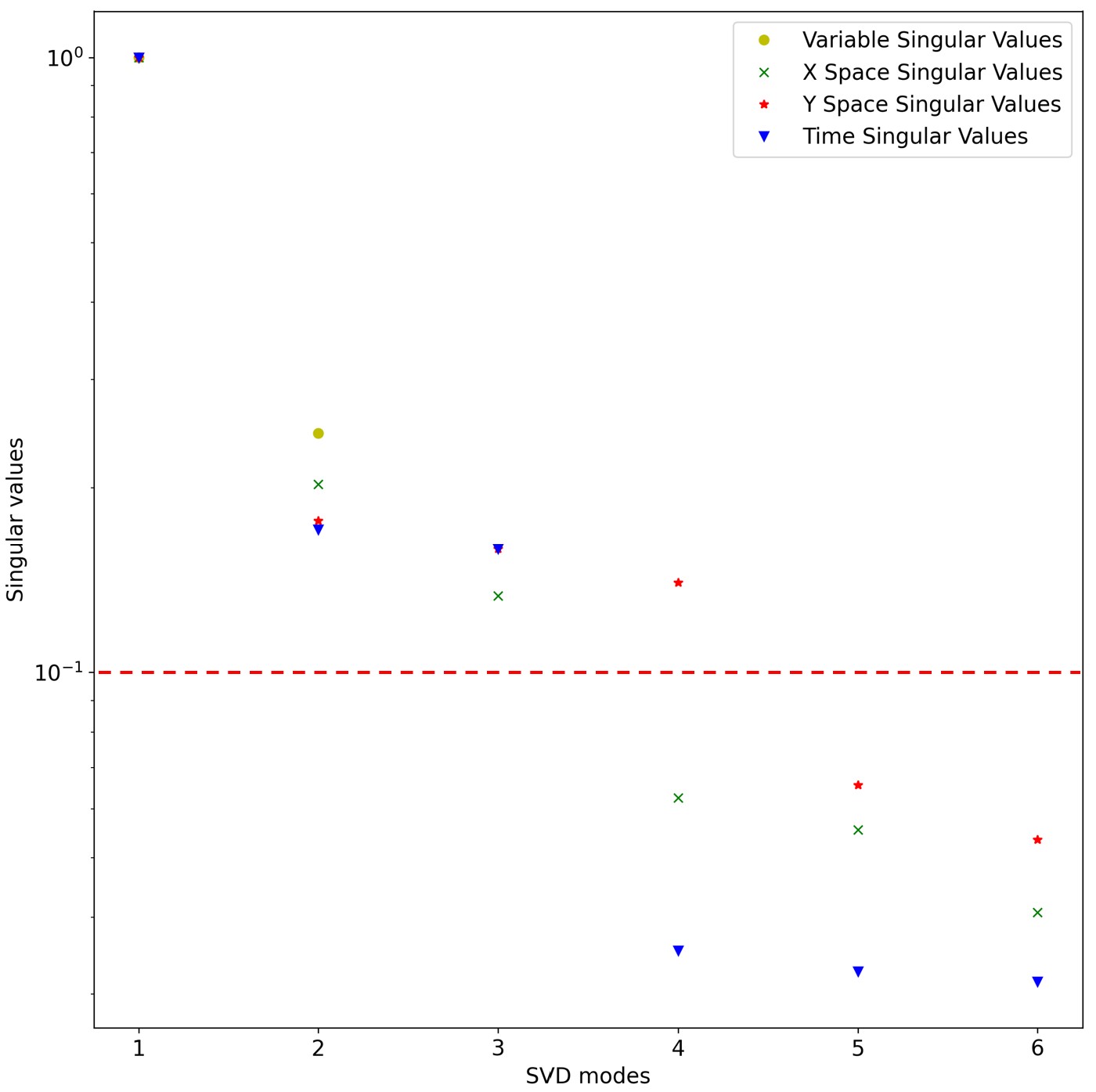}
    \caption{Decay of the retained singular values of the turbulent cylinder after applying LC-HOSVD-DLinear.}
    \label{fig:lchosvd2DCYLmodes}
\end{figure}

We now follow the same steps, this time using the LC-HOSVD-DLinear model. First, the singular values decay of each component are analyzed in fig. \ref{fig:lchosvd2DCYLmodes}, followed by a study of the frequencies of the first 4 modes in fig. \ref{fig:lchosvd2DCYLmodes}.

\begin{figure}[H]
    \centering
    \includegraphics[width=1\textwidth, angle=0]{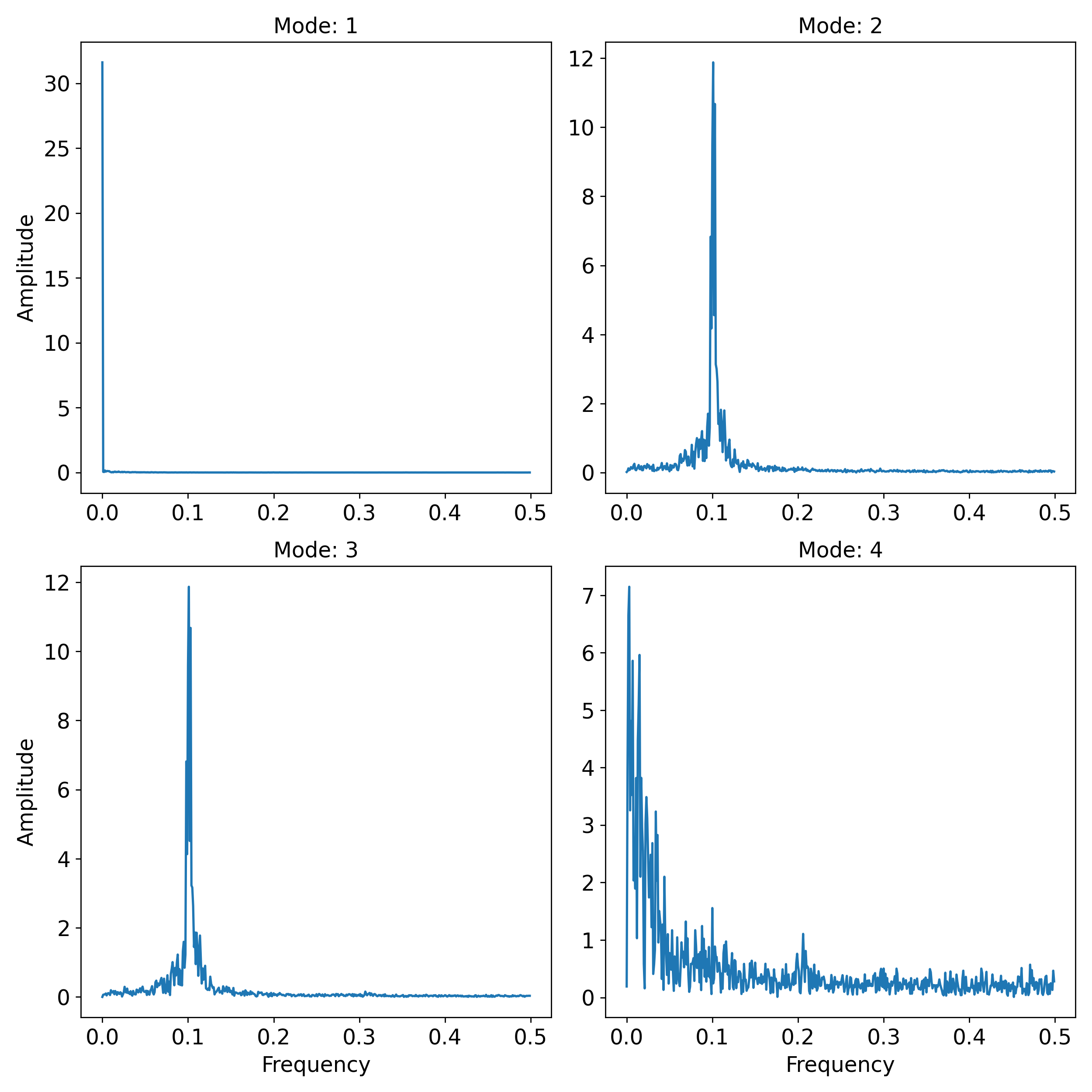}
    \caption{Frequencies of the first 4 temporal modes of the turbulent cylinder for LC-HOSVD-DLinear.}
    \label{fig:lchosvd2DCYLfrequencies}
\end{figure}

The X space and time singular values of fig. \ref{fig:lchosvd2DCYLmodes} both show a similar trend in the singular values decay, where two groups of 3 modes are clearly identifiable, separated by the $1e^{-1}$ threshold. On the other hand, the Y space modes are relatively close and don't present a dramatic decay even though, using the same threshold, we can divide these into two groups, a first group with the first 4 modes, which define the coherent structures, and the second group with the last 2 modes. The reason why the singular values of the X space and time decay so drastically is due to the fact that velocity field principally varies in the horizontal axis, following the direction of the flow, and changes over time, contrary to the Y space, where changes in the flow are less significant. When analyzing the frequencies associated to the first 4 modes in fig. \ref{fig:lchosvd2DCYLfrequencies}, observe that, for the first 3 modes, similar to the frequencies in fig. \ref{fig:lcsvd2DCYLfrequencies} a main frequency is clearly identifiable, immediate for the mean flow (mode 1), and at harmonic 0.1 for modes 2 and 3, which represent the wake. The main difference between LC-SVD-DLinear and LC-HOSVD-DLinear can be seen for mode 4 where, for this last model, since HOSVD has been used to filter noise, the spectral space of this mode has been slightly modified, now with an immediate dominant frequency, and subsequent ones at harmonic 0.1 and 0.2. However, low frequencies introduce perturbations in this modes signal, giving the illusion of white noise, but LC-HOSVD-DLinear is capable of understanding the existing periodicity in the data.

\begin{figure}[H]
    \centering
    \includegraphics[width=1\textwidth, angle=0]{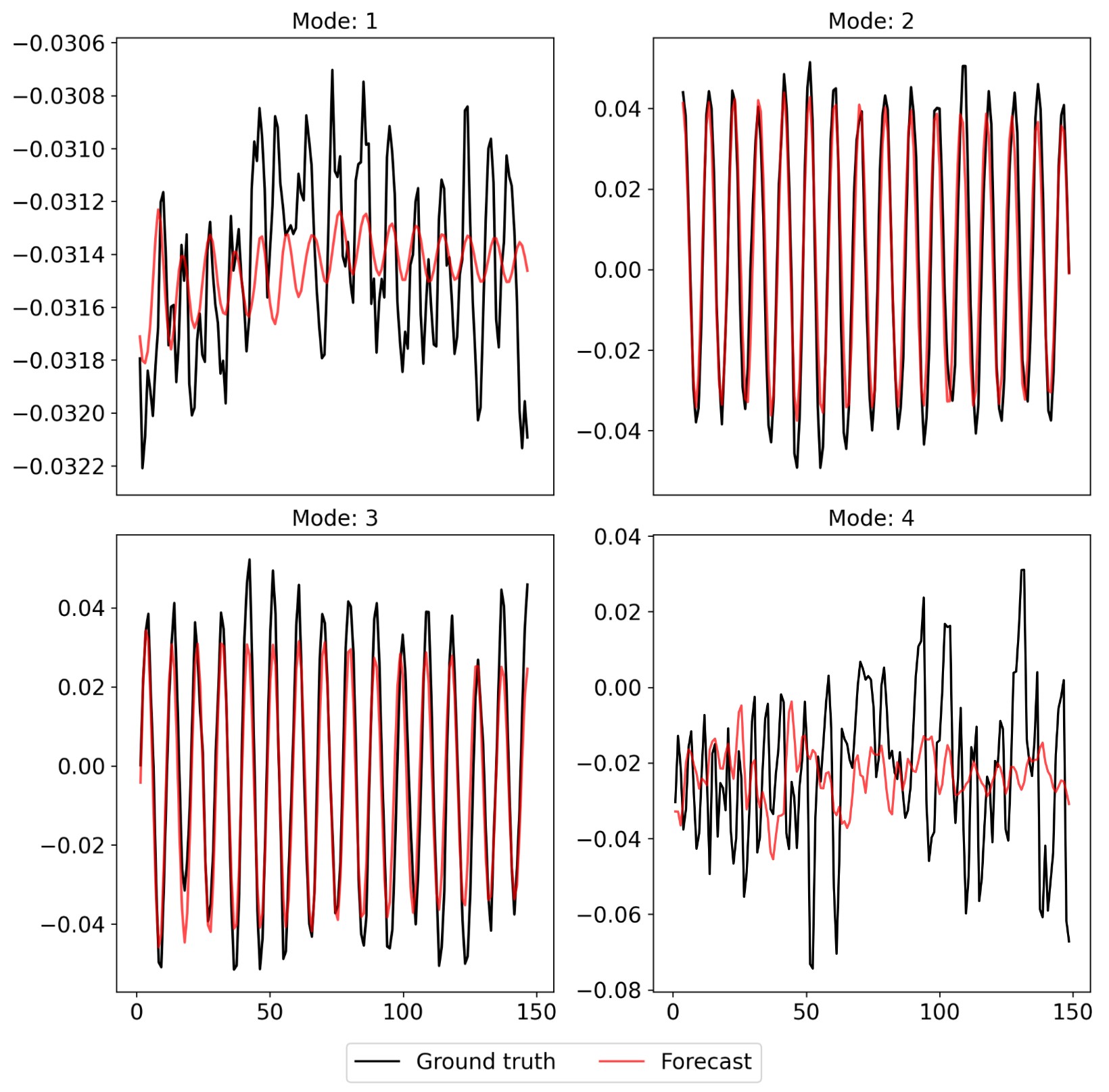}
    \caption{Comparison between the ground truth and forecast temporal coefficient values for the first 4 modes of the turbulent cylinder test case using LC-HOSVD-DLinear.}
    \label{fig:lchosvd2600test}
\end{figure}

Figure \ref{fig:lchosvd2600test} displays the LC-HOSVD-DLinear forecast results for the first 4 modes of the dataset. The forecast of the temporal coefficient values corresponding to the test set are $MAE = 0.314$, and $MSE = 0.153$, slightly lower than those for LC-SVD-DLinear. 

In fig. \ref{fig:lchosvd2600test} it is apparent that, since noise has been filtered out more effectively in LC-HOSVD-DLinear, patterns of the first mode, which were slightly more complex to identify and basically considered noise surrounding the mean for LC-SVD-DLinear (fig. \ref{fig:lcsvd2600test}), are now clearer. Now, the model is capable of capturing some periodicity in the data. However, mode 4 continues to present complexity due to the high spectral complexity, and the underlying periodic temporal pattern is predicted, to an extent. Reconstruction of the test dataset snapshot using these predicted temporal modes results in a reconstruction error of $RRMSE = 10.571\%$, a value which is lower compared to LC-SVD-DLinear.

\begin{figure}[H]
    \centering
    \includegraphics[width=1\textwidth, angle=0]{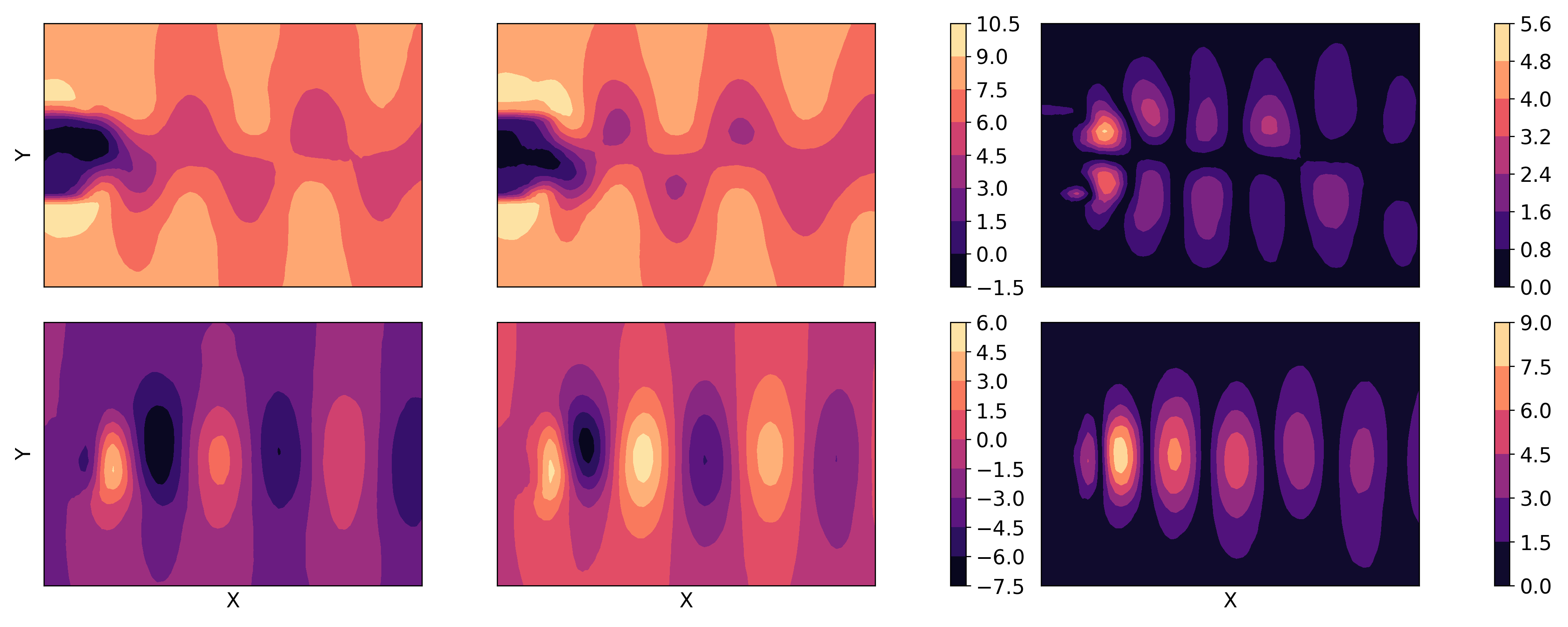}
    \caption{From left to right and top to bottom: the ground truth, forecast and relative error of both, of the streamwise and normal velocity components of the turbulent cylinder at $Re = 2600$, for snapshot $z = 83$, when using LC-HOSVD-DLinear.}
    \label{fig:lchosvd2DCYLcont}
\end{figure}

The highest error snapshot is $z = 83$, with a Wasserstein distance of $w = 0.28$ and reconstruction MAE of $MAE_{rec} = 1.05 m/s$, both of these values signifying an improvement when using LC-HOSVD-DLinear over LC-SVD-DLinear for turbulent data. The reconstruction results for this snapshot are displayed in fig. \ref{fig:lchosvd2DCYLcont}. 

For the highest error snapshot from fig. \ref{fig:lchosvd2DCYLcont}, a minuscule data drift causes the model to predict velocity values slightly higher than the ground truth, in the high velocity regions surrounding the wake. This drift is easier to identify when visualizing the overlapping data distributions for the ground truth and reconstruction of snapshot $z = 83$ in fig. \ref{fig:lcsvd2DCYLdist}.

\begin{figure}[H]
    \centering
    \includegraphics[width=1\textwidth, angle=0]{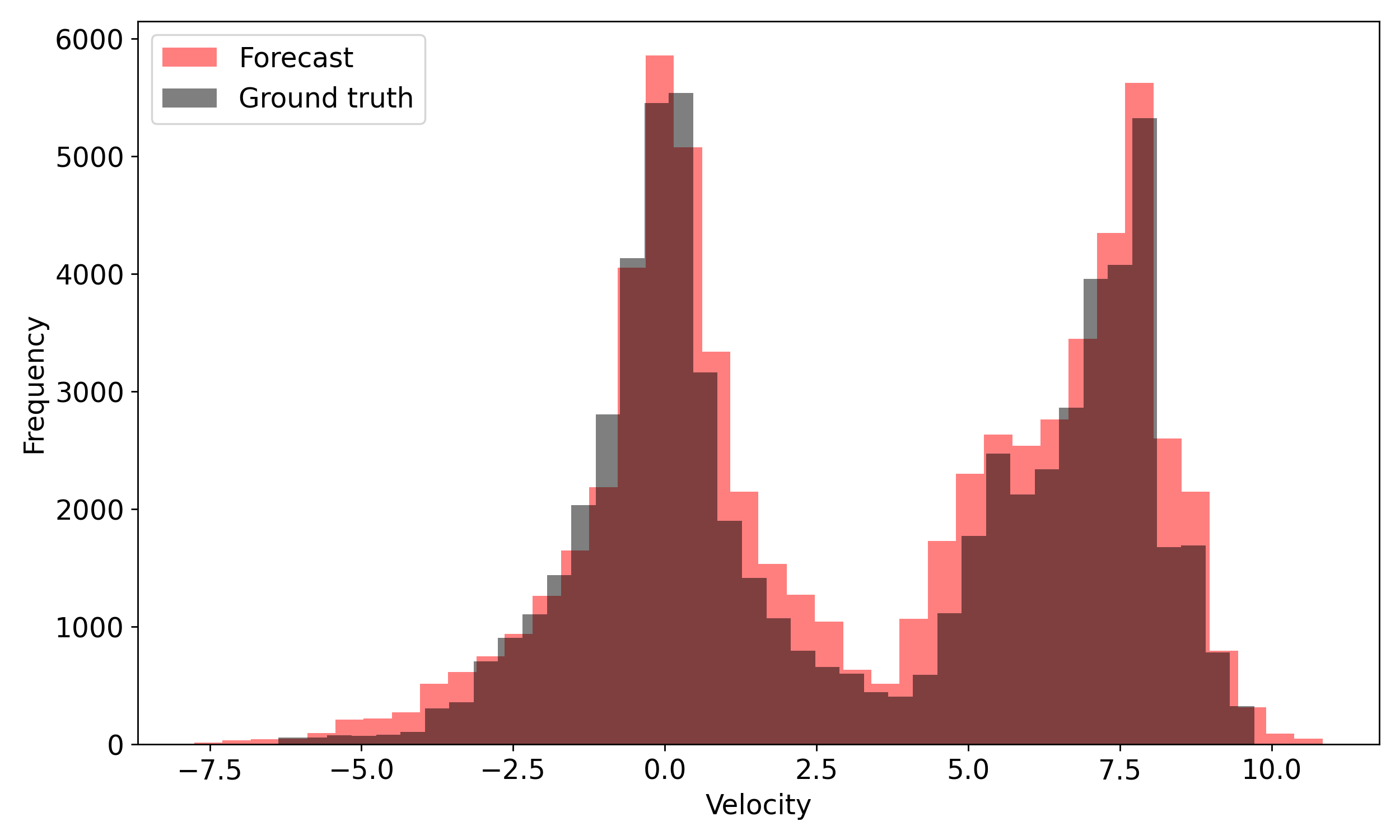}
    \caption{Comparison between the ground truth and forecast data distributions for snapshot $z = 83$ of the turbulent cylinder at $Re = 2600$ upon applying LC-HOSVD-DLinear.}
    \label{fig:lchosvd2DCYLdist}
\end{figure}

Verifying the previous statement, the analysis of data distributions in fig. \ref{fig:lchosvd2DCYLdist} shows a slight overestimation for positive velocity values, and an underestimation of the velocity values the region close to 0. Once again, the 200 snapshots consecutive to the test set are reconstructed using the forecast temporal coefficient values, with snapshot 100 being displayed in fig. \ref{fig:lcsvd2DCYLforecast}.

\begin{figure}[H]
    \centering
    \includegraphics[width=1\textwidth, angle=0]{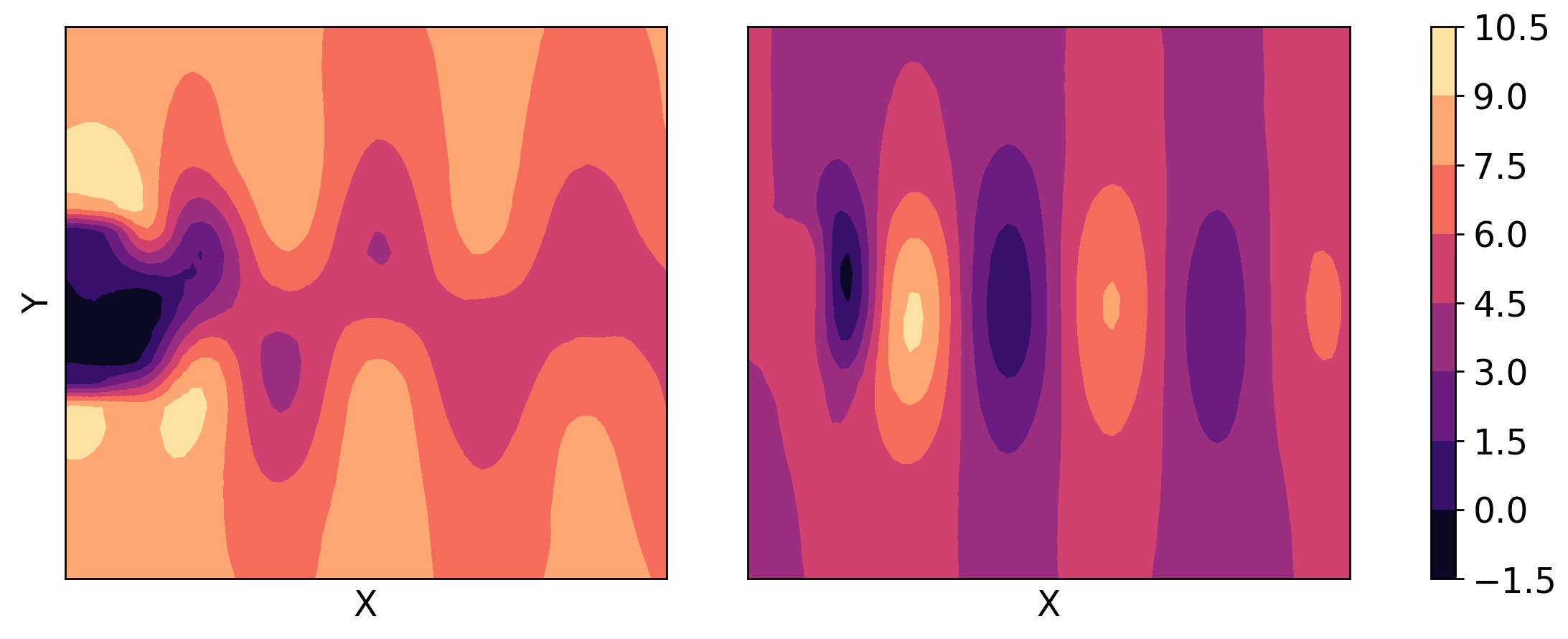}
    \caption{From left to right: forecast of the streamwise and normal velocity components of turbulent cylinder at $Re = 2600$, for snapshot $z = 100$ applying LC-HOSVD-DLinear.}
    \label{fig:lchosvd2DCYLforecast}
\end{figure}

Finally, \ref{fig:lcsvd2DCYLlatest} presents the last snapshot of the forecast sequence made by the LC-HOSVD-DLinear model.

\begin{figure}[H]
    \centering
    \includegraphics[width=1\textwidth, angle=0]{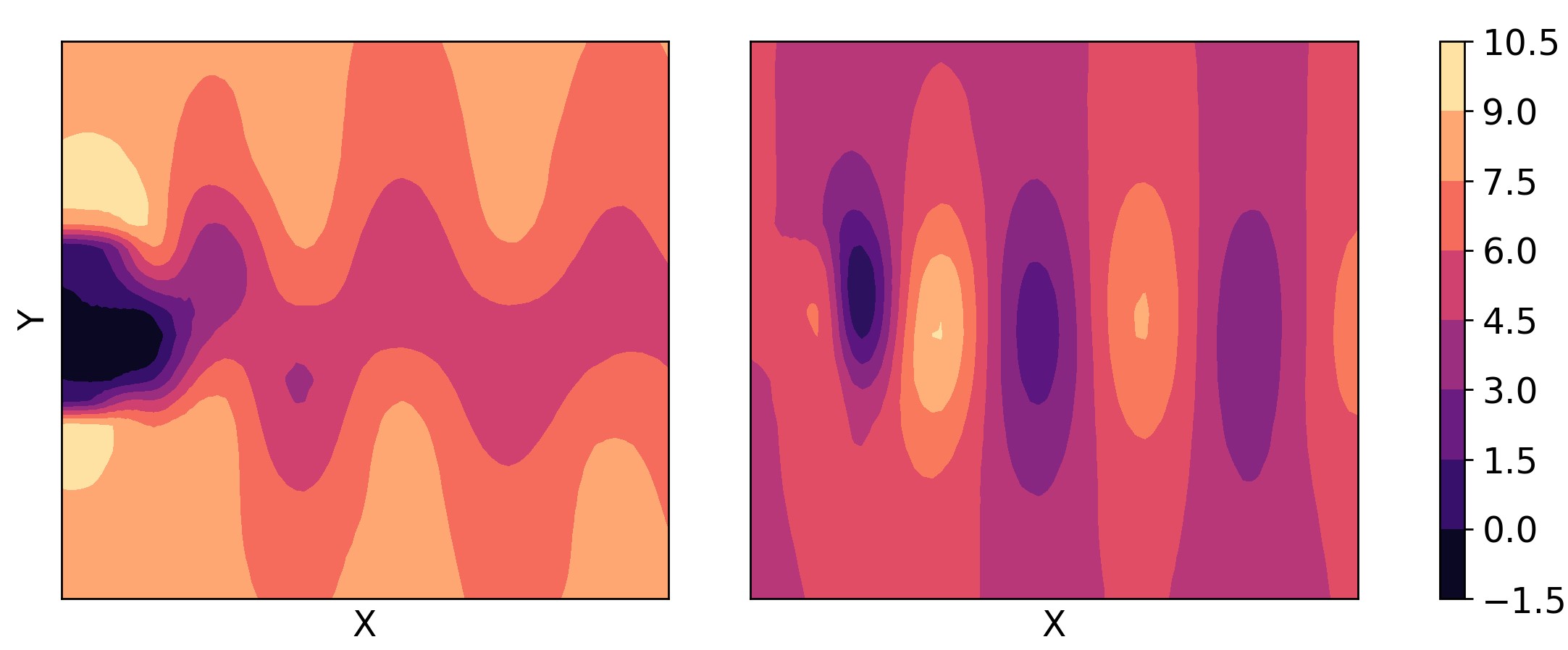}
    \caption{From left to right: forecast of the streamwise and normal velocity components of turbulent cylinder at $Re = 2600$, for snapshot $z = 200$ applying LC-HOSVD-DLinear.}
    \label{fig:lchosvd2DCYLlatest}
\end{figure}


Similar to LC-SVD-DLinear, velocity ranges in fig, \ref{fig:lchosvd2DCYLforecast} and \ref{fig:lchosvd2DCYLlatest} have maintained stable, since the range of values for the velocity field has not varied 200 snapshots later, which is expected since, in this case, we have used HOSVD to clean the data. 

\section{Conclusions\label{sec:conclusions}}
This article has presented a new methodology, which consists of two hybrid neural networks product of the combination of the data resolution enhancement techniques LC-SVD and LC-HOSVD with the DLinear model, to forecast high-resolution snapshots using an under-resolved (or high-resolution) input dataset. The hybrid models input data with any type of resolution. They create a high-resolution denoised version of the input data using one of two previously mentioned SVD-based data assimilation methods, with the temporal coefficients matrix being used by the DLinear model to forecast future values. The architecture is optimal, since all calculations are performed on the low-resolution data, drastically reducing the computational cost, and only at the end of the pipeline is the newly predicted data reconstructed into high-resolution snapshots using the denoised data created at the beginning.

Based on the ease for the model to identify patterns in the temporal coefficients, the output sequence length can vary considerably. For example, most of the temporal coefficients extracted from turbulent data contain chaos and, therefore, reduce the model capability to produce a large forecast sequence before diverging to the trend of each mode while, in the case of the temporal coefficients from laminar datasets, periodic trends are clearly identifiable in the most relevant modes, so the forecast sequence can be up-to 30 times the size of the input sequence.

The previous statement has been demonstrated by using the model to forecast future snapshots of the two test cases, which consist of a numerical simulation of a laminar flow passing a three-dimensional cylinder, and an experimental turbulent flow passing a circular cylinder. The effects of the number of retained modes on the models performance accuracy has also been tested proving that, for laminar datasets, the models performance is consistent, independently to the number of retained modes while, for turbulent data, the models accuracy is inversely proportional to the number of modes. By analyzing the frequencies, it has been possible to find the relationship between the models forecast performance of the main modes, relating peak results to periodicity (and unique frequencies) in the data.

The hybrid model presented in this article will be incorporated into a future release of ModelFLOWs-app, which can be downloaded from the official ModelFLOWs-app software website \cite{ModelFLOWsappWeb}.

\section[]{Acknowledgements}
The authors acknowledge the MODELAIR and ENCODING projects that have received funding from the European Union’s Horizon Europe research and innovation programme under the Marie Sklodowska-Curie grant agreement No. 101072559 and 101072779, respectively. The results of this publication reflect only the author(s) view and do not necessarily reflect those of the European Union. The
European Union can not be held responsible for them.  The authors acknowledge the grants TED2021- 129774B-C21 and PLEC2022-009235 funded by MCIN/AEI/ 10.13039/501100011033 and by the European Union “NextGenerationEU”/PRTR and the grant PID2023-147790OB-I00 funded by MCIU/AEI/10.13039/501100011033/FEDER, UE.



\bibliographystyle{elsarticle-num-names} 

\begin{thebibliography}{53}
\expandafter\ifx\csname natexlab\endcsname\relax\def\natexlab#1{#1}\fi
\providecommand{\url}[1]{\texttt{#1}}
\providecommand{\href}[2]{#2}
\providecommand{\path}[1]{#1}
\providecommand{\DOIprefix}{doi:}
\providecommand{\ArXivprefix}{arXiv:}
\providecommand{\URLprefix}{URL: }
\providecommand{\Pubmedprefix}{pmid:}
\providecommand{\doi}[1]{\href{http://dx.doi.org/#1}{\path{#1}}}
\providecommand{\Pubmed}[1]{\href{pmid:#1}{\path{#1}}}
\providecommand{\bibinfo}[2]{#2}
\ifx\xfnm\relax \def\xfnm[#1]{\unskip,\space#1}\fi
\bibitem[{Mortensen and Langtangen(2016)}]{mortensen2016high}
\bibinfo{author}{M.~Mortensen}, \bibinfo{author}{H.~P. Langtangen},
\newblock \bibinfo{title}{High performance python for direct numerical
  simulations of turbulent flows},
\newblock \bibinfo{journal}{Computer Physics Communications}
  \bibinfo{volume}{203} (\bibinfo{year}{2016}) \bibinfo{pages}{53--65}.
\bibitem[{Wu et~al.(2020)Wu, Kenway, Mader, Jasa, and
  Martins}]{wu2020pyoptsparse}
\bibinfo{author}{N.~Wu}, \bibinfo{author}{G.~Kenway}, \bibinfo{author}{C.~A.
  Mader}, \bibinfo{author}{J.~Jasa}, \bibinfo{author}{J.~R. Martins},
\newblock \bibinfo{title}{pyoptsparse: A python framework for large-scale
  constrained nonlinear optimization of sparse systems},
\newblock \bibinfo{journal}{Journal of Open Source Software}
  \bibinfo{volume}{5} (\bibinfo{year}{2020}) \bibinfo{pages}{2564}.
\bibitem[{Abad{\'\i}a-Heredia et~al.(2022)Abad{\'\i}a-Heredia,
  L{\'o}pez-Mart{\'\i}n, Carro, Arribas, P{\'e}rez, and
  Le~Clainche}]{abadia2022predictive}
\bibinfo{author}{R.~Abad{\'\i}a-Heredia},
  \bibinfo{author}{M.~L{\'o}pez-Mart{\'\i}n}, \bibinfo{author}{B.~Carro},
  \bibinfo{author}{J.~I. Arribas}, \bibinfo{author}{J.~M. P{\'e}rez},
  \bibinfo{author}{S.~Le~Clainche},
\newblock \bibinfo{title}{A predictive hybrid reduced order model based on
  proper orthogonal decomposition combined with deep learning architectures},
\newblock \bibinfo{journal}{Expert Systems with Applications}
  \bibinfo{volume}{187} (\bibinfo{year}{2022}) \bibinfo{pages}{115910}.
\bibitem[{Corrochano et~al.(2023)Corrochano, Freitas, Parente, and
  Clainche}]{corrochano2023predictive}
\bibinfo{author}{A.~Corrochano}, \bibinfo{author}{R.~S. Freitas},
  \bibinfo{author}{A.~Parente}, \bibinfo{author}{S.~L. Clainche},
\newblock \bibinfo{title}{A predictive physics-aware hybrid reduced order model
  for reacting flows},
\newblock \bibinfo{journal}{arXiv preprint arXiv:2301.09860}
  (\bibinfo{year}{2023}).
\bibitem[{D{\'\i}az et~al.(2023)D{\'\i}az, Corrochano, L{\'o}pez-Mart{\'\i}n,
  and Clainche}]{diaz2023deep}
\bibinfo{author}{P.~D{\'\i}az}, \bibinfo{author}{A.~Corrochano},
  \bibinfo{author}{M.~L{\'o}pez-Mart{\'\i}n}, \bibinfo{author}{S.~L. Clainche},
\newblock \bibinfo{title}{Deep learning combined with singular value
  decomposition to reconstruct databases in fluid dynamics},
\newblock \bibinfo{journal}{arXiv preprint arXiv:2305.08832}
  (\bibinfo{year}{2023}).
\bibitem[{Parente and Sutherland(2013)}]{parente2013principal}
\bibinfo{author}{A.~Parente}, \bibinfo{author}{J.~C. Sutherland},
\newblock \bibinfo{title}{Principal component analysis of turbulent combustion
  data: Data pre-processing and manifold sensitivity},
\newblock \bibinfo{journal}{Combustion and flame} \bibinfo{volume}{160}
  (\bibinfo{year}{2013}) \bibinfo{pages}{340--350}.
\bibitem[{Scherl et~al.(2020)Scherl, Strom, Shang, Williams, Polagye, and
  Brunton}]{scherl2020robust}
\bibinfo{author}{I.~Scherl}, \bibinfo{author}{B.~Strom}, \bibinfo{author}{J.~K.
  Shang}, \bibinfo{author}{O.~Williams}, \bibinfo{author}{B.~L. Polagye},
  \bibinfo{author}{S.~L. Brunton},
\newblock \bibinfo{title}{Robust principal component analysis for modal
  decomposition of corrupt fluid flows},
\newblock \bibinfo{journal}{Physical Review Fluids} \bibinfo{volume}{5}
  (\bibinfo{year}{2020}) \bibinfo{pages}{054401}.
\bibitem[{{Le Clainche} and Vega(2017)}]{LeClaincheVega17}
\bibinfo{author}{S.~{Le Clainche}}, \bibinfo{author}{J.~M. Vega},
\newblock \bibinfo{title}{Higher order dynamic mode decomposition},
\newblock \bibinfo{journal}{SIAM J. Appl. Dyn. Sys.} \bibinfo{volume}{16 (2)}
  (\bibinfo{year}{2017}) \bibinfo{pages}{882--925}.
\bibitem[{Le~Clainche et~al.(2019)Le~Clainche, Moreno-Ramos, Taylor, and
  Vega}]{le2019new}
\bibinfo{author}{S.~Le~Clainche}, \bibinfo{author}{R.~Moreno-Ramos},
  \bibinfo{author}{P.~Taylor}, \bibinfo{author}{J.~M. Vega},
\newblock \bibinfo{title}{New robust method to study flight flutter testing},
\newblock \bibinfo{journal}{Journal of Aircraft} \bibinfo{volume}{56}
  (\bibinfo{year}{2019}) \bibinfo{pages}{336--343}.
\bibitem[{Corrochano et~al.(2023)Corrochano, D’Alessio, Parente, and
  Le~Clainche}]{corrochano2023higher}
\bibinfo{author}{A.~Corrochano}, \bibinfo{author}{G.~D’Alessio},
  \bibinfo{author}{A.~Parente}, \bibinfo{author}{S.~Le~Clainche},
\newblock \bibinfo{title}{Higher order dynamic mode decomposition to model
  reacting flows},
\newblock \bibinfo{journal}{International Journal of Mechanical Sciences}
  \bibinfo{volume}{249} (\bibinfo{year}{2023}) \bibinfo{pages}{108219}.
\bibitem[{Mata et~al.(2023)Mata, Abad{\'i}a-Heredia, Lopez-Martin, P{\'e}rez,
  and Clainche}]{MataLeon2023}
\bibinfo{author}{L.~Mata}, \bibinfo{author}{R.~Abad{\'i}a-Heredia},
  \bibinfo{author}{M.~Lopez-Martin}, \bibinfo{author}{J.~M. P{\'e}rez},
  \bibinfo{author}{S.~L. Clainche},
\newblock \bibinfo{title}{Forecasting through deep learning and modal
  decomposition in two-phase concentric jets}  (\bibinfo{year}{2023}).
  \href{http://arxiv.org/abs/2212.12731}{{\tt arXiv:2212.12731}}.
\bibitem[{Huang et~al.(2022)Huang, Li, Huang, Ke, Lin, and
  Wang}]{huang2022predictions}
\bibinfo{author}{Z.~Huang}, \bibinfo{author}{T.~Li},
  \bibinfo{author}{K.~Huang}, \bibinfo{author}{H.~Ke},
  \bibinfo{author}{M.~Lin}, \bibinfo{author}{Q.~Wang},
\newblock \bibinfo{title}{Predictions of flow and temperature fields in a
  t-junction based on dynamic mode decomposition and deep learning},
\newblock \bibinfo{journal}{Energy} \bibinfo{volume}{261}
  (\bibinfo{year}{2022}) \bibinfo{pages}{125228}.
\bibitem[{Mu{\~n}oz et~al.(2023)Mu{\~n}oz, Dave, D'Alessio, Bontempi, Parente,
  and Le~Clainche}]{munoz2023extraction}
\bibinfo{author}{E.~Mu{\~n}oz}, \bibinfo{author}{H.~Dave},
  \bibinfo{author}{G.~D'Alessio}, \bibinfo{author}{G.~Bontempi},
  \bibinfo{author}{A.~Parente}, \bibinfo{author}{S.~Le~Clainche},
\newblock \bibinfo{title}{Extraction and analysis of flow features in planar
  synthetic jets using different machine learning techniques},
\newblock \bibinfo{journal}{Available at SSRN 4429450}  (\bibinfo{year}{2023}).
\bibitem[{Eivazi et~al.(2022)Eivazi, Le~Clainche, Hoyas, and
  Vinuesa}]{ae_modal}
\bibinfo{author}{H.~Eivazi}, \bibinfo{author}{S.~Le~Clainche},
  \bibinfo{author}{S.~Hoyas}, \bibinfo{author}{R.~Vinuesa},
\newblock \bibinfo{title}{{Towards extraction of orthogonal and parsimonious
  non-linear modes from turbulent flows}},
\newblock \bibinfo{journal}{Expert Syst. Appl.} \bibinfo{volume}{202}
  (\bibinfo{year}{2022}) \bibinfo{pages}{117038}.
\bibitem[{Soto-Valle et~al.(2020)Soto-Valle, Bartholomay, Alber, Manolesos,
  Nayeri, and Paschereit}]{soto2020determination}
\bibinfo{author}{R.~Soto-Valle}, \bibinfo{author}{S.~Bartholomay},
  \bibinfo{author}{J.~Alber}, \bibinfo{author}{M.~Manolesos},
  \bibinfo{author}{C.~N. Nayeri}, \bibinfo{author}{C.~O. Paschereit},
\newblock \bibinfo{title}{Determination of the angle of attack on a research
  wind turbine rotor blade using surface pressure measurements},
\newblock \bibinfo{journal}{Wind Energy Science} \bibinfo{volume}{5}
  (\bibinfo{year}{2020}) \bibinfo{pages}{1771--1792}.
\bibitem[{Fang and Hong(2018)}]{fang2018particle}
\bibinfo{author}{C.~Fang}, \bibinfo{author}{L.~Hong},
\newblock \bibinfo{title}{Particle image velocimetry for combustion
  measurements: Applications and developments},
\newblock \bibinfo{journal}{Chinese Journal of Aeronautics}
  \bibinfo{volume}{31} (\bibinfo{year}{2018}) \bibinfo{pages}{1407--1427}.
\bibitem[{Woodward et~al.(2023)Woodward, Tian, Lin, Mohan, Hader, Fasel,
  Chertkov, and Livescu}]{woodward2023data}
\bibinfo{author}{M.~Woodward}, \bibinfo{author}{Y.~Tian},
  \bibinfo{author}{Y.~T. Lin}, \bibinfo{author}{A.~T. Mohan},
  \bibinfo{author}{C.~Hader}, \bibinfo{author}{H.~F. Fasel},
  \bibinfo{author}{M.~Chertkov}, \bibinfo{author}{D.~Livescu},
\newblock \bibinfo{title}{Data-driven mori-zwanzig: Reduced order modeling of
  sparse sensors measurements for boundary layer transition},
\newblock in: \bibinfo{booktitle}{AIAA AVIATION 2023 Forum},
  \bibinfo{year}{2023}, p. \bibinfo{pages}{4256}.
\bibitem[{Botez(2018)}]{botez2018morphing}
\bibinfo{author}{R.~Botez},
\newblock \bibinfo{title}{Morphing wing, uav and aircraft multidisciplinary
  studies at the laboratory of applied research in active controls, avionics
  and aeroservoelasticity larcase},
\newblock \bibinfo{journal}{Aerospace Lab}  (\bibinfo{year}{2018})
  \bibinfo{pages}{1--11}.
\bibitem[{Siano et~al.(2017)Siano, Frosina, and Senatore}]{siano2017diagnostic}
\bibinfo{author}{D.~Siano}, \bibinfo{author}{E.~Frosina},
  \bibinfo{author}{A.~Senatore},
\newblock \bibinfo{title}{Diagnostic process by using vibrational sensors for
  monitoring cavitation phenomena in a getoror pump used for automotive
  applications},
\newblock \bibinfo{journal}{Energy Procedia} \bibinfo{volume}{126}
  (\bibinfo{year}{2017}) \bibinfo{pages}{1115--1122}.
\bibitem[{Hetherington et~al.(2023)Hetherington, Corrochano,
  Abad{\'\i}a-Heredia, Lazpita, Mu{\~n}oz, D{\'\i}az, Moira,
  L{\'o}pez-Mart{\'\i}n, and Clainche}]{hetherington2023modelflows}
\bibinfo{author}{A.~Hetherington}, \bibinfo{author}{A.~Corrochano},
  \bibinfo{author}{R.~Abad{\'\i}a-Heredia}, \bibinfo{author}{E.~Lazpita},
  \bibinfo{author}{E.~Mu{\~n}oz}, \bibinfo{author}{P.~D{\'\i}az},
  \bibinfo{author}{E.~Moira}, \bibinfo{author}{M.~L{\'o}pez-Mart{\'\i}n},
  \bibinfo{author}{S.~L. Clainche},
\newblock \bibinfo{title}{Model{FLOW}s-app: data-driven post-processing and
  reduced order modelling tools},
\newblock \bibinfo{journal}{arXiv preprint arXiv:2305.17150}
  (\bibinfo{year}{2023}) \bibinfo{pages}{Software available at
  \href{https://github.com/modelflows/ModelFLOWs--app}{https://github.com/modelflows/ModelFLOWs--app}}.
\bibitem[{Zhou et~al.(2015)Zhou, Soh, Jiang, and Wu}]{zhou2015compressed}
\bibinfo{author}{H.~Zhou}, \bibinfo{author}{Y.~C. Soh},
  \bibinfo{author}{C.~Jiang}, \bibinfo{author}{X.~Wu},
\newblock \bibinfo{title}{Compressed representation learning for fluid field
  reconstruction from sparse sensor observations},
\newblock in: \bibinfo{booktitle}{2015 International Joint Conference on Neural
  Networks (IJCNN)}, \bibinfo{organization}{IEEE}, \bibinfo{year}{2015}, pp.
  \bibinfo{pages}{1--6}.
\bibitem[{Erichson et~al.(2019)Erichson, Mathelin, Yao, Brunton, Mahoney, and
  Kutz}]{erichson2019shallow}
\bibinfo{author}{N.~B. Erichson}, \bibinfo{author}{L.~Mathelin},
  \bibinfo{author}{Z.~Yao}, \bibinfo{author}{S.~L. Brunton},
  \bibinfo{author}{M.~W. Mahoney}, \bibinfo{author}{J.~N. Kutz},
\newblock \bibinfo{title}{Shallow learning for fluid flow reconstruction with
  limited sensors and limited data},
\newblock \bibinfo{journal}{arXiv preprint arXiv:1902.07358}
  (\bibinfo{year}{2019}).
\bibitem[{de~Silva et~al.(2021)de~Silva, Manohar, Clark, Brunton, Brunton, and
  Kutz}]{pysensors}
\bibinfo{author}{B.~de~Silva}, \bibinfo{author}{K.~Manohar},
  \bibinfo{author}{E.~Clark}, \bibinfo{author}{B.~Brunton},
  \bibinfo{author}{S.~Brunton}, \bibinfo{author}{N.~Kutz},
\newblock \bibinfo{title}{Pysensors: A python package for sparse sensor
  placement},
\newblock \bibinfo{journal}{J. Open Source Software} \bibinfo{volume}{6}
  (\bibinfo{year}{2021}) \bibinfo{pages}{2828}.
\bibitem[{Arciniega-Ceballos et~al.(2012)Arciniega-Ceballos,
  Alatorre-Ibarguengoitia, Perton, Sanchez-Sesma, and
  Dingwell}]{arciniega2012deciphering}
\bibinfo{author}{A.~Arciniega-Ceballos},
  \bibinfo{author}{M.~Alatorre-Ibarguengoitia}, \bibinfo{author}{M.~Perton},
  \bibinfo{author}{F.~Sanchez-Sesma}, \bibinfo{author}{D.~Dingwell},
\newblock \bibinfo{title}{Deciphering seismic signatures of physical processes
  in dynamic complex systems: an experimental approach},
\newblock in: \bibinfo{booktitle}{AGU Fall Meeting Abstracts}, volume
  \bibinfo{volume}{2012}, \bibinfo{year}{2012}, pp. \bibinfo{pages}{V31G--07}.
\bibitem[{Wong et~al.(2016)Wong, Sahoo, and McFarland}]{wong2016integrated}
\bibinfo{author}{K.~Wong}, \bibinfo{author}{S.~Sahoo},
  \bibinfo{author}{M.~McFarland},
\newblock \bibinfo{title}{Integrated high temperature sensors for advanced
  propulsion materials},
\newblock in: \bibinfo{booktitle}{52nd AIAA/SAE/ASEE Joint Propulsion
  Conference}, \bibinfo{year}{2016}, p. \bibinfo{pages}{4849}.
\bibitem[{Umargono et~al.(2019)Umargono, Suseno, and Gunawan}]{umargono2019k}
\bibinfo{author}{E.~Umargono}, \bibinfo{author}{J.~E. Suseno},
  \bibinfo{author}{S.~Gunawan},
\newblock \bibinfo{title}{K-means clustering optimization using the elbow
  method and early centroid determination based-on mean and median},
\newblock in: \bibinfo{booktitle}{Proceedings of the International Conferences
  on Information System and Technology},
  \bibinfo{organization}{SCITEPRESS—Science and Technology Publications
  Setubal, Portugal}, \bibinfo{year}{2019}, pp. \bibinfo{pages}{234--240}.
\bibitem[{Sirovich(1987)}]{Sirovich87}
\bibinfo{author}{L.~Sirovich},
\newblock \bibinfo{title}{Turbulence and the dynamic of coherent structures,
  parts i--iii},
\newblock \bibinfo{journal}{Q. Appl. Math.} \bibinfo{volume}{45(3)}
  (\bibinfo{year}{1987}) \bibinfo{pages}{561}.
\bibitem[{Lumley(1967)}]{Lumley}
\bibinfo{author}{J.~L. Lumley},
\newblock \bibinfo{title}{The structure of inhomogeneous turbulent flows},
\newblock \bibinfo{journal}{In: Yaglam, A.M., Tatarsky, V.I. (eds.) Proceedings
  of the International Colloquium on the Fine Scale Structure of the Atmosphere
  and Its Influence on Radio Wave Propagation. Doklady Akademii Nauk SSSR,
  Nauka, Moscow}  (\bibinfo{year}{1967}).
\bibitem[{{Le Clainche} et~al.(2020){Le Clainche}, Izvassarov, Rosti, Brandt,
  and Tammisola}]{LeClaincheetalJFM20}
\bibinfo{author}{S.~{Le Clainche}}, \bibinfo{author}{D.~Izvassarov},
  \bibinfo{author}{M.~Rosti}, \bibinfo{author}{L.~Brandt},
  \bibinfo{author}{O.~Tammisola},
\newblock \bibinfo{title}{Coherent structures in the turbulent flow of an
  elastoviscoplastic fluid, doi:10.1017/jfm.2020.31},
\newblock \bibinfo{journal}{J. Fluid Mech.}  (\bibinfo{year}{2020})
  \bibinfo{pages}{A5}.
\bibitem[{Parente and Sutherland(2013)}]{PCA}
\bibinfo{author}{A.~Parente}, \bibinfo{author}{J.~C. Sutherland},
\newblock \bibinfo{title}{Principal component analysis of turbulent combustion
  data: Data pre-processing and manifold sensitivity},
\newblock \bibinfo{journal}{Combustion and flame} \bibinfo{volume}{160}
  (\bibinfo{year}{2013}) \bibinfo{pages}{340--350}.
\bibitem[{Rap{\'u}n et~al.(2017)Rap{\'u}n, Terragni, and Vega}]{Lupod}
\bibinfo{author}{M.-L. Rap{\'u}n}, \bibinfo{author}{F.~Terragni},
  \bibinfo{author}{J.~Vega},
\newblock \bibinfo{title}{{LUPOD}: Collocation in {POD} via {LU}
  decomposition},
\newblock \bibinfo{journal}{J. Comp. Phys.} \bibinfo{volume}{335}
  (\bibinfo{year}{2017}) \bibinfo{pages}{1--20}.
\bibitem[{Manohar et~al.(2018)Manohar, Brunton, Kutz, and
  Brunton}]{pysensorsMethod}
\bibinfo{author}{K.~Manohar}, \bibinfo{author}{B.~Brunton},
  \bibinfo{author}{N.~Kutz}, \bibinfo{author}{S.~Brunton},
\newblock \bibinfo{title}{Data-driven sparse sensor placement for
  reconstruction},
\newblock \bibinfo{journal}{IEEE Contr. Syst. Magaz.} \bibinfo{volume}{38}
  (\bibinfo{year}{2018}) \bibinfo{pages}{63--86}.
\bibitem[{Businger and Golub(1965)}]{R1pys}
\bibinfo{author}{P.~Businger}, \bibinfo{author}{G.~H. Golub},
\newblock \bibinfo{title}{Linear least squares solutions by house-holder
  transformations},
\newblock \bibinfo{journal}{Numer. Math.} \bibinfo{volume}{7}
  (\bibinfo{year}{1965}) \bibinfo{pages}{269–276}.
\bibitem[{Sommariva and Vianello(2009)}]{R2pys}
\bibinfo{author}{A.~Sommariva}, \bibinfo{author}{M.~Vianello},
\newblock \bibinfo{title}{Computing approximate fekete points by qr
  factorizations of vandermonde matrices},
\newblock \bibinfo{journal}{Comput. Math. Applicat.} \bibinfo{volume}{57}
  (\bibinfo{year}{2009}) \bibinfo{pages}{1324–1336}.
\bibitem[{Heck et~al.(1998)Heck, Olkin, and Naghshineh}]{R3pys}
\bibinfo{author}{L.~P. Heck}, \bibinfo{author}{J.~Olkin},
  \bibinfo{author}{K.~Naghshineh},
\newblock \bibinfo{title}{Transducer placement for broadband active vibration
  control using a novel multidimensional qr factorization},
\newblock \bibinfo{journal}{J. Vibration Acoust.} \bibinfo{volume}{120}
  (\bibinfo{year}{1998}) \bibinfo{pages}{663–670}.
\bibitem[{Seshadri et~al.(2017)Seshadri, Narayan, and Mahadevan}]{R4pys}
\bibinfo{author}{P.~Seshadri}, \bibinfo{author}{A.~Narayan},
  \bibinfo{author}{S.~Mahadevan},
\newblock \bibinfo{title}{Effectively sampled quadratures for least squares
  polynomial approximations},
\newblock \bibinfo{journal}{SIAM/ASA J. Uncertain. Quantif.}
  \bibinfo{volume}{5} (\bibinfo{year}{2017}) \bibinfo{pages}{1003–1023}.
\bibitem[{Kuraria et~al.(2018)Kuraria, Jharbade, and
  Soni}]{kuraria2018centroid}
\bibinfo{author}{A.~Kuraria}, \bibinfo{author}{N.~Jharbade},
  \bibinfo{author}{M.~Soni},
\newblock \bibinfo{title}{Centroid selection process using wcss and elbow
  method for k-mean clustering algorithm in data mining},
\newblock \bibinfo{journal}{International Journal of Scientific Research in
  Science, Engineering and Technology}  (\bibinfo{year}{2018})
  \bibinfo{pages}{190--195}.
\bibitem[{Vega and {Le Clainche}(2020)}]{VegaLeClaincheBook20}
\bibinfo{author}{J.~Vega}, \bibinfo{author}{S.~{Le Clainche}},
\newblock \bibinfo{title}{Higher order dynamic mode decomposition and its
  applications},
\newblock \bibinfo{journal}{BOOK: Elsevier}  (\bibinfo{year}{2020}).
\bibitem[{Jackson(1987)}]{jackson1987finite}
\bibinfo{author}{C.~Jackson},
\newblock \bibinfo{title}{A finite-element study of the onset of vortex
  shedding in flow past variously shaped bodies},
\newblock \bibinfo{journal}{Journal of fluid Mechanics} \bibinfo{volume}{182}
  (\bibinfo{year}{1987}) \bibinfo{pages}{23--45}.
\bibitem[{Barkley and Henderson(1996)}]{barkley1996three}
\bibinfo{author}{D.~Barkley}, \bibinfo{author}{R.~D. Henderson},
\newblock \bibinfo{title}{Three-dimensional floquet stability analysis of the
  wake of a circular cylinder},
\newblock \bibinfo{journal}{Journal of Fluid Mechanics} \bibinfo{volume}{322}
  (\bibinfo{year}{1996}) \bibinfo{pages}{215--241}.
\bibitem[{Nek(????)}]{Nek5000}
\bibinfo{title}{{Nek5000}}, \URLprefix \url{https://nek5000.mcs.anl.gov/}.
\bibitem[{Towne et~al.(2023)Towne, Dawson, Br{\`e}s, Lozano-Dur{\'a}n,
  Saxton-Fox, Parthasarathy, Jones, Biler, Yeh, Patel
  et~al.}]{towne2023database}
\bibinfo{author}{A.~Towne}, \bibinfo{author}{S.~T. Dawson},
  \bibinfo{author}{G.~A. Br{\`e}s}, \bibinfo{author}{A.~Lozano-Dur{\'a}n},
  \bibinfo{author}{T.~Saxton-Fox}, \bibinfo{author}{A.~Parthasarathy},
  \bibinfo{author}{A.~R. Jones}, \bibinfo{author}{H.~Biler},
  \bibinfo{author}{C.-A. Yeh}, \bibinfo{author}{H.~D. Patel}, et~al.,
\newblock \bibinfo{title}{A database for reduced-complexity modeling of fluid
  flows},
\newblock \bibinfo{journal}{AIAA Journal}  (\bibinfo{year}{2023})
  \bibinfo{pages}{1--26}.
\bibitem[{Rodriguez(2020)}]{rodriguez2020development}
\bibinfo{author}{J.~Rodriguez}, \bibinfo{title}{Development of a test section
  featuring a flat plate conditioned for the study of fully developed turbulent
  boundary layers using PIV}, Ph.D. thesis, \bibinfo{year}{2020}.
\bibitem[{Br{\`e}s et~al.(2018)Br{\`e}s, Jordan, Jaunet, Le~Rallic, Cavalieri,
  Towne, Lele, Colonius, and Schmidt}]{bres2018importance}
\bibinfo{author}{G.~A. Br{\`e}s}, \bibinfo{author}{P.~Jordan},
  \bibinfo{author}{V.~Jaunet}, \bibinfo{author}{M.~Le~Rallic},
  \bibinfo{author}{A.~V. Cavalieri}, \bibinfo{author}{A.~Towne},
  \bibinfo{author}{S.~K. Lele}, \bibinfo{author}{T.~Colonius},
  \bibinfo{author}{O.~T. Schmidt},
\newblock \bibinfo{title}{Importance of the nozzle-exit boundary-layer state in
  subsonic turbulent jets},
\newblock \bibinfo{journal}{Journal of Fluid Mechanics} \bibinfo{volume}{851}
  (\bibinfo{year}{2018}) \bibinfo{pages}{83--124}.
\bibitem[{Vreman(2004)}]{vreman2004}
\bibinfo{author}{A.~Vreman},
\newblock \bibinfo{title}{An eddy-viscosity subgrid-scale model for turbulent
  shear flow: Algebraic theory and applications},
\newblock \bibinfo{journal}{Physics of fluids} \bibinfo{volume}{16}
  (\bibinfo{year}{2004}) \bibinfo{pages}{3670--3681}.
\bibitem[{Freund(1997)}]{freund1997}
\bibinfo{author}{J.~B. Freund},
\newblock \bibinfo{title}{Proposed inflow/outflow boundary condition for direct
  computation of aerodynamic sound},
\newblock \bibinfo{journal}{AIAA journal} \bibinfo{volume}{35}
  (\bibinfo{year}{1997}) \bibinfo{pages}{740--742}.
\bibitem[{Mani(2012)}]{mani2012}
\bibinfo{author}{A.~Mani},
\newblock \bibinfo{title}{Analysis and optimization of numerical sponge layers
  as a nonreflective boundary treatment},
\newblock \bibinfo{journal}{Journal of Computational Physics}
  \bibinfo{volume}{231} (\bibinfo{year}{2012}) \bibinfo{pages}{704--716}.
\bibitem[{Zaman(1998)}]{zaman1998}
\bibinfo{author}{K.~Zaman},
\newblock \bibinfo{title}{Asymptotic spreading rate of initially compressible
  jets—experiment and analysis},
\newblock \bibinfo{journal}{Physics of Fluids} \bibinfo{volume}{10}
  (\bibinfo{year}{1998}) \bibinfo{pages}{2652--2660}.
\bibitem[{Zaman(1999)}]{zaman1999}
\bibinfo{author}{K.~Zaman},
\newblock \bibinfo{title}{Spreading characteristics of compressible jets from
  nozzles of various geometries},
\newblock \bibinfo{journal}{Journal of Fluid mechanics} \bibinfo{volume}{383}
  (\bibinfo{year}{1999}) \bibinfo{pages}{197--228}.
\bibitem[{Mendez et~al.(2020)Mendez, Hess, Watz, and
  Buchlin}]{mendez2020multiscale}
\bibinfo{author}{M.~A. Mendez}, \bibinfo{author}{D.~Hess},
  \bibinfo{author}{B.~B. Watz}, \bibinfo{author}{J.-M. Buchlin},
\newblock \bibinfo{title}{Multiscale proper orthogonal decomposition (m{POD})
  of {TR-PIV} data—a case study on stationary and transient cylinder wake
  flows},
\newblock \bibinfo{journal}{Measurement Science and Technology}
  \bibinfo{volume}{31} (\bibinfo{year}{2020}) \bibinfo{pages}{094014}.
\bibitem[{Williamson(1996)}]{williamson1996vortex}
\bibinfo{author}{C.~H. Williamson},
\newblock \bibinfo{title}{Vortex dynamics in the cylinder wake},
\newblock \bibinfo{journal}{Annual review of fluid mechanics}
  \bibinfo{volume}{28} (\bibinfo{year}{1996}) \bibinfo{pages}{477--539}.
\bibitem[{mem(????)}]{memprof}
\bibinfo{title}{memory profiler},
\newblock \bibinfo{journal}{Available at
  \href{https://pypi.org/project/memory-profiler/}{https://pypi.org/project/memory-profiler/}}.
\bibitem[{Mod(2023)}]{ModelFLOWsappWeb}
\bibinfo{title}{Model{FLOW}s-app},
\newblock \bibinfo{journal}{developed by ModelFLOWs research group.}
  \bibinfo{volume}{{Software available at
  \href{https://modelflows.github.io/modelflowsapp/}{https://modelflows.github.io/modelflowsapp/}}}
  (\bibinfo{year}{2023}).

\bibitem[{PyTorch()}]{PyTorch}
\newblock \bibinfo{title}{PyTorch},
\newblock \bibinfo{journal}{Available at \href{https://pytorch.org/}{https://pytorch.org/}}.

\bibitem[{Optuna()}]{Optuna}
\newblock \bibinfo{title}{Optuna},
\newblock \bibinfo{journal}{Available at \href{https://optuna.org/}{https://optuna.org/}}.

\bibitem[{Hetherington and Clainche(2023)}]{hetherington2023low}
\bibinfo{author}{A.~Hetherington and S.~Le Clainche},
\newblock \bibinfo{title}{Low-cost singular value decomposition with optimal sensor placement},
\newblock \bibinfo{journal}{arXiv preprint arXiv:2311.09791},
\bibinfo{year}{2023}.

\bibitem[{Zeng et al.(2023)}]{zeng2023transformers}
\bibinfo{author}{A.~Zeng, M.~Chen, L.~Zhang, and Q.~Xu},
\newblock \bibinfo{title}{Are transformers effective for time series forecasting?},
\newblock \bibinfo{journal}{Proceedings of the AAAI conference on artificial intelligence},
\bibinfo{volume}{37} (\bibinfo{year}{2023}),
\bibinfo{number}{9}, \bibinfo{pages}{11121--11128}.

\bibitem[{Liu et al.(2023)}]{liu2023multi}
\bibinfo{author}{J.~Liu, C.~Gong, S.~Chen, and N.~Zhou},
\newblock \bibinfo{title}{Multi-step-ahead wind speed forecast method based on outlier correction, optimized decomposition, and dlinear model},
\newblock \bibinfo{journal}{Mathematics},
\bibinfo{volume}{11} (\bibinfo{year}{2023}),
\bibinfo{number}{12}, \bibinfo{pages}{2746},
\bibinfo{publisher}{MDPI}.

\bibitem[{Iuliano and Quagliarella(2013)}]{iuliano2013proper}
\bibinfo{author}{E.~Iuliano, D.~Quagliarella},
\newblock \bibinfo{title}{Proper orthogonal decomposition, surrogate modelling and evolutionary optimization in aerodynamic design},
\newblock \bibinfo{journal}{Computers \& Fluids},
\bibinfo{volume}{84} (\bibinfo{year}{2013}),
\bibinfo{pages}{327--350},
\bibinfo{publisher}{Elsevier}.

\bibitem[{Freitag et~al.(2018)}]{freitag2018recurrent}
\bibinfo{author}{S.~Freitag, B.~T. Cao, J.~Nini{\'c}, G.~Meschke},
\newblock \bibinfo{title}{Recurrent neural networks and proper orthogonal decomposition with interval data for real-time predictions of mechanised tunnelling processes},
\newblock \bibinfo{journal}{Computers \& Structures},
\bibinfo{volume}{207} (\bibinfo{year}{2018}),
\bibinfo{pages}{258--273},
\bibinfo{publisher}{Elsevier}.

\bibitem[{Guo and Hesthaven(2019)}]{guo2019data}
\bibinfo{author}{M.~Guo, J.~S. Hesthaven},
\newblock \bibinfo{title}{Data-driven reduced order modeling for time-dependent problems},
\newblock \bibinfo{journal}{Computer Methods in Applied Mechanics and Engineering},
\bibinfo{volume}{345} (\bibinfo{year}{2019}),
\bibinfo{pages}{75--99},
\bibinfo{publisher}{Elsevier}.

\bibitem[{Guemes et~al.(2019)}]{guemes2019sensing}
\bibinfo{author}{A.~Guemes, S.~Discetti, A.~Ianiro},
\newblock \bibinfo{title}{Sensing the turbulent large-scale motions with their wall signature},
\newblock \bibinfo{journal}{Physics of Fluids},
\bibinfo{volume}{31} (\bibinfo{year}{2019}),
\bibinfo{number}{12},
\bibinfo{publisher}{AIP Publishing}.

\bibitem[{Discetti et~al.(2018)}]{discetti2018estimation}
\bibinfo{author}{S.~Discetti, M.~Raiola, A.~Ianiro},
\newblock \bibinfo{title}{Estimation of time-resolved turbulent fields through correlation of non-time-resolved field measurements and time-resolved point measurements},
\newblock \bibinfo{journal}{Experimental Thermal and Fluid Science},
\bibinfo{volume}{93} (\bibinfo{year}{2018}),
\bibinfo{pages}{119--130},
\bibinfo{publisher}{Elsevier}.

\bibitem[{Guastoni et~al.(2021)}]{guastoni2021convolutional}
\bibinfo{author}{L.~Guastoni, A.~Güemes, A.~Ianiro, S.~Discetti, P.~Schlatter, H.~Azizpour, R.~Vinuesa},
\newblock \bibinfo{title}{Convolutional-network models to predict wall-bounded turbulence from wall quantities},
\newblock \bibinfo{journal}{Journal of Fluid Mechanics},
\bibinfo{volume}{928} (\bibinfo{year}{2021}),
\bibinfo{pages}{A27},
\bibinfo{publisher}{Cambridge University Press}.

\bibitem[{Tucker(1966)}]{Tucker66}
\bibinfo{author}{L.~R. Tucker},
\newblock \bibinfo{title}{Some mathematical notes on three-mode factor
  analysis},
\newblock \bibinfo{journal}{Psychometrika} \bibinfo{volume}{16}
  (\bibinfo{year}{1966}) \bibinfo{pages}{279--311}.

\bibitem[{De~Lathawer et~al.(2000{\natexlab{a}})De~Lathawer, De~Moor, and
  Vandewalle}]{DeLathawer}
\bibinfo{author}{L.~De~Lathawer}, \bibinfo{author}{B.~De~Moor},
  \bibinfo{author}{J.~Vandewalle},
\newblock \bibinfo{title}{On the best rank-$1$ and rank-$(r_1,r_2, \ldots,
  r_n)$ approximation of higher-order tensors},
\newblock \bibinfo{journal}{SIAM J. Matrix. Anal. Appl.}
  (\bibinfo{year}{2000}{\natexlab{a}}) \bibinfo{pages}{1324--1342}.

\bibitem[{De~Lathawer et~al.(2000{\natexlab{b}})De~Lathawer, De~Moor, and
  Vandewalle}]{DeLathawer0}
\bibinfo{author}{L.~De~Lathawer}, \bibinfo{author}{B.~De~Moor},
  \bibinfo{author}{J.~Vandewalle},
\newblock \bibinfo{title}{A multilinear singular value decomposition},
\newblock \bibinfo{journal}{SIAM J. Matrix. Anal. Appl.}
  (\bibinfo{year}{2000}{\natexlab{b}}) \bibinfo{pages}{1253--1278}.

\bibitem[{Zhou et~al.(2022)}]{fedformer}
\bibinfo{author}{T.~Zhou, Z.~Ma, Q.~Wen, X.~Wang, L.~Sun, R.~Jin},
\newblock \bibinfo{title}{Fedformer: Frequency enhanced decomposed transformer for long-term series forecasting},
\newblock \bibinfo{booktitle}{Proceedings of the International Conference on Machine Learning},
\bibinfo{pages}{27268--27286},
\bibinfo{year}{2022},
\bibinfo{organization}{PMLR}.

\bibitem[{Wu et~al.(2021)}]{autoformer}
\bibinfo{author}{H.~Wu, J.~Xu, J.~Wang, M.~Long},
\newblock \bibinfo{title}{Autoformer: Decomposition transformers with auto-correlation for long-term series forecasting},
\newblock \bibinfo{journal}{Advances in Neural Information Processing Systems},
\bibinfo{volume}{34},
\bibinfo{pages}{22419--22430},
\bibinfo{year}{2021}.













  
\end{thebibliography}

\end{document}